\documentclass[reqno,12pt]{amsart}

\usepackage{amsmath,amssymb,amsthm,amsfonts,amsxtra}
\usepackage{fullpage}

\usepackage{graphicx}
\usepackage{float}
\usepackage{subfig}
\usepackage{placeins}

\usepackage{url}

\numberwithin{equation}{section}
\begin{document}

\newtheorem{thm}{Theorem}[section]
\newtheorem{cor}{Corollary}[section]
\newtheorem{lem}{Lemma}[section]
\newtheorem{prop}{Proposition}[section]
\theoremstyle{definition}
\newtheorem{exmp}{Example}

\def\Ref#1{Ref.~\cite{#1}}

\def\Rnum{{\mathbb R}}
\def\sgn{{\mathop{\hbox{\rm sgn}}}}
\def\const{\text{const.}}

\def\max{\text{max}}
\def\min{\text{min}}
\def\inflect{\text{inflect}}

\def\sech{{\mathop{\hbox{\rm sech}}}}
\def\arctanh{{\mathop{\hbox{\rm arctanh}}}}
\def\cn{{\mathop{\hbox{\rm cn}}}}
\def\sn{{\mathop{\hbox{\rm sn}}}}

\def\Esp{{\mathcal E}}

\tolerance=10000
\allowdisplaybreaks[4]
\predisplaypenalty=0

\title{Travelling wave solutions on a non-zero background for the generalized Korteweg-de Vries equation}

\author{
Stephen C. Anco$^1$
\lowercase{\scshape{and}}
HamidReza Nayeri$^2$
\lowercase{\scshape{and}}  
Elena Recio$^{3}$
\\\\
\lowercase{\scshape{
${}^1$Department of Mathematics and Statistics\\
Brock University\\
St. Catharines, ON L2S3A1, Canada}}
\\\\
\lowercase{\scshape{
${}^2$Department of Physics\\
Brock University\\
St. Catharines, ON L2S3A1, Canada}} 
\\\\
\lowercase{\scshape{
${}^3$Department of Mathematics\\
University of C\'adiz\\
Puerto Real, C\'adiz, Spain, 11510}}\\
}

\thanks{$^1$sanco@brocku.ca, $^2$hn15ir@brocku.ca, $^3$elena.recio@uca.es}

\begin{abstract}
For the generalized $p$-power Korteweg-de Vries equation,
all non-periodic travelling wave solutions with non-zero boundary conditions 
are explicitly classified for all integer powers $p\geq 1$.
These solutions are shown to consist of:
bright solitary waves and static humps on a non-zero background for odd $p$;
dark solitary waves on a non-zero background and kink (shock) waves for even $p$
in the defocusing case;
pairs of bright/dark solitary waves on a non-zero background,
and also bright and dark heavy-tail waves (with power decay) on a non-zero background, for even $p$ in the focusing case. 
An explicit physical parameterization is given for each type of solution
in terms of the wave speed $c$, background size $b$, and wave height/depth $h$.
The allowed kinematic region for existence of the solutions is derived, 
and other main kinematic features are discussed. 
Analytical formulas are presented in the higher power cases $p=3,4$,
which are compared to the integrable cases $p=1,2$. 
\end{abstract}

\maketitle

\section{Introduction}\label{sec:intro}

Many different nonlinear dispersive wave phenomena are described by 
the Korteweg-de Vries (KdV) equation and its nonlinear generalizations
\begin{equation}\label{gkdv}
u_t +\alpha u^p u_x +\beta u_{xxx}=0,
\quad
p\geq 1
\end{equation}
called the gKdV equation,
where $p$ is the nonlinearity power, 
$\alpha$ is the interaction coefficient,
and $\beta$ is the dispersion coefficient.
Here $t$ is time, $x$ is position, and $u$ is the wave amplitude. 

The KdV equation has the weakest nonlinearity, $p=1$,
and was first derived from the theory of shallow water waves \cite{KordeVri,Joh} .
The next stronger nonlinearity, $p=2$, is the mKdV equation.
It can be derived from the theory of internal water waves \cite{Joh}
and is connected to the KdV equation by a Miura transformation \cite{Miu}.
Both of these equations are integrable systems (see e.g. \Ref{AblCla})
possessing a bi-Hamiltonian structure and soliton solutions, 
and they are now known to arise in numerous other physical applications
(see \Ref{Cri,DauPey} and references therein)
such as 
Alfven waves and ion-acoustic waves in plasmas, 
pressure waves in bubbly liquids,
atmospheric waves,
voltage waves in electrical transmission lines, 
compression waves in granular chains and molecular peptide chains,
heat pulses in solids,
and traffic flow models. 

Higher-order nonlinearities, $p\geq 2$, are of physical interest in the study of
anharmonic lattice waves \cite{Ono}, 
internal waves in stratified fluids \cite{GriPelPol},
and ion acoustic waves in plasmas \cite{Mam,MusSha}.
Solutions $u(t,x)$ have interesting different behaviour depending on $p$,
and there has been considerable mathematical interest in this behaviour 
such as stability of solitary waves \cite{MarMer2008},
global (long-time) existence of initial-value solutions \cite{MasSeg,FarLinPas,Str}, 
and blow-up of solutions for critical and supercritical nonlinearities \cite{Mer,MarMerRap2014}. 

Solitary wave solutions have been studied for the gKdV equation for arbitrary $p$,
and the resulting generalization of the familiar $\sech$-form solitary wave of the KdV equation 
is well known \cite{Zab,Wei,HerTak,Li}:
\begin{equation}\label{gkdv-solitary}
u(t,x) =
(\tfrac{(p+2)(p+1)}{2}c/\alpha)^{1/p}\, \sech^{2/p}\big(\tfrac{p}{2}\sqrt{c\beta}\,(x-ct)\big)
\end{equation}
where $c=\const$ is the wave speed. 
In contrast,
very little study has been done on other types of
non-periodic travelling wave solutions of the gKdV equation \eqref{gkdv}, 
specifically solitary waves on a non-zero background and kink (shock) waves.

A \emph{solitary wave on a background} is a travelling wave whose amplitude
decays to a non-zero value exponentially in $|x|$.
This type of solitary wave arises when non-zero boundary conditions are considered on $u(t,x)$ as $|x|\rightarrow \infty$.
In keeping with standard terminology,
when the peak of a solitary wave is above/below the background,
it is called a \emph{bright}/\emph{dark} wave. 

A \emph{kink wave} (shock wave) is a travelling wave whose amplitude
exponentially approaches two different values for $x\rightarrow \pm \infty$.
Relative to the point on the wave where its convexity changes, 
one asymptotic end is ``bright'' while the other asymptotic end is ``dark''. 

Solitary waves with non-zero boundary conditions have been found previously
for the KdV and mKdV equations by two different methods:
direct integration of the travelling wave ODE \cite{JefKak,Mar}; 
and use of the inverse scattering transform \cite{Au-YeuFunAu1983,Au-YeuFunAu1984,KamSpiKon}
which comes from the special integrability properties of those equations. 
For the gKdV equation,
solitary waves with non-zero boundary conditions 
have been mentioned briefly by Zabusky \cite{Zab},
going back to his original work with Kruskal on the soliton solutions of the KdV equation,
but no detailed study was carried out. 

In the present paper,
we undertake a full study of solitary waves on a non-zero background and kink (shock) waves
for the gKdV equation \eqref{gkdv} for all integer powers $p\geq 1$.

Several main results are obtained:
\begin{itemize}
\item
Solitary waves on a non-zero background 
are parameterized in a physical form in terms of
their wave speed $c$, background size $b$, and wave height/depth $h$,
for all powers $p\geq 1$. 
\item
The kinematic region in $(c,b)$ as well as in $(h,b)$ for existence of these waves is derived.
\item 
For all odd powers $p$,
solitary waves can move in an opposite direction when $b<0$
in comparison to the direction of the solitary wave with zero background,
and a special case is a static hump with $c=0$. 
\item
For all even powers $p$,
there are both bright and dark solitary waves. 
In the (defocusing) case $\alpha/\beta <0$, 
these waves propagate in the opposite direction
compared to the (focusing) case $\alpha/\beta >0$. 
\item
A certain limit of the (focusing case) solitary waves
yields a \emph{heavy-tail wave} that has a power decay instead of an exponential decay. 
\item
Analytical expressions for solitary wave solutions, including the heavy-tail wave solutions, are obtained for $p=3,4$ in terms of elliptic functions.
\end{itemize}
and also 
\begin{itemize}
\item
Kink waves, which exist only for even powers $p$,
are parameterized in terms of their wave speed $c$ and asymptotic value $\pm b$.\item
An explicit relation between $c$ and $b$ for existence of these waves is derived.
\item
Explicit kink wave solutions are obtained for $p=4$ in terms of elementary functions.
\end{itemize}  

The rest of the paper is organized as follows.

In section~\ref{sec:prelims},
we discuss the conserved integrals for mass, momentum, and energy of the gKdV equation
and use them to obtain a first-order separable ODE for all travelling wave solutions
by applying a symmetry multi-reduction method \cite{AncGan2020}. 
We also discuss how to evaluate the conserved mass, momentum, and energy of these solutions 
directly from the ODE (without the need to know the solutions in an explicit form). 

In section~\ref{sec:gkdv-classify},
we study the gKdV travelling wave ODE qualitatively by a standard nonlinear oscillator method \cite{JefKak,Gol}
and classify all types of (non-periodic) travelling waves with non-zero boundary conditions for odd and even $p$.
A summary of this analysis is listed in Table~\ref{table:types}. 

In section~\ref{sec:gkdv-odd-p},
we consider odd powers $p$ and study the solitary wave solutions
by using a parameterization in terms of $b$ and $h$. 
We derive the solitary waves for $p=3$ analytically in terms of elliptic functions
and compared them to the KdV case. 

In section~\ref{sec:gkdv-even-p-defocus}, 
we consider even powers $p$ in the defocusing case 
and study the solitary wave solutions and kink (shock) solutions
by again using a parameterization in terms of $b$ and $h$.
For $p=4$, we derive solitary wave solutions in terms of elliptic functions,
and explicit kink (shock) solutions in terms of elementary functions. 
We compare these new solutions to their counterparts in the defocusing mKdV case. 

In section~\ref{sec:gkdv-even-p-focus}, 
we consider even powers $p$ in the focusing case. 
We first study the pairs of bright/dark solitary wave solutions, 
using the previous parameterization,
and then we explain how heavy-tail waves arise in a special limit. 
For $p=4$, we derive bright/dark solitary wave solutions and heavy-tail wave solutions
in terms of elliptic functions. 
We compare these new solutions to their counterparts in the focusing mKdV case. 

Some technical steps in the derivation of the solutions for general odd and even $p$ 
are put in an Appendix.

Lastly, we make some concluding remarks in section~\ref{sec:conclusion}.

\section{Conservation laws and travelling wave ODE}\label{sec:prelims}

We will first review the point symmetries and local conservation laws
of the gKdV equation \eqref{gkdv} for arbitrary $p\neq 0$.
(There are additional symmetries and conservation laws
in the integrable cases $p=1,2$, which will not be relevant here.)
Next, we will derive the ODE for travelling wave solutions
and show how the conservation laws yield
first integrals that reduce this ODE to a first-order separable differential equation. 
Finally, we will discuss the nonlinear-oscillator formulation of the ODE,
which is starting point for our subsequent study of its solutions
for all odd and even powers $p\geq 1$. 

The gKdV equation for arbitrary $p\neq 0$ is invariant 
under a group of point symmetries that are generated by
time translations $t\rightarrow t+\epsilon$ 
and space translations $x\rightarrow x+\epsilon$,
with parameter $\epsilon\in\Rnum$;
scaling $x\rightarrow \lambda x$, $t\rightarrow \lambda^3 t$, $u\rightarrow \lambda^{-2/p} u$,
with parameter $\lambda\in\Rnum^+$;
and separate reflections $t\rightarrow -t$ and $x\rightarrow -x$,
as well as $u\rightarrow -u$ when $p$ is an even integer.

For odd powers $p$, 
the sign of $\alpha/\beta$ is inessential 
because it can be changed by the reflection $u\rightarrow -u$.
In contrast, for even powers $p$,
the sign cannot be changed, 
because $u\rightarrow -u$ is a symmetry of the equation,
and the two distinct possibilities for the sign of $\alpha/\beta$ 
distinguish different types of behaviour:
$\alpha/\beta <0$ is called the \emph{defocusing} case,
while $\alpha/\beta >0$ is called the \emph{focusing} case,
which correspond to the sign property of the conserved energy, 
as shown next.

The local conservation laws of the gKdV equation for arbitrary $p\neq 0$
consist of \cite{AncBlu2002a} mass, momentum, and energy.
These have the form of a continuity equation
\begin{equation}
(D_t T + D_x \Phi)|_\Esp =0
\end{equation}  
holding for all gKdV solutions, denoted by the space $\Esp$, 
where $T$ denotes a conserved density and $\Phi$ denotes a spatial flux,
which are functions of $t$, $x$, $u$ and derivatives of $u$. 
Here $D_t$ and $D_x$ are total derivatives with respect to $t,x$, 
which indicate the use of the chain rule in the differentiations. 
(If a specific solution $u(t,x)$ is substituted into the conservation law, 
then these differentiations become partial derivatives.)

The densities and fluxes for mass, momentum, and energy are given by 
\begin{gather}
\label{gkdv-mass}
T=u,
\quad
\Phi = \tfrac{1}{p+1}\alpha u^{p+1} +\beta u_{xx} ;
\\
\label{gkdv-mom}
T=u^2,
\quad
\Phi = \tfrac{2}{p+2}\alpha u^{p+2} -\beta u_x^2 + 2\beta uu_{xx} ;
\\
\label{gkdv-ener}
T=\tfrac{1}{2}u_x^2 -\tfrac{1}{(p+1)(p+2)}(\alpha/\beta) u^{p+2}, 
\quad
\Phi = -\tfrac{1}{2}\beta (u_{xx} +\tfrac{1}{p+1}(\alpha/\beta)u^{p+1})^2 -u_t u_x .
\end{gather}
Each of these local conservation laws gives rise to a conserved integral
\begin{equation}
\frac{d}{dt}\int_\Rnum T\;dx = -\Phi\Big|_{-\infty}^{+\infty} =0
\end{equation}
for gKdV solutions $u(t,x)$ with sufficient decay as $|x|\rightarrow \infty$.
The conserved mass, momentum, and energy integrals are respectively given by 
\begin{align}
\mathcal{M} & = \int_\Rnum u\;dx , 
\label{mass}
\\
\mathcal{P} & = \int_\Rnum u^2\; dx , 
\label{mom}
\\
\mathcal{E} & = \int_\Rnum \tfrac{1}{2}u_x^2 -\tfrac{1}{(p+1)(p+2)}(\alpha/\beta) u^{p+2}\; dx . 
\label{ener}
\end{align}
Their names come from the physical meaning that they have in the KdV case $p=1$
for the situation of shallow water waves \cite{Joh}
where $u$ represents both the velocity and the height of the wave (in scaled units).

Note that the energy integral \eqref{ener} is
non-negative when $\alpha/\beta<0$ 
and otherwise has an indefinite sign when $\alpha/\beta>0$. 
The sign property of the energy is well-known to be
crucially related to the long-time behaviour of initial-value solutions \cite{Tao}. 
When the sign is indefinite, conservation of energy does not preclude blow-up of $|u|$ and $|u_x|$ 
caused by concentration of the energy density, which is referred to as focusing. 
When the sign is non-negative, conservation of energy prevents such blow-up, 
which is referred to as defocusing. 

Travelling waves are solutions of the form 
\begin{equation}\label{travellingwave}
u=U(\xi),
\quad
\xi=x-ct,
\quad
c=\const
\end{equation}
which are invariant under the combined translation symmetry
$t\rightarrow t+\epsilon$ and $x\rightarrow x+\epsilon c$.
A \emph{solitary wave} is a travelling wave that is localized in the sense that it decays to zero exponentially in $x$.
If a travelling wave is not exponentially localized
but instead decays to zero as a power of $x$,
it is called a \emph{heavy-tail wave}.
This type of travelling wave can emerge from rational solutions,
as happens for $p=2$. 
However, for higher even powers, 
there are heavy-tail waves that have a non-rational form. 

In the integrable cases $p=1,2$, it is well known that 
all solitary waves preserve their speed and shape in interactions. 
Solitary waves with these features are called solitons.

Substitution of the travelling wave expression \eqref{travellingwave}
into the gKdV equation \eqref{gkdv} yields a third-order nonlinear ODE
\begin{equation}\label{gkdv-travelwave-ODE}
\beta U''' + (\alpha U^p-c)U'=0 .
\end{equation}
First integrals of this ODE can be derived directly from 
the conservation laws \eqref{gkdv-mass}, \eqref{gkdv-mom}, \eqref{gkdv-ener}
evaluated for travelling waves.
This is seen by noting that the total derivatives become derivatives with respect to the travelling wave variable $\xi$:
$D_t = -c\frac{d}{d\xi}$ and $D_x = \frac{d}{d\xi}$,
and hence, $D_t T + D_x \Phi = \frac{d}{d\xi}(\Phi -cT)$. 
Thus, we obtain a first integral 
\begin{equation}
\Psi = \Phi -cT =C=\const
\end{equation}  
holding for all travelling wave solutions \eqref{travellingwave}.
This yields 
\begin{align}
\label{gkdv-FI1}
\beta U'' +\tfrac{1}{p+1}\alpha U^{p+1} -cU =C_1 = \const, 
\\
\label{gkdv-FI2}
2\beta UU'' -\beta U'{}^2 + \tfrac{2}{p+2}\alpha U^{p+2} -cU^2 =C_2 = \const, 
\\
-\tfrac{1}{2}\beta (U'' +\tfrac{1}{p+1}(\alpha/\beta)U^{p+1})^2 +\tfrac{1}{2}c U'{}^2 +\tfrac{1}{(p+1)(p+2)}(\alpha/\beta)c U^{p+2} = C_3 = \const.
\label{gkdv-FI3}
\end{align}
These three first integrals have the physical meaning that they describe the spatial flux of mass, momentum, and energy in a reference frame moving with the travelling wave.
Since none of them contain $\xi$ explicitly, and the ODE is third-order, 
they cannot be functionally independent. 
It is easy to see that they are related by $cC_2 + C_1^2 = -2\beta C_3$. 

The reduction of conservation laws has a symmetry explanation, 
which is worth noting. 
Just as travelling waves are invariant solutions under the Galilean symmetry 
$t\rightarrow t+\epsilon$ and $x\rightarrow x+\epsilon c$, 
first integrals are invariant conservation laws. 
In general, they can be found by a systematic method \cite{AncGan2020}
by seeking multipliers of a certain form adapted to the symmetry.

\subsection{Travelling wave ODE}

When the first integrals \eqref{gkdv-FI1} and \eqref{gkdv-FI2} are combined,
they yield a reduction of the travelling wave ODE to a first-order separable differential equation
\begin{equation}\label{gkdvODE}
\tfrac{1}{2}\beta U'{}^2 +\tfrac{1}{(p+2)(p+1)}\alpha U^{p+2} -\tfrac{1}{2}cU^2-C_1 U +\tfrac{1}{2}C_2 =0 . 
\end{equation}
This ODE \eqref{gkdvODE} has the form of a nonlinear oscillator equation
$\tfrac{1}{2} U'{}^2 +V(U) = E$
where $\tfrac{1}{2} U'{}^2$ is the kinetic energy, 
$V(U)=\tfrac{1}{(p+2)(p+1)}(\alpha/\beta) U^{p+2} -\tfrac{1}{2}(c/\beta) U^2 - M U$
is the potential,
and $E=-\tfrac{1}{2}C_2/\beta$ is the oscillator energy,
with $M=C_1/\beta$ being a free parameter.
Every solution $U(\xi)$ will correspond to an oscillator motion 
in which $U$ goes between the turning points given by roots of $V(U)=E$ 
where $U'$ is zero. 

Hereafter, for convenience,
by scaling $t$ and $x$ we will put $\beta =1$;
further, by scaling $u$ we will put $\alpha = 1$ when $p$ is odd and $\alpha =\pm 1$ when $p$ is even. 
This yields the scaled (dimensionless) form of the gKdV equation, 
\begin{equation}\label{scaled-gkdv}
u_t +\sigma u^p u_x + u_{xxx}=0,
\quad
\sigma = \begin{cases}
1, & p \text{ is odd }
\\
\pm 1, & p \text{ is even }
\end{cases}
\end{equation}
where, in the same terminology used for the gKdV equation, 
$\sigma=1$ is called the focusing case,
and $\sigma=-1$ is called the defocusing case.
The corresponding scaled form of the travelling wave ODE \eqref{gkdvODE}
in nonlinear oscillator form looks like 
\begin{gather}
\tfrac{1}{2} U'{}^2 +V(U) = E,
\label{scaled-oscil-ode}
\\
V(U)=\tfrac{1}{(p+2)(p+1)}\sigma U^{p+2} -\tfrac{1}{2}c U^2 - M U . 
\label{scaled-oscil-V}
\end{gather}

All solutions of the travelling wave ODE can be classified by studying
the shape of the potential function \eqref{scaled-oscil-V}. 
This approach \cite{JefKak,Gol} is an alternative to a phase-plane analysis
and has the advantage that the qualitative analysis of the potential
is very straightforward and intuitive.
It has been used successfully in recent work \cite{PrzAnc2017} to derive all travelling wave solutions of a highly nonlinear fourth-order wave equation. 

Solitary wave solutions will arise when one turning point coincides with a local maximum of $V(U)$;
kink wave solutions will arise when two turning points coincide with a pair of local maximums of $V(U)$ with the same height. 
For both types of solution,
the turning points give the minimum and maximum wave amplitudes, $U_\min$ and $U_\max$, of $U(\xi)$. 

When $U_\max$ is a peak of the wave $U(\xi)$, 
then the background on which the wave propagates is $b=U_\min$; 
when $U_\min$ is a trough (inverted peak) of the wave $U(\xi)$, 
then the background on which the wave propagates is $b=U_\max$. 
A standard terminology in the physics and applied mathematics literature
for these two types of solitary waves are, respectively,
\emph{bright} and \emph{dark}. 
The height/depth of the wave is $h=U_\max-U_\min>0$. 

Each solution $U(\xi)$ is implicitly given by the quadrature
\begin{equation}\label{quadrature}
\int^{U}_{U_0} \frac{dU}{\sqrt{E-V(U)}} = \pm\sqrt{2} (\xi-\xi_0)
\end{equation}
where $U_0=U(\xi_0)$ is an arbitrary constant between $U_\min$ and $U_\max$,
and where $\xi_0$ can be chosen to be $0$ by translation invariance. 
A convenient choice of $U_0$ is,  
for solitary waves, the turning point at which $V(U_0)$ is a local maximum,
and for kink waves, the point at which $V'(U_0)$ is a local maximum.

\subsection{Conserved quantities for travelling waves}

The mass, momentum, and energy for travelling wave solutions are given by substitution of $u=U(\xi)$
into the respective conserved integrals \eqref{mass}, \eqref{mom}, \eqref{ener}. 
It may be possible to evaluate these integrals directly 
if the explicit form of a solution is known. 
Alternatively, they can be expressed in the form of a quadrature by using the 
nonlinear oscillator form \eqref{scaled-oscil-ode}--\eqref{scaled-oscil-V} of 
the travelling wave ODE. 
The quadrature then can be evaluated without the need to know solutions in an explicit form. 

This ODE method consists of converting the integration with respect to $\xi$ over $\Rnum$
into an equivalent integration with respect to $U$ where the transformed integration domain is given by the monotonically increasing or decreasing parts of the solution $U(\xi)$. 
For a solution that has a single peak $U=b\pm h$ on a background $b$, 
the conserved integral of a density $T$ can thereby be expressed as 
\begin{equation}
\int_\Rnum T|_{u=U(x-ct)}\;dx = \int_\Rnum T|_{u=U(\xi)}\;d\xi
= \int_{b\pm h}^{b} 2\,T|_{u=U} \frac{dU}{U'}
= \int_{b\pm h}^{b} \frac{\sqrt{2}\,T|_{u=U}}{\sqrt{E-V(U)}}\;dU
\end{equation}
through use of the ODE. 
This integral, however, will in general be infinite when $b\neq 0$,
due to the contribution from the background. 
Nevertheless, 
a finite conserved integral can be obtained by 
subtraction of the constant background term in the conserved density, 
$T|_{u=b}$. 

This yields the following conserved integral expression
\begin{equation}
\mathcal{C} = \int_\Rnum \big( T|_{u=U(\xi)} - T|_{u=b} \big)\;d\xi
= \sqrt{2}\int_{b\pm h}^{b} \frac{T|_{u=U}-T|_{u=b}}{\sqrt{E-V(U)}}\; dU . 
\end{equation}
Note that $\mathcal{C}$ can be evaluated 
without the solution $u=U(\xi)$ being given in an explicit form. 

The corresponding conserved integrals for mass, momentum, and energy are given by
\begin{align}
\mathcal{M} & = \sqrt{2}\int_{b\pm h}^{b} \frac{U-b}{\sqrt{E-V(U)}}\;dU,
\label{U-mass}
\\
\mathcal{P} & = \sqrt{2}\int_{b\pm h}^{b} \frac{U^2 -b^2}{\sqrt{E-V(U)}}\;dU,
\label{U-mom}
\\
\mathcal{E} & = \sqrt{2}\int_{b\pm h}^{b} \bigg( \sqrt{E-V(U)} 
-\tfrac{1}{(p+1)(p+2)}\frac{\sigma(U^{p+2}-b^{p+2})}{\sqrt{E-V(U)}} \bigg)\;dU. 
\label{U-ener}
\end{align}
It is easy to adjust the integration domain in these integrals
if $u=U(\xi)$ does not have a single peak.

\section{gKdV equation with non-zero boundary conditions}\label{sec:gkdv-classify}

We will now classify all travelling wave solutions with non-zero boundary conditions for the gKdV equation \eqref{scaled-gkdv}
by studying the potential \eqref{scaled-oscil-V}
in the nonlinear oscillator form of the travelling wave ODE \eqref{scaled-oscil-ode}.
An outline of this analysis for the KdV case $p=1$ and the defocusing mKdV case $p=1$ can be found in \Ref{JefKak}. 

When $p$ is odd, then similarly to the KdV case, 
a potential well exists only if $V(U)$ has at least one local minimum and one local maximum. 
In this situation, the travelling waves of interest will be 
solitary waves on zero/non-zero backgrounds,
which arise for $E=V_\max$. 
As we will see later, 
no kink waves exist because $V(U)$ turns out to never have 
a pair of local maximums with equal heights. 

When $p$ is even, then similarly to the mKdV case,
in the focusing case a potential well always exists because $V(U)$ is upward, 
whereas in the defocusing case a potential well exists only if $V(U)$ has at least two local maximums, because $V(U)$ is downward. 
Solitary waves on zero/non-zero backgrounds will arise 
whenever $V(U)$ has at least one local minimum and one local maximum,
and heavy-tail waves on a non-zero background will arise in the limiting case
when $V(U)$ has a local minimum and an inflection. 
Kink waves will arise 
whenever $V(U)$ has at least one pair of local maximums with equal heights. 

The critical points of $V(U)$ are determined by 
\begin{equation}\label{gkdv-V'}
0= V'(U)=\tfrac{\sigma}{p+1}U^{p+1}-cU-M,
\quad
\sgn(V''(U)) = \sgn(\sigma U^p -c)   . 
\end{equation}
To proceed, we look at the separate cases of odd and even powers $p$
to find the shape of the potential well
and to classify the types of solutions of interest in each case.
The derivation and properties of these solutions will then be discussed in subsequent sections.

\subsection{Odd-power gKdV potential}\label{subsec:gkdv-odd}

For odd $p\geq 1$,  
the critical points of the potential \eqref{scaled-oscil-V} are the real roots of 
the polynomial \eqref{gkdv-V'} with even degree $p+1\geq 2$. 
To determine the number and type of these critical points, 
we will write the polynomial in a scaled form in terms of 
\begin{equation}
\tilde U = \sgn(c) U/|c|^{1/p}, 
\quad
\tilde M= M/c^{1+1/p} , 
\end{equation}
giving 
$0=\tfrac{1}{2(q+1)} \tilde U^{2q+2} -\tilde U-\tilde M$
where $p=2q+1$, $q=0,1,2,\ldots$. 
This equation can be rearranged 
\begin{equation}\label{gkdv-oddp-V'}
\tfrac{1}{2(q+1)}\tilde U^{2q+2}=\tilde U+\tilde M.
\end{equation}
where the intersection points of the two sides are precisely the scaled critical points of the potential. 
Note the left side is an non-negative even-power (convex) function whose slope is $\tilde U^{2q+1}$ 
while the right side is a linear function whose slope is $1$. 
Therefore, graphically, the number of intersections is either two, one, or none
depending on $\tilde M$. 
There will be one intersection when $\tilde U+\tilde M$ is tangent to $\tfrac{1}{2(q+1)}\tilde U^{2q+2}$,
which occurs when their slopes are equal: $1=\tilde U^{2q+1}$. 
This intersection is given by $\tilde U=1$ and $\tilde M= -\tfrac{2q+1}{2(q+1)}$. 
Consequently, 
if $\tilde M < -\tfrac{2q+1}{2(q+1)}$, there will be no intersection,
whereas if $\tilde M > -\tfrac{2q+1}{2(q+1)}$, there will be two intersections, 
which lie to the right and left of $\tilde U=1$. 

Since $p$ is odd, we have 
\begin{equation}
c^{1+1/p}= (c^{p+1})^{1/p}\geq 0,
\quad
c^{1/p} = \sgn(c) |c|^{1/p},
\quad
\sgn(\tilde M)=\sgn(M) .
\end{equation}
Hence for $M\leq -\tfrac{p}{p+1}c^{1+1/p}$,
the potential $V(U)$ has at most one critical point,
and in this case there is no potential well. 
Hereafter we consider the case
\begin{equation}\label{gkdv-oddp-Mcond}
M> -\tfrac{p}{p+1}c^{1+1/p},
\end{equation}
whereby $V(U)$ has two critical points, $U=U_1$ and $U=U_2$, 
which consist of a local maximum and a local minimum 
because $V''(U)=U^p -c$ changes sign at $U=\sgn(c) |c|^{1/p}$
which lies between the critical points. 
In particular, 
\begin{equation}\label{gkdv-oddp-critpoints}
U_1 <\sgn(c) |c|^{1/p} < U_2, 
\quad
V''(U_1)<0,
\quad
V''(U_2)>0,
\end{equation}
where the minimum point $U_2$ sits to the right of the maximum point $U_1$. 

\begin{figure}[h]
\centering
\includegraphics[trim=2cm 14cm 6cm 1cm,clip,width=0.48\textwidth]{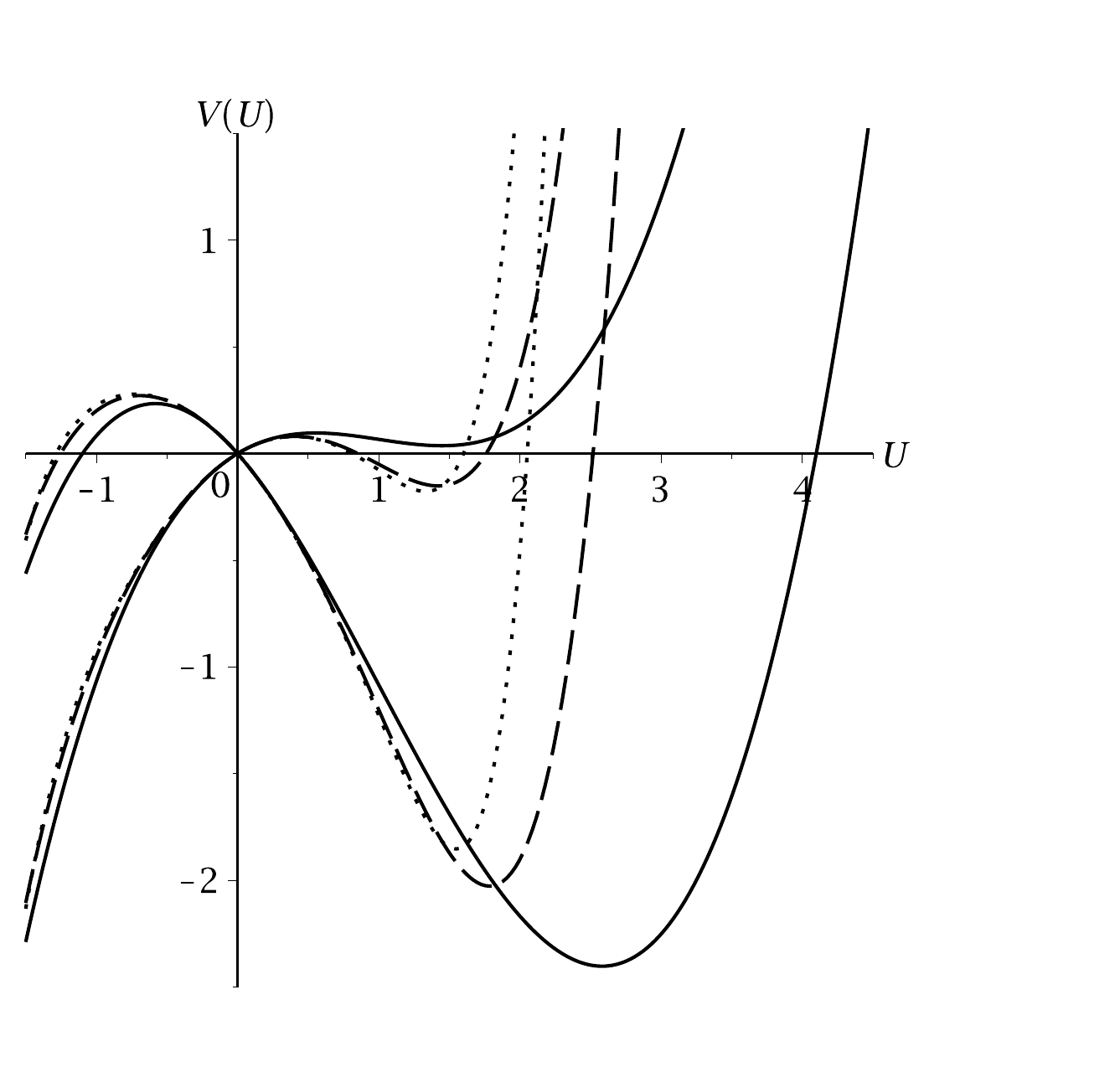} 
\quad
\includegraphics[trim=2cm 14cm 6cm 1cm,clip,width=0.48\textwidth]{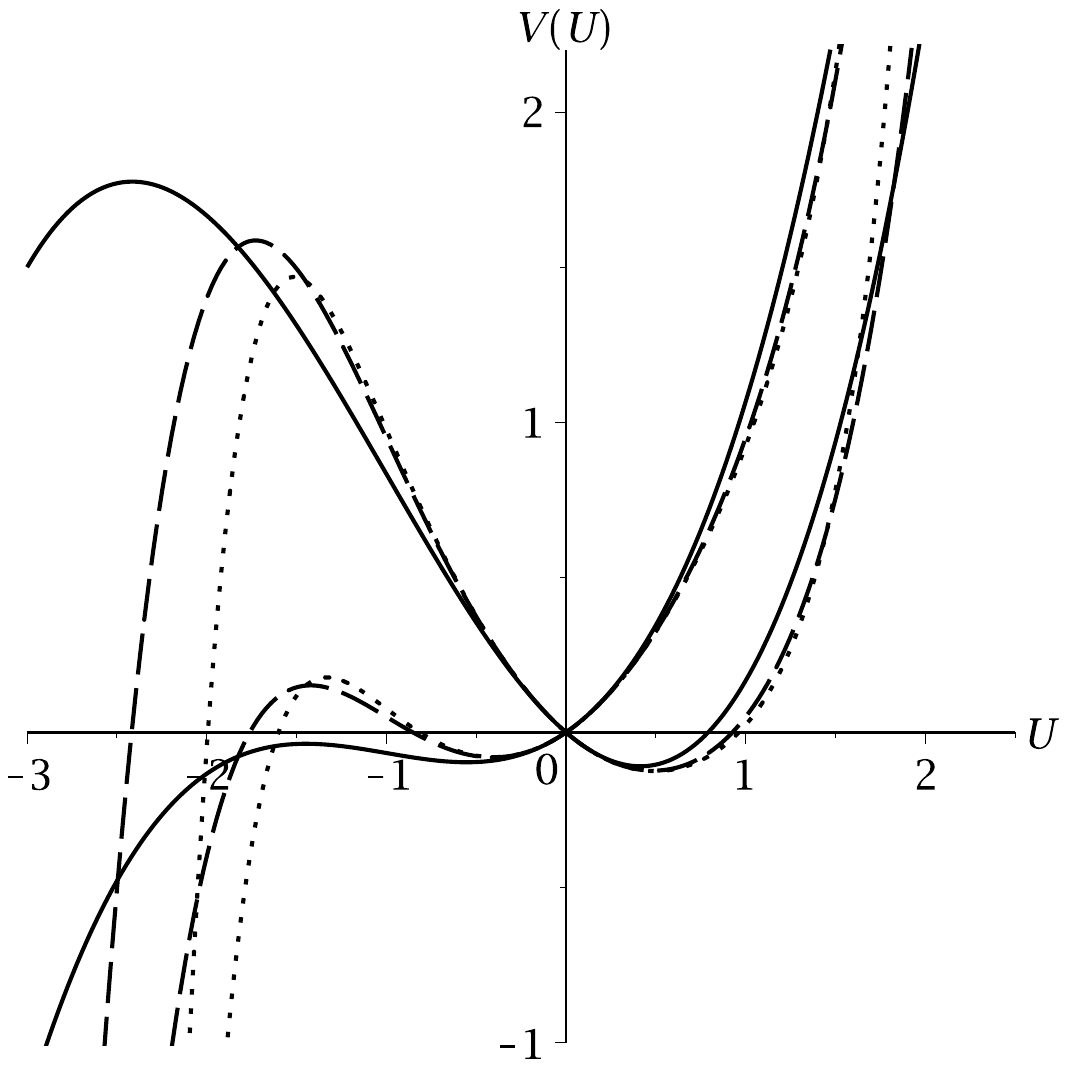}
\caption{gKdV potential for positive $c$ (left) and negative $c$ (right), with $M>0$ and $0>M>-\tfrac{p}{p+1}|c|^{1+1/p}$,
for different odd powers: $p=1$ solid; $p=3$ dash; $p=5$ dot.}\label{gkvdfig-potential-odd-p}
\end{figure}

Thus, when $M$ satisfies the condition \eqref{gkdv-oddp-Mcond}, 
there is a potential well defined by 
\begin{equation}\label{gkdv-oddp-potentialwell}
V(U)=\tfrac{1}{(p+1)(p+2)} U^{p+2}-\tfrac{1}{2}cU^2-MU \leq E_\max ,
\quad
E_\max= V_\max=V(U_1) . 
\end{equation}
One rim is the maximum at $U_1$,
which is an asymptotic turning point, 
and the other rim is a turning point.
See Fig.~\ref{gkvdfig-potential-odd-p}.
This potential supports bright solitary waves
but not dark solitary waves or kink waves.
Also, it does not support heavy-tail waves.
Interestingly, there is no sign restriction on $c$ when $U_1 <0$, 
and hence solitary waves on a negative background can propagate
in either direction or can be static.

For later, note that the potential \eqref{gkdv-oddp-potentialwell} 
changes sign, $V\to -V$, under the reflection
\begin{equation}\label{odd-p-V-reflect}
(U,M,c)\to (-U,M,-c) . 
\end{equation}

\subsection{Even-power gKdV potential}\label{subsec:gkdv-even}

For even $p\geq 2$, 
the critical points of the potential \eqref{scaled-oscil-V} are the real roots of 
the polynomial \eqref{gkdv-V'} with odd degree $p+1\geq 3$. 
We will write this polynomial in a scaled form in terms of 
\begin{equation}
\tilde U = U/|c|^{1/p}, 
\quad
\tilde M= M/|c|^{1+1/p}, 
\end{equation}
which gives
$0= \tfrac{\sigma}{2q+1} \tilde U^{2q+1} -\sgn(c)\tilde U-\tilde M$
where $p=2q$, $q=1,2,3,\ldots$, 
and where $\sigma =\pm 1$ in the focusing and defocusing cases, respectively. 
This equation can be rearranged so that we have 
a linear function on one side and an odd-power (non convex) function on the other side:
\begin{equation}\label{gkdv-evenp-V'}
\tfrac{\sigma}{2q+1}\tilde U^{2q+1}=\sgn(c)\tilde U+\tilde M.
\end{equation}
The intersection points of the two sides are precisely the scaled critical points of the potential. 
Note the left side has slope $\sigma \tilde U^{2q}$ 
while the right side has $\sgn(c)$.
Therefore, graphically, we have one, two, or three different intersections,
$\tilde U_*$, depending on $\tilde M$ and $\sgn(c)$.
Two intersections occur when $\sgn(c)\tilde U+\tilde M$ is tangent to $\tfrac{\sigma}{2q+1}\tilde U^{2q+1}$,
which requires that the slopes on the left and right sides are equal: 
$\sigma\tilde U^{2q}= \sgn(c)$,
and hence $\sigma=\sgn(c)$. 
One of these two intersections is the tangential intersection point 
given by $\tilde U_*=\pm 1$, 
with $\tilde M=\mp\tfrac{2q\sigma}{2q+1}$. 
As a consequence, 
if $|\tilde M| <\tfrac{2q}{2q+1}$ and $\sigma=\sgn(c)$, 
there will be three intersections, 
with one point lying in an interval
$-1<\tilde U_*<0$ when $\sigma\tilde M>0$
or $0<\tilde U_*<1$ when $\sigma\tilde M<0$, 
while the other two points lie in separate intervals $\tilde U_*>1$ and $\tilde U_*<-1$.
In contrast, 
if $|\tilde M| > \tfrac{2q}{2q+1}$ or $\sigma\neq\sgn(c)$, 
there will be only one intersection. 

Hence when $|M|> \tfrac{p}{p+1}|c|^{1+1/p}$ or $\sigma\neq\sgn(c)$, 
the potential $V(U)$ has a single critical point,
which consists of a local minimum if $\sigma=1$
or a local maximum if $\sigma=-1$. 
In the latter situation, no potential well exists, 
while in the former situation, 
there is a potential well but it does not support either solitary wave solutions or kink wave solutions. 

Next we consider the conditions
\begin{equation}\label{gkdv-evenp-Mcond}
|M| < \tfrac{p}{p+1}|c|^{1+1/p},
\quad
\sgn(c) = \sigma , 
\end{equation}
whereby the potential $V(U)$ has three critical points
$U=U_1,U_2,U_3$. 

\begin{figure}[h]
\centering
\includegraphics[trim=2cm 14cm 6cm 1cm,clip,width=0.48\textwidth]{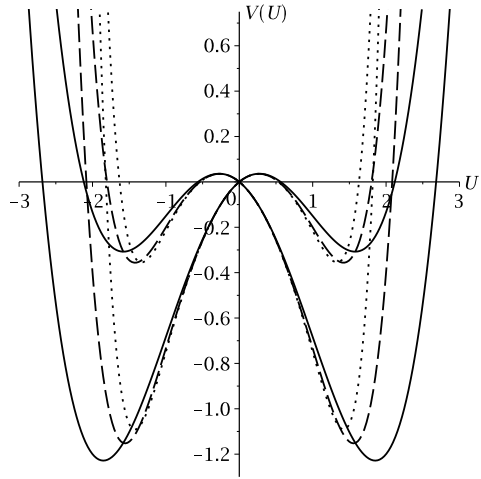} 
\caption{gKdV focusing potential ($c>0$), with $0<M<\tfrac{p}{p+1}c^{1+1/p}$ and $0>M>-\tfrac{p}{p+1}c^{1+1/p}$, 
for different even powers: $p=2$ solid; $p=4$ dash; $p=6$ dot.}\label{gkvdfig-potential-even-p-focus}
\end{figure}

In the focusing case, $\sigma=1$, 
$V(U)$ is upward and defines a potential well 
\begin{equation}\label{gkdv-evenp-focus-potentialwell}
V(U)=\tfrac{1}{(p+1)(p+2)} U^{p+2}-\tfrac{1}{2}cU^2-MU \leq E_\max=\infty . 
\end{equation}
This potential well has two local minimums and a local maximum
\begin{equation}\label{gkdv-evenp-focus-critpoints}
(+)\qquad
\begin{aligned}
& U_1<-c^{1/p}, 
\quad
U_3> c^{1/p},
\quad
V''(U_1)>0,
\quad
V''(U_3)>0,
\\
& 
|U_2| <c^{1/p}, 
\quad
V''(U_2)<0,
\end{aligned}
\end{equation}
with 
\begin{equation}\label{gkdv-evenp-focus-c-M-cond}
c> ((1+1/p)|M|)^{p/(p+1)} >0, 
\end{equation}
where the maximum point $U_2$ sits between the two minimum points $U_1$ and $U_3$.
These critical points also can be shown to satisfy
\begin{equation}\label{gkdv-evenp-focus-critpoints-rel}
\sgn(U_2)=-\sgn(M),
\quad
\sgn(U_3 -|U_1|) = \sgn(M)
\end{equation}
by a refinement of the graphical analysis of $V(U)$, 
using continuity in $M$.
The shape of $V(U)$ thereby indicates that the maximum point is an asymptotic turning point, 
and hence the potential well supports pairs of bright/dark solitary waves.
Kink waves are not supported since $V(U)$ never has two local maximums. 
See Fig.~\ref{gkvdfig-potential-even-p-focus}. 
Note that, under the reflection \eqref{even-p-V-reflect}, 
the critical points transform as $(U_1,U_2,U_3)\to(-U_3,-U_2,-U_1)$.

\begin{figure}[h]
\centering
\includegraphics[trim=2cm 14cm 6cm 1cm,clip,width=0.48\textwidth]{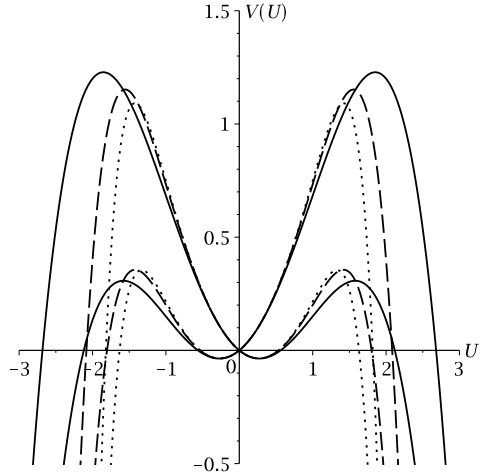} 
\caption{gKdV defocusing potential ($c<0$), with $0<M<\tfrac{p}{p+1}|c|^{1+1/p}$ and $0>M>-\tfrac{p}{p+1}|c|^{1+1/p}$, 
for different even powers: $p=2$ solid; $p=4$ dash; $p=6$ dot.}\label{gkvdfig-potential-even-p-defocus}
\end{figure}

In the defocusing case, $\sigma =-1$, 
$V(U)$ is downward and has two local maximums and a local minimum. 
In particular, when $M\neq 0$, we have 
\begin{equation}\label{gkdv-evenp-defocus-critpoints-Mnot0}
(-)\qquad
\begin{aligned}
& 
\sgn(M)\,U_1>|c|^{1/p}, 
\quad
\sgn(M)\,U_3<-|c|^{1/p},
\\
& 
V''(U_1)<0,
\quad
V''(U_3)<0,
\quad
V(U_1)<V(U_3), 
\\
&
|U_2| <|c|^{1/p} , 
\quad
V''(U_2)>0,
\end{aligned}
\end{equation}
with
\begin{equation}\label{gkdv-evenp-defocus-c-M-cond}
c< -\big((1+\tfrac{1}{p})|M|\big)^{p/(p+1)} <0, 
\end{equation}
where the minimum point $U_2$ sits between the two maximum points $U_1$ and $U_3$,
and where the relative heights of the maximum points directly follows 
from the term $-MU$ in the expression \eqref{scaled-oscil-V} for the potential.
Similarly to the focusing case,
these critical points also satisfy
\begin{equation}\label{gkdv-evenp-defocus-critpoints-rel}
\sgn(U_2)=\sgn(M),
\quad
|U_3| > |U_1| . 
\end{equation}
The potential well is defined by 
\begin{equation}\label{gkdv-evenp-defocus-potentialwell}
V(U)=-\tfrac{1}{(p+1)(p+2)} U^{p+2}-\tfrac{1}{2}cU^2-MU \leq E_\max =V(U_1)
\end{equation}
which has a rim on each side. 
One rim is the maximum at $U_1$,
which is an asymptotic turning point, 
and the other rim is a turning point. 
Solitary waves, but not kink waves, 
are supported by this potential well.
The solitary waves are bright when $M<0$ and dark when $M>0$.
See Fig.~\ref{gkvdfig-potential-even-p-defocus}. 

A useful observation for later is that 
both the focusing potential \eqref{gkdv-evenp-focus-potentialwell} and the defocusing potential \eqref{gkdv-evenp-defocus-potentialwell} 
are invariant, $V\to V$, under the reflection 
\begin{equation}\label{even-p-V-reflect}
(U,M,c)\to (-U,-M,c) . 
\end{equation}

In the special case $M=0$, 
the potential is symmetric about $U=0$, and we have 
\begin{equation}\label{gkdv-evenp-defoc-critpoints-Mis0}
(-)\qquad
\begin{aligned}
& U_1=-U_3 = ((p+1)|c|)^{1/p}, 
\quad
V''(U_1)<0,
\quad
V''(U_3)<0,
\quad
V(U_1)=V(U_3), 
\\
&
U_2 =0, 
\quad
V''(U_2)>0 . 
\end{aligned}
\end{equation}
The potential well is defined by 
\begin{equation}\label{gkdv-evenp-defocus-Mis0-potentialwell}
V(U)=-\tfrac{1}{(p+1)(p+2)} U^{p+2}-\tfrac{1}{2}cU^2 \leq E_\max =V(U_1)
\end{equation}
with $-U_1\leq U\leq U_1$.
Kink waves, but not solitary waves, are supported by this potential well. 

\begin{figure}[h]
\centering
\includegraphics[trim=2cm 14cm 6cm 1cm,clip,width=0.48\textwidth]{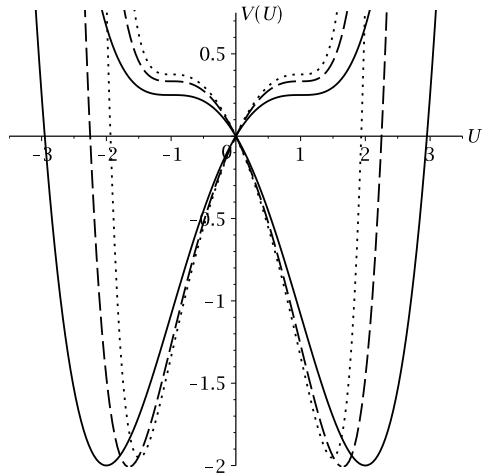} 
\caption{gKdV focusing potential ($c>0$) with $|M|=\tfrac{p}{p+1}|c|^{1+1/p}$ 
for different even powers: $p=2$ solid; $p=4$ dash; $p=6$ dot.}\label{gkvdfig-potential-even-p-focus-inflect}
\end{figure}

Finally,
we also consider the limiting case of the conditions \eqref{gkdv-evenp-Mcond}
given by 
\begin{equation}\label{gkdv-evenp-Mlimit}
|M| =\tfrac{p}{p+1}|c|^{1+1/p},
\quad
\sgn(c) = \sigma , 
\end{equation}
whereby the potential $V(U)$ has only two critical points. 
This occurs when a local minimum and a local maximum coalesce 
into an inflection. 
In the focusing case, 
the potential well \eqref{gkdv-evenp-focus-potentialwell}
will now have a local minimum and an inflection
\begin{equation}\label{gkdv-evenp-foc-Mlimit-critpoints}
(+)\qquad
\begin{aligned}
& \sgn(M)\,U_0>c^{1/p},
\quad
V''(U_0)>0,
\\
& 
U_2=-\sgn(M)\,c^{1/p}, 
\quad
V''(U_2)=0,
\end{aligned}
\end{equation}
which arises from either $U_1=U_2$ when $\sgn(M)=1$, or $U_3=U_2$ when $\sgn(M)=-1$. 
This potential well supports a heavy-tail wave,
which is bright when $M>0$ and dark when $M<0$
and which has 
\begin{equation}
c = \big((1+\tfrac{1}{p})|M|\big)^{p/(p+1)} > 0. 
\end{equation}
See Fig.~\ref{gkvdfig-potential-even-p-focus-inflect}.
In defocusing case, 
the potential well \eqref{gkdv-evenp-defocus-potentialwell}
will similarly have a local maximum and an inflection,
but it will not support any bounded solutions. 

\subsection{Summary of types of travelling wave solutions with non-zero boundary conditions}

The different types of allowed non-periodic travelling wave solutions with non-zero boundary conditions are summarized in Table~\ref{table:types}. 
In particular, this classification is complete, namely, 
no other types of bounded non-periodic travelling waves exist. 

\begin{table}[!htbp]
\centering
\caption{Types of gKdV travelling waves with non-zero boundary conditions in 
the potential $V(U)=\tfrac{1}{(p+2)(p+1)}\sigma U^{p+2} -\tfrac{1}{2}c U^2 - M U$.}
\label{table:types}
\begin{tabular}{|c|r|c|c|c|}
\hline
$p$ & $\sigma$ & Type of Solution & $c$ & Conditions on $M$ \\
\hline \hline
odd 
& $1$
& bright solitary wave
& $\gtrless 0$
& $M> -\tfrac{p}{p+1}|c|^{1+1/p}$
\\
\hline
even 
& $-1$
& bright solitary wave
& $<0$
& $0>M >-\tfrac{p}{p+1}|c|^{1+1/p}$
\\
\hline
even 
& $-1$
& dark solitary wave
& $<0$
& $0<M<\tfrac{p}{p+1}|c|^{1+1/p}$
\\
\hline
even
& $1$
& bright \& dark solitary waves
& $>0$
& $|M| < \tfrac{p}{p+1}c^{1+1/p}$
\\
\hline
even
& $-1$
& kink (shock) wave
& $<0$
& $M=0$ 
\\
\hline
even
& $1$
& dark/bright heavy-tail wave
& $>0$
& $|M| =\tfrac{p}{p+1}c^{1+1/p}$
\\
\hline
odd
& $1$
& static hump
& $0$
& $M>0$
\\
\hline
\end{tabular}
\end{table}

\section{Odd-power gKdV travelling waves with non-zero boundary conditions}\label{sec:gkdv-odd-p}

The analysis in the preceding section has shown that
when the nonlinearity power $p$ is an odd integer then 
the gKdV equation supports solitary waves on non-zero backgrounds, 
but not any other kind of bounded non-periodic waves.

We will now study the kinematic properties of the solitary wave solutions 
and derive analytical formulas in the cases $p=1,3$. 
The discussion will use a physical parameterization given by the speed $c$, background $b$, and height/depth $h$ of the solitary waves. 
The results for $p=3$ are new. 
For $p=1$, which is the KdV case, 
the solitary wave solution with $b\neq 0$ is known in the literature \cite{JefKak,Au-YeuFunAu1983}, 
however its kinematics using a physical parameterization has been less well studied. 

To proceed,
we start from the gKdV potential \eqref{gkdv-oddp-potentialwell} for an arbitrary odd power $p$,
under the condition \eqref{gkdv-oddp-Mcond} 
which gives the existence of a local maximum at $U_1$ and a local minimum at $U_2$.
These critical points are the roots of the polynomial \eqref{gkdv-V'} with $\sigma=1$ in terms of $(c,M)$, 
and they have the properties \eqref{gkdv-oddp-critpoints}.

\begin{figure}[h]
\centering
\includegraphics[trim=2cm 12cm 5cm 1cm,clip,width=0.45\textwidth]{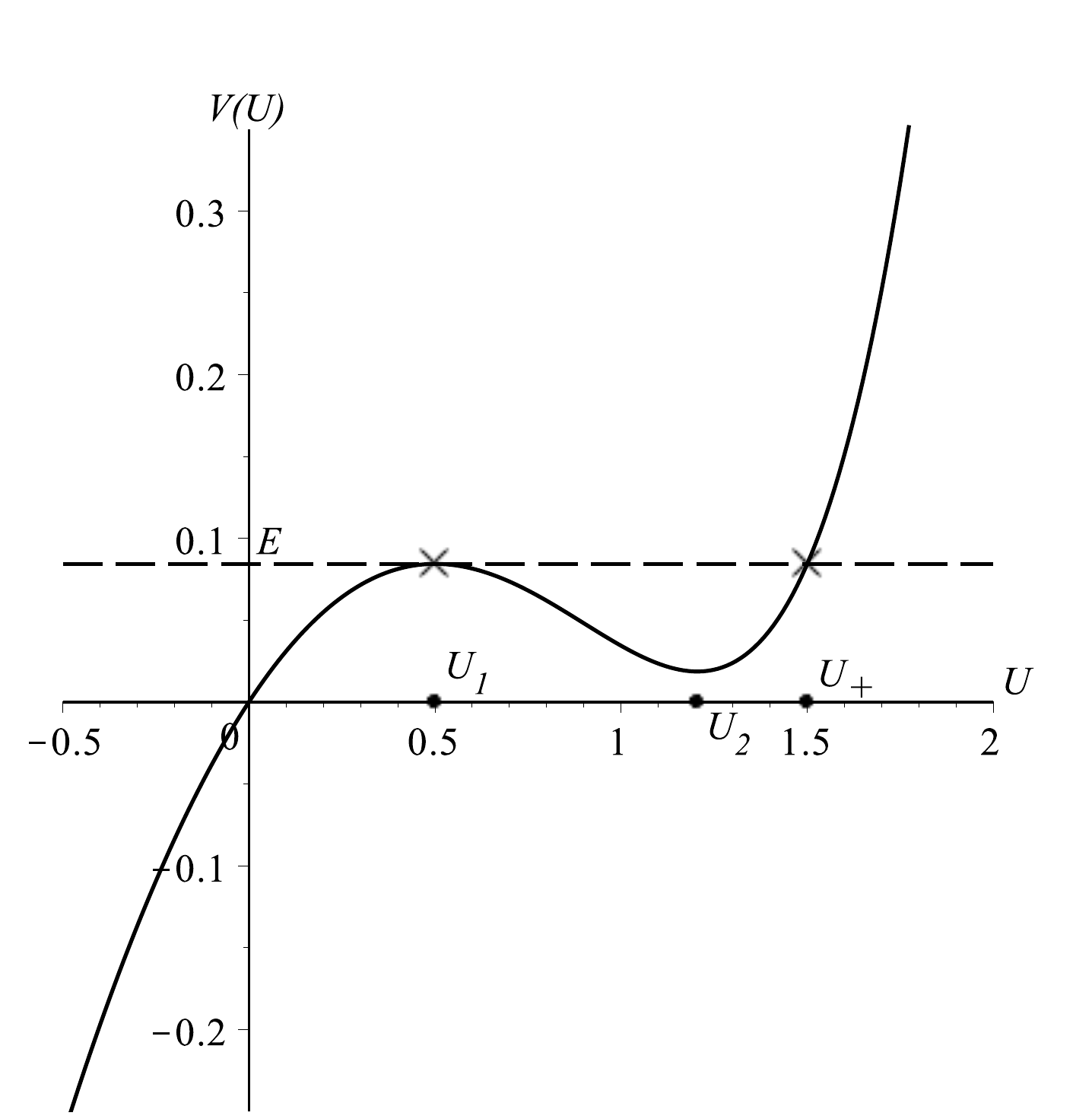} 
\quad
\includegraphics[trim=2cm 12cm 5cm 1cm,clip,width=0.45\textwidth]{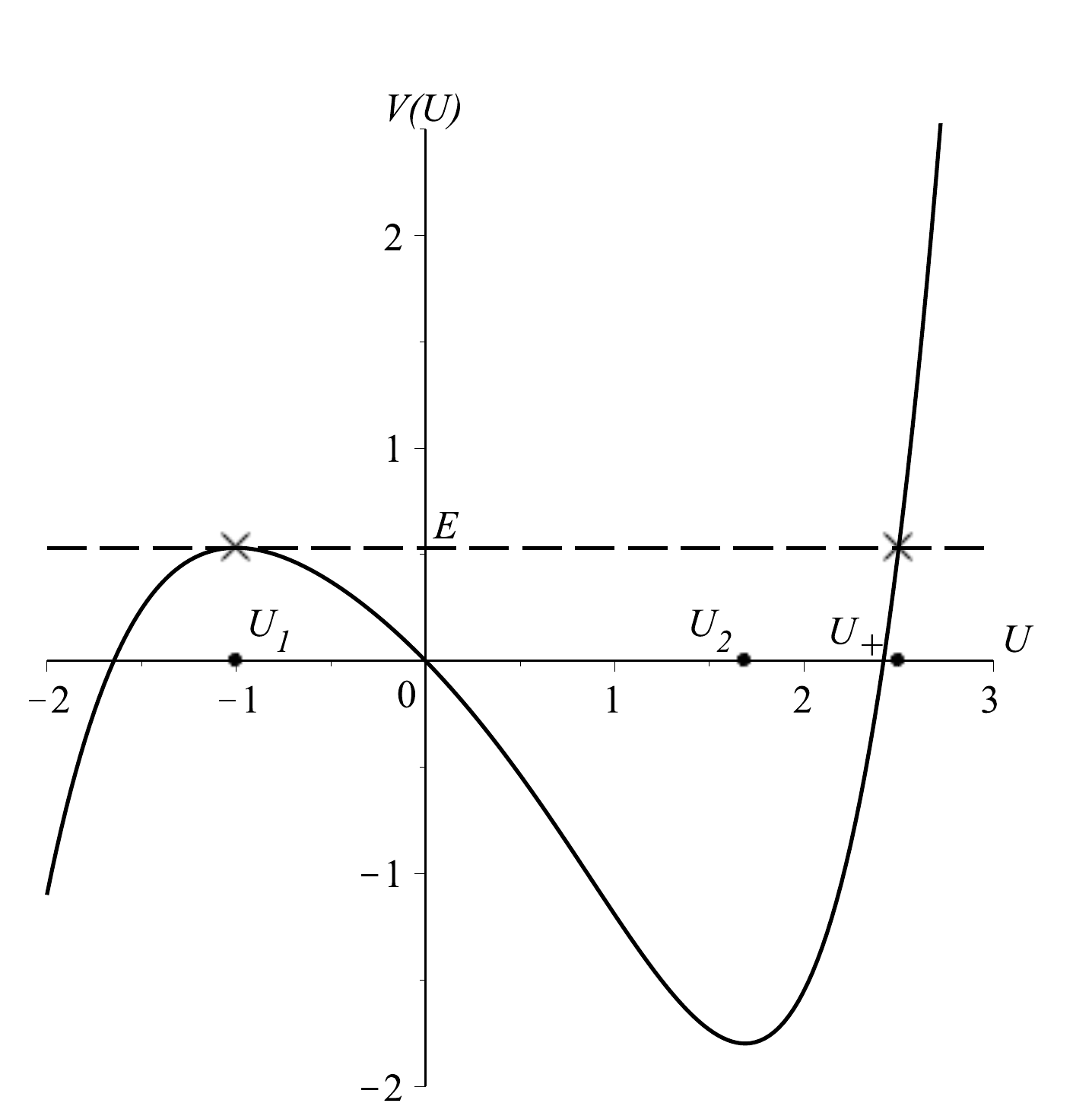}
\caption{gKdV potential well for $p=3$, with $c>0$ (left) and $c<0$ (right).}\label{gkvdfig-potentialwell-pis3}
\end{figure}

Solitary wave solutions $U(\xi)$ are obtained from
the resulting nonlinear oscillator equation \eqref{scaled-oscil-ode}
by taking $E=V_\max=V(U_1)$. 
Then the potential \eqref{scaled-oscil-V} has the factorization 
\begin{equation}\label{gkdv-oddp-energyeqn}
V(U_1) - V(U)
=(U-U_1)^2(U_+-U) W(U), 
\end{equation}
where 
\begin{equation}\label{gkdv-oddp-tp}
U_+ >U_2
\end{equation}
is the turning point, 
and where $W(U)$ is a polynomial that has even degree $p-1$ 
and that is positive on $U_1\leq U\leq U_+$. 
This potential well is shown in Fig.~\ref{gkvdfig-potentialwell-pis3}. 

Hence,
the quadrature \eqref{quadrature} for $U(\xi)$ is given by 
\begin{equation}\label{gkdv-oddp-solitary-quadrature}
\int^{U_+}_{U} \frac{dU}{(U-U_1)\sqrt{(U_+-U)W(U)}} = \sqrt{2}\, |\xi|,
\quad
U_1\leq U\leq U_+ .
\end{equation}
When $p=3$, the integral can be evaluated in terms of elliptic functions
as shown later, 
but for $p\geq 5$, the integral has no explicit evaluation in general.
For $p=1$, $W(U)$ is a positive constant, 
and so the integral has a straightforward evaluation in terms of elementary functions,
yielding the KdV soliton (on a background).

\subsection{Kinematics}\label{sec:solitarywave-oddp}

For any odd power $p\geq 1$, 
the quadrature \eqref{gkdv-oddp-solitary-quadrature}
implicitly defines a family of the solitary wave solutions $U(\xi)$, 
parameterized by $(c,M)$.
This family has not been previously studied when the background is non-zero and the power is greater than $3$. 
Here we will examine the main kinematic properties. 

In this family, 
the wave peak is $U_+$ and the background (asymptote) is $b=U_1$,
and so the wave height is $h=U_+-U_1>0$,
while $U(\xi)$ is an even function of $\xi$.
The width of the wave is proportional to $w=2/\sqrt{c-U_1^p}$,
which is obtained from an asymptotic expansion of the quadrature as $U\to U_1$,
combined with $V''(U_1)=2(U_+-U_1)W(U_1)=U_1^p-c$ from the factorization \eqref{gkdv-oddp-energyeqn}.
The background $b=U_1$ can be positive or negative, 
with the speed $c$ satisfying the kinematic condition $c>b^p$.
Thus, for $b>0$, the speed is necessarily positive,
whereas for $b<0$, the speed is allowed to be negative, since $p$ is odd. 
In both cases, the speed must satisfy 
\begin{equation}
c> \sgn(b) |b|^p ,
\quad
b\gtrless 0 . 
\end{equation}
Here $U_1$ and $U_+$ are implicitly given in terms of $(c,M)$
by the polynomial equations
\begin{equation}
\tfrac{1}{p+1}U_1^{p+1} - cU_1 -M=0,
\quad
\tfrac{1}{(p+1)(p+2)}(U_+^{p+2} -U_1^{p+2}) -\tfrac{1}{2}c(U_+^2 - U_1^2) -M(U_+ -U_1) =0
\end{equation}
which come from $V'(U_1)=0$ and $V(U_1)=V(U_+)$. 
These equations can be used to obtain an equivalent implicit physical parameterization 
in terms of $(c,b)$: 
\begin{equation}\label{gkdv-oddp-solitary-b-c}
M= \tfrac{1}{p+1} b^{p+1} -c b,
\quad
U_1=b,
\quad
U_+ = bz, 
\quad
b^p (\tfrac{1}{p+2}S_{p+2}(z)  -1) -\tfrac{p+1}{2}c(z -1) =0,
\end{equation}
where 
\begin{equation}\label{Spoly}
S_n(z)= \sum_{j=1}^{n} z^{n-j}= (z^n -1)/(z-1),
\quad
n\in\mathbb{Z}^+
\end{equation}  
is a polynomial of degree $n-1$ in $z$. 

An explicit, useful physical parameterization of the solution family 
can be achieved by writing out the quadrature \eqref{gkdv-oddp-solitary-quadrature} 
in terms of $(b,h)$. 
As shown in the Appendix, 
this yields
\begin{equation}\label{gkdv-oddp-solitary-bh-quadrature}
\int^{h+b}_{U} \frac{dU}{(U-b)\sqrt{(h+b-U)W(U)}} = \sqrt{2}\, |\xi|,
\quad
b\leq U\leq h+b
\end{equation}
with
\begin{equation}\label{gkdv-oddp-solitary-bh-W}
W(U) =
\tfrac{1}{(p+1)(p+2)}|b|^{p-1}\big( 
R_{p}(1+h/b) + R_{p-1}(1+h/b)(U/b) + \cdots + R_{1}(1+h/b)(U/b)^{p-1}
\big)
\end{equation}
where
\begin{equation}\label{Rpoly}
R_n(z)= \sum_{j=1}^{n} j z^{n-j}= (z^{n+1} -(n+1)z+ n)/(z-1)^2,
\quad
n\in\mathbb{Z}^+
\end{equation}  
is a polynomial of degree $n-1$ in $z$. 
Moreover, the speed and the width are given by the explicit expressions
\begin{equation}\label{gkdv-oddp-solitary-bh-c}
c = \tfrac{2}{(p+1)(p+2)}|b|^{p-1}b R_{p+1}(1+h/b)
= \tfrac{2}{(p+1)(p+2)} ((h+b)^{p+2} -(p+2)hb^{p+1} -b^{p+2})/h^2
\end{equation}
and 
\begin{equation}\label{gkdv-oddp-solitary-bh-w}
\begin{aligned}
w & = 2/\sqrt{|b|^{p-1}b(\tfrac{2}{(p+1)(p+2)} R_{p+1}(1+h/b) -1)}
\\
&= \sqrt{2(p+1)(p+2)}h/( 2((h+b)^{p+2}-b^{p+2}) -2(p+2)hb^{p+1} -(p+1)(p+2)h^2b^p )
\end{aligned}
\end{equation}
which satisfy the relation
\begin{equation}\label{gkdv-oddp-solitary-bh-cwrel}
w^2(c-b^p)=4 . 
\end{equation}

In this explicit parameterization,
the kinematic features of the solitary wave solution
are directly tied to the properties of the polynomial $R_{p+1}(1+h/b)$.
Specifically, as a function of $h/b$, 
it has an odd degree $p$,
is monotonic increasing,
has a negative root $h/b=-r_{p+1}$ with $2< r_{p+1} \leq 3$ depending on $p$
(with $r_2=3$ when $p=1$),
and reaches the value $\tfrac{(p+1)(p+2)}{2}$ at $h/b=0$. 

As a consequence,
the solitary wave solution \eqref{gkdv-oddp-solitary-bh-quadrature}
is well-defined 
with no restriction on the parameters $(h,b)$ other than $h>0$.
The wave speed has the following properties:
$c$ is an increasing function of $h$; 
for any fixed $b>0$, 
$c$ is positive, 
whereas for any fixed $b<0$,
$c$ is negative for $h\ll |b|$ and goes through zero when $h=r_{p+1}|b|$. 
In both cases, $w$ decreases to $0$ as $h$ increases, 
and increases to $\infty$ as $h$ goes to $0$. 

When $c\neq 0$, the solution $u=U(\xi)$ is 
a \emph{bright solitary wave on a positive/negative background}. 
In the special case when $c=0$,
the solution $u=U(\xi)$ describes a \emph{static hump} 
on a negative background.
Its height is $h=r_{p+1}|b|$ and its width is $w = 2/\sqrt{|b|^{p}}$. 

The well-known solitary wave on a zero background, $b=0$, 
can be obtained from the quadrature \eqref{gkdv-oddp-solitary-bh-quadrature}
in the limit $b\to0$ as follows.
First, note $W(U) \to \tfrac{1}{(p+1)(p+2)} h^{p-1} S_{p}(U/h)$
since $R_n(z) \to z^{n-1}$ as $|z|\to \infty$,
where $S_n(z)= \sum_{j=1}^{n} z^{j-1} = (z^n-1)/(z-1)$
is a polynomial of degree $n-1$ in $z$
for $n\in\mathbb{Z}^+$.
In this limit, the quadrature becomes 
\begin{equation}
\begin{aligned}
& \int^{h}_{U} \frac{dU}{U\sqrt{(1-U/h) S_p(U/h)}}
  = \int^{h}_{U} \frac{dU}{U\sqrt{1-(U/h)^p}}
  = \tfrac{2}{p}\arctanh\big(\sqrt{1-(U/h)^p}\big)
\\
&= \tfrac{2}{(p+1)(p+2)}h^{p} |\xi|,
\quad
0\leq U\leq h .
\end{aligned}
\end{equation}
This yields the explicit solution 
\begin{equation}\label{gkdv-oddp-soliton}
U(\xi) =  h\, \sech^{2/p}\Big(\tfrac{p}{\sqrt{2(p+1)(p+2)}}h^{p/2} \xi\Big)
\end{equation}
which is a solitary wave with zero background, $b=0$.
Its speed and width can be obtained from the limit $b\to 0$
of relations \eqref{gkdv-oddp-solitary-bh-c} and \eqref{gkdv-oddp-solitary-bh-w}:
\begin{equation}\label{gkdv-oddp-soliton-cw}
c=\tfrac{2}{(p+2)(p+1)} h^p>0,
\quad
w= \tfrac{\sqrt{2(p+1)(p+2)}}{p}h^{-p/2} . 
\end{equation}
These expressions reproduce the scaled form of the solitary wave \eqref{gkdv-solitary} with $\alpha=\beta=1$.

It is interesting to observe that the solitary wave on a negative background $b<0$ 
can propagate in the opposite direction, $c<0$,
compared to the zero-background soliton. 
In contrast,
on a positive background $b>0$, 
the solitary wave propagates only with positive speed, $c>0$.

\subsection{KdV ($\boldsymbol{p=1}$) solitary waves on a background}

In the KdV case, $p=1$, 
the critical points $U_1,U_2$, and the turning point $U_+$, can be found explicitly,
and the corresponding factorization of the potential \eqref{gkdv-oddp-energyeqn} has 
$W=\tfrac{1}{6}$. 

\begin{figure}[h]
\centering
\includegraphics[trim=3cm 14cm 8cm 2cm,clip,width=0.32\textwidth]{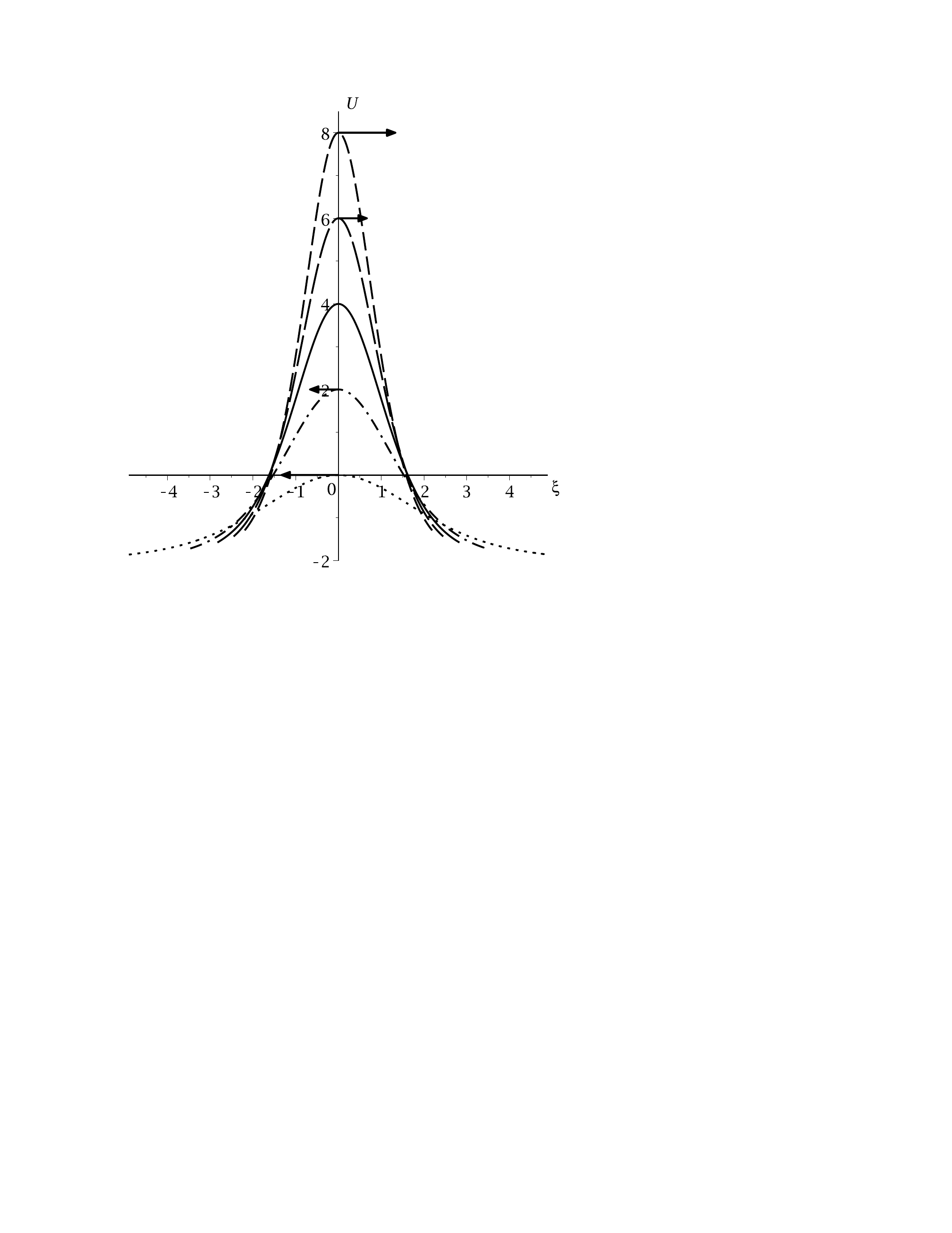}
\quad
\includegraphics[trim=3cm 15cm 8cm 3.5cm,clip,width=0.46\textwidth]{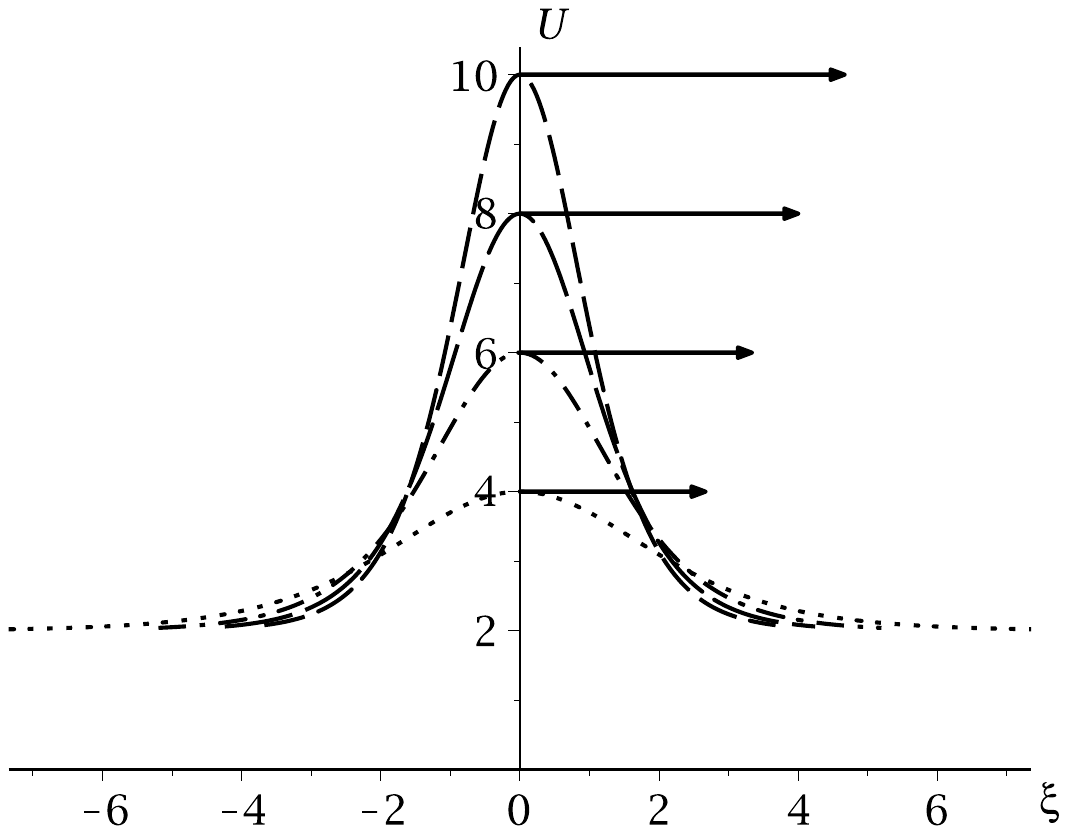}
\caption{ KdV solitary waves with different heights on a negative background (left) and a positive background (right). Arrows indicate direction and speed of the waves.}\label{gkdvfig-kdv-profile-heights}
\end{figure}

\begin{figure}[h]
\centering
\includegraphics[trim=4cm 15cm 10cm 2cm,clip,width=0.3\textwidth]{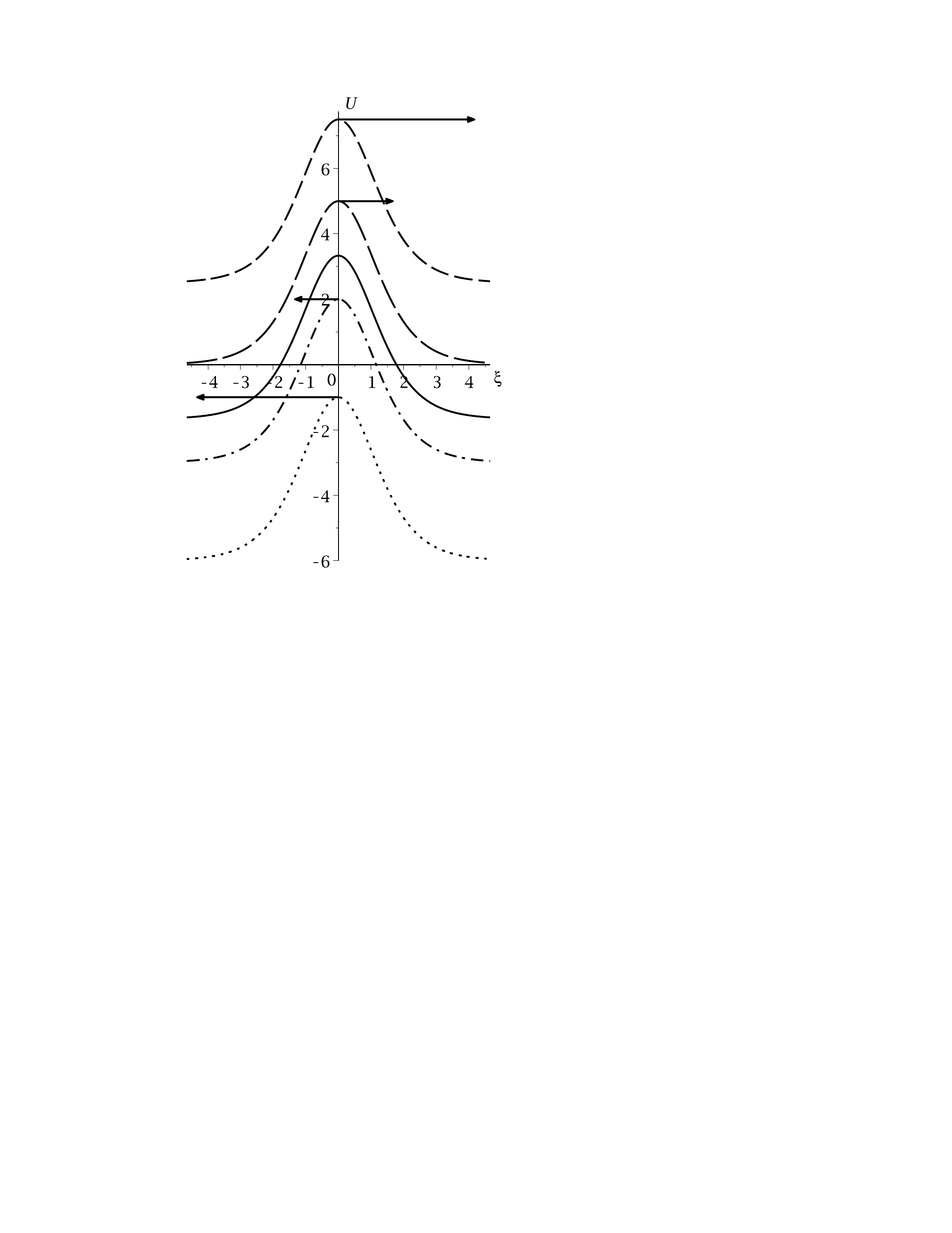}
\caption{KdV solitary waves with the same height on different backgrounds. Arrows indicate direction and speed of the waves.}\label{gkdvfig-kdv-profile-backgrounds}
\end{figure}

The quadrature \eqref{gkdv-oddp-solitary-bh-quadrature} 
for the explicit physical parameterization of the solitary wave solution yields 
\begin{equation}\label{kdv-solitary} 
U(\xi)= h\,\sech^2\big(\tfrac{1}{6}\sqrt{3h}\,\xi\big) + b
\end{equation}
in terms of the height $h>0$ and the background $b$, where 
\begin{equation}
c= \tfrac{1}{3}h + b
\end{equation}
is the speed. 
This shows that the solution consists of 
linearly superimposing the zero-background soliton profile 
on an arbitrary background, $b$, 
and adjusting the speed by adding $b$ to it.
See Figs.~\ref{gkdvfig-kdv-profile-heights} and~\ref{gkdvfig-kdv-profile-backgrounds}.

The solution family \eqref{kdv-solitary} describes
a \emph{bright solitary wave on a background},
parameterized by $(b,h)$. 
It was first derived with a non-physical parameterization \cite{JefKak,Au-YeuFunAu1983}
and has been studied recently in the context of 
soliton mean field theory in hydrodynamics \cite{MaiAndFraElHoe}. 

Because $h$ can be expressed explicitly in terms of the speed $c$ and the background $b$, 
the alternative physical parameterization \eqref{gkdv-oddp-solitary-quadrature} and \eqref{gkdv-oddp-solitary-b-c}
also has an explicit form
\begin{equation}\label{kdv-solitary-physical}
U(\xi) =3(c-b) \sech^2\big(\tfrac{1}{2}\sqrt{c-b}\,\xi\big)+b, 
\quad
c-b >0. 
\end{equation}
Here the background $b$ can be positive, negative, or zero, 
with the speed $c$ satisfying the kinematic condition $c>b$. 
Alternatively, the speed can be taken to be positive, negative, or zero,
with the background satisfying $b<c$.
In the limit $c\to b$,
the height goes to zero while the width goes to infinity,
namely the wave flattens to become $U=0$. 
The well-known KdV soliton on a zero background is obtained when $b=0$,
which implies $c>0$. 

An interesting special case is solutions with zero speed, $c=0$. 
These solutions 
\begin{equation}\label{kdv-statichump}
U(\xi) =3|b| \sech^2\big(\tfrac{1}{2}\sqrt{|b|}\,\xi\big)-|b|
\end{equation}
represent \emph{static humps} on a negative background $b<0$. 

In all cases, 
the height $h=3(c-b)$ and the width $w = 2/\sqrt{|b|-|c|}$ satisfy the scaling relation 
$hw^2=12$
which is independent of both the speed and the background. 
Note this is the same relation that holds for the soliton on a zero background. 

Finally, 
it is interesting to evaluate the conserved integrals \eqref{U-mass}--\eqref{U-ener} 
for mass, momentum, and energy of the solitary wave \eqref{kdv-solitary-physical}. 
We straightforwardly obtain 
\begin{align}
\mathcal{M}=12\sqrt{c-b} , 
\quad
\mathcal{P} =12c\sqrt{c-b} , 
\quad
\mathcal{E} =\tfrac{6}{5}((c-b)^2 +5c^2)\sqrt{c-b} . 
\end{align}
For $b=0$, these conserved integrals reduce to the well-known mass, momentum, and energy of the KdV soliton,
which obey the free-particle relations $\mathcal{P}=c\mathcal{M}$ and $\mathcal{E}\propto \mathcal{P}^2/\mathcal{M}$. 
In the case $b\neq0$, 
only the free-particle relation $\mathcal{P}=c\mathcal{M}$ continues to hold.
Note that $\mathcal{E}$ can be expressed as a linear combination of $\mathcal{P}^2/\mathcal{M}$ and $\mathcal{M}^5$.

\subsection{Solitary wave solutions for $\boldsymbol{p=3}$}

The physically parameterized quadrature \eqref{gkdv-oddp-solitary-bh-quadrature} 
for the case $p=3$
will now be evaluated in terms of elliptic functions,
and the resulting solitary wave solution will be compared to the KdV case $p=1$.

From equation \eqref{gkdv-oddp-solitary-bh-W},
we have
\begin{equation}\label{gkdv-pis3-solitary-W}
W(U)= \tfrac{1}{20}(U^2 +(3b+h)U +6b^2+ 4hb+h^2) . 
\end{equation}
To evaluate the quadrature \eqref{gkdv-oddp-solitary-bh-quadrature} in this case, 
we use a suitable change of variable given by a standard method \cite{AbrSte,Law}
to transform it into a sum of an elementary integral and two elliptic integrals
which have a straightforward evaluation in terms of the elliptic functions
$\cn$, $\sn$, and $\Pi$.
After simplifications, we obtain
\begin{subequations}\label{gkdv-pis3-solitarywave}
\begin{gather}
\begin{aligned}
&  \frac{\mu_2-h}{2h(\mu_2+h)\sqrt{\mu_2}}\Pi\bigg( \frac{(\mu_2+h)^2}{4\mu_2 h};\sn^{-1}\bigg(\frac{2\sqrt{\mu_2}\sqrt{b+h -U}}{\mu_2+b+h-U}, \frac{2\mu_2 +5b +3h}{4\mu_2}\bigg), \frac{2\mu_2 +5b +3h}{4\mu_2} \bigg)
\\&\qquad
+ \frac{1}{(\mu_2+h)\sqrt{\mu_2}}\cn^{-1}\bigg( \frac{\mu_2-b-h +U}{\mu_2+b+h-U},\frac{2\mu_2 +5b +3h}{4\mu_2} \bigg)
\\&\qquad
+ \frac{1}{2\mu_1\sqrt{h}}\arctanh\bigg( \frac{4\sqrt{5}\mu_1 \sqrt{h}\sqrt{b+h-U}\sqrt{W(U)}}{\mu_1^2 (b+h-U) +20h W(U)} \bigg)
= \frac{1}{\sqrt{10}}|\xi|,
\end{aligned}
\\
\begin{aligned}
& b\leq U\leq b+h,
\end{aligned}    
\\
\begin{aligned}    
&
\mu_1 = \sqrt{10b^2+5bh+h^2}, 
\quad
\mu_2 = \sqrt{10b^2+10bh+3h^2}
\end{aligned}    
\end{gather}
\end{subequations}
which is an implicit algebraic expression for the solitary wave solution $U(\xi)$
on a background $b$ with height $h$.
See \Ref{AbrSte} for notation of the elliptic functions.
For $\sn^{-1}$ and $\cn^{-1}$, the primary branch is used. 

\begin{figure}[h]
\centering
\includegraphics[trim=2cm 16cm 6cm 2cm,clip,width=0.48\textwidth]{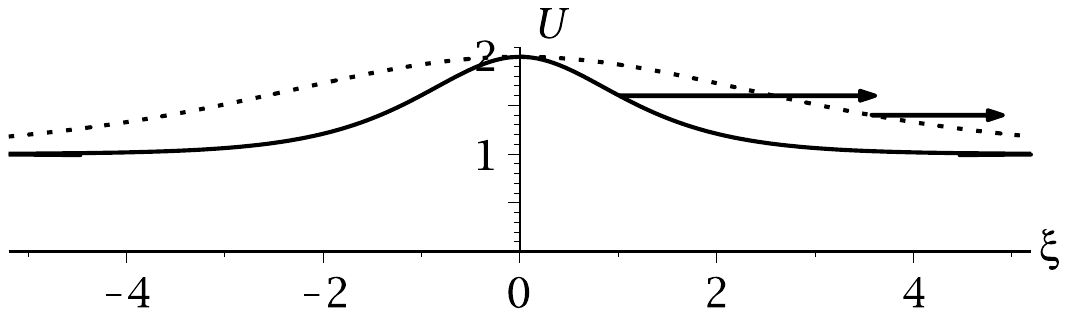}
\quad
\includegraphics[trim=2cm 16cm 6cm 2cm,clip,width=0.48\textwidth]{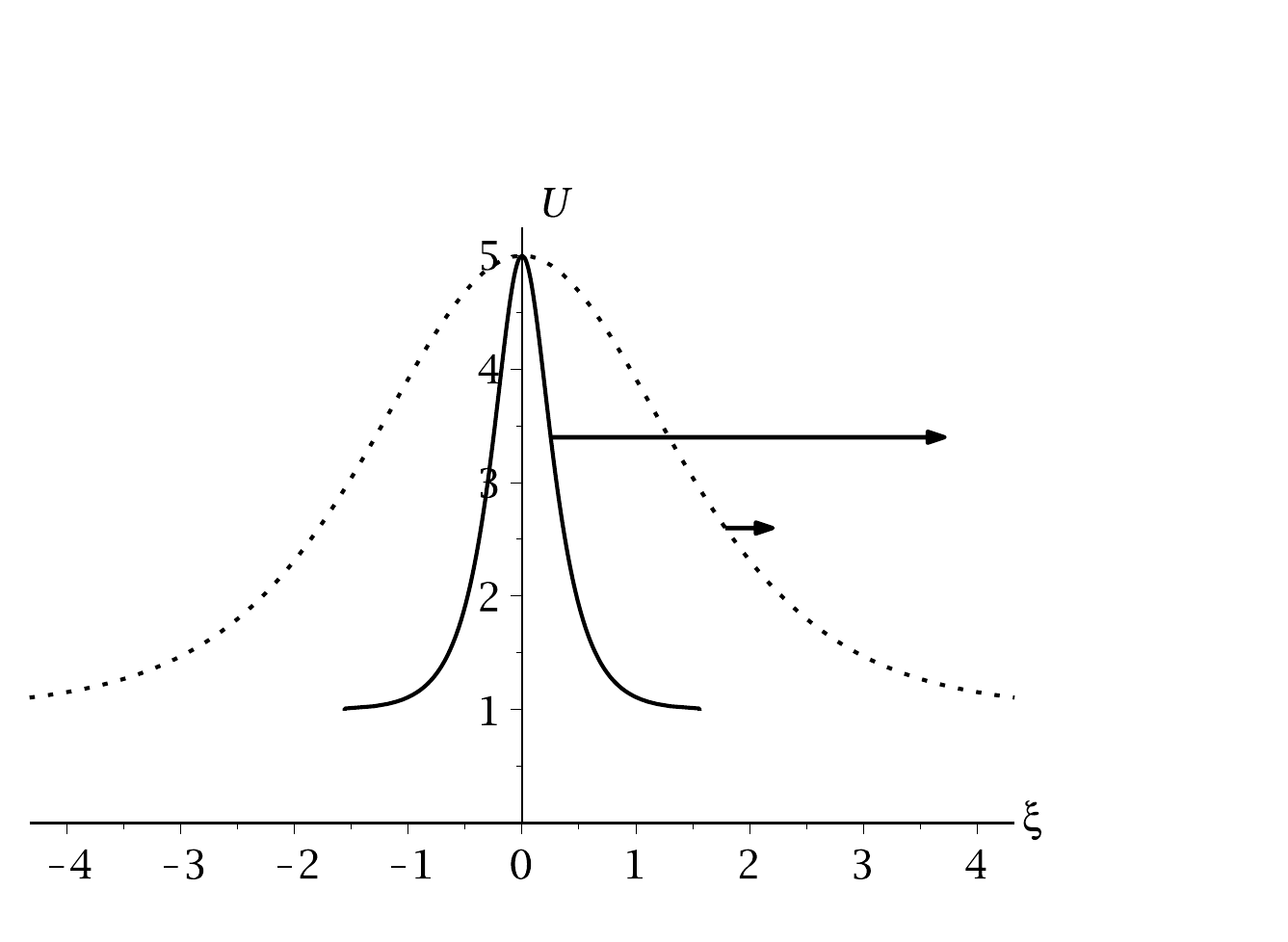}
\quad
\includegraphics[trim=2cm 16cm 6cm 2cm,clip,width=0.48\textwidth]{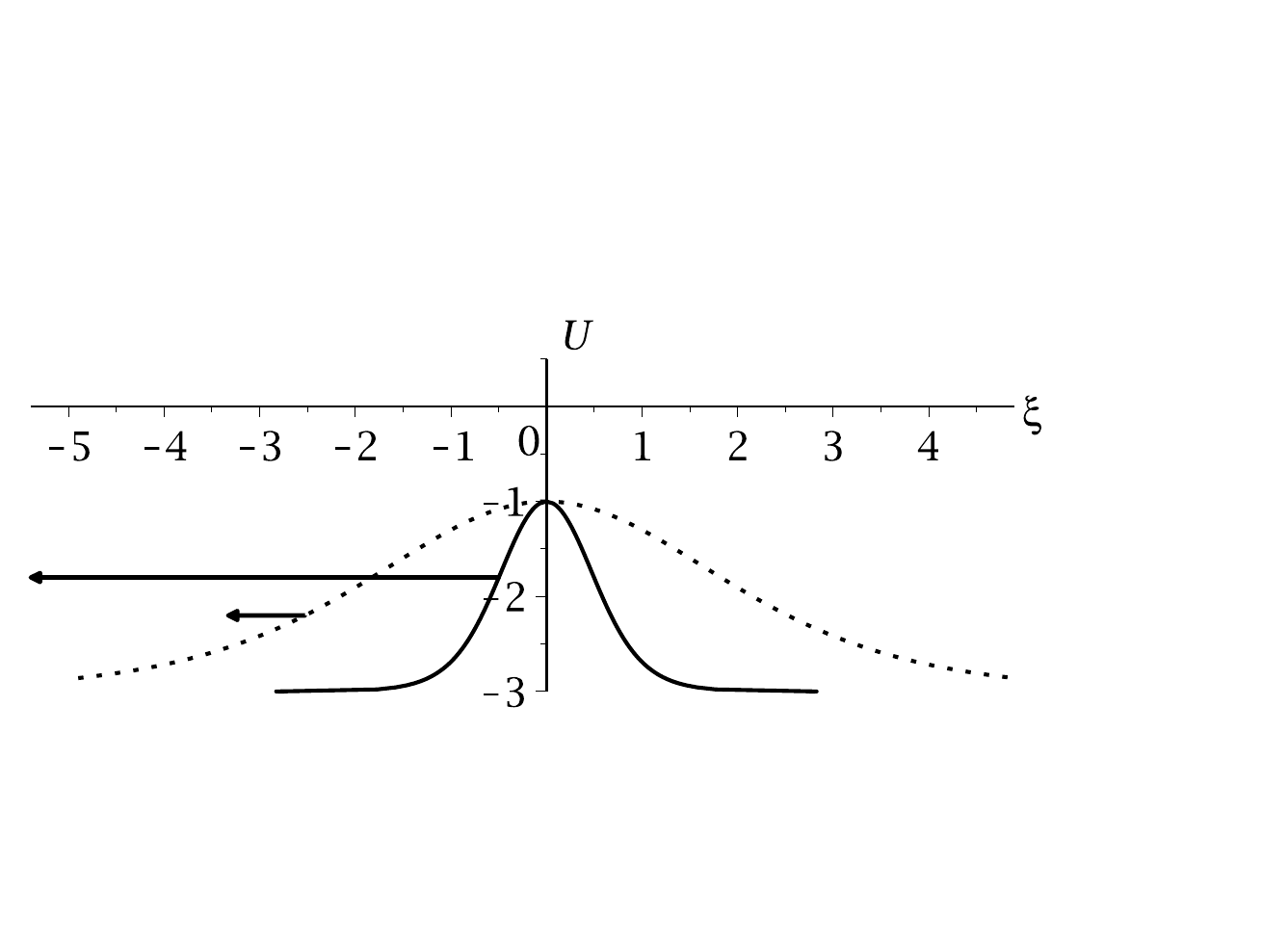}
\quad
\includegraphics[trim=2cm 16cm 6cm 2cm,clip,width=0.48\textwidth]{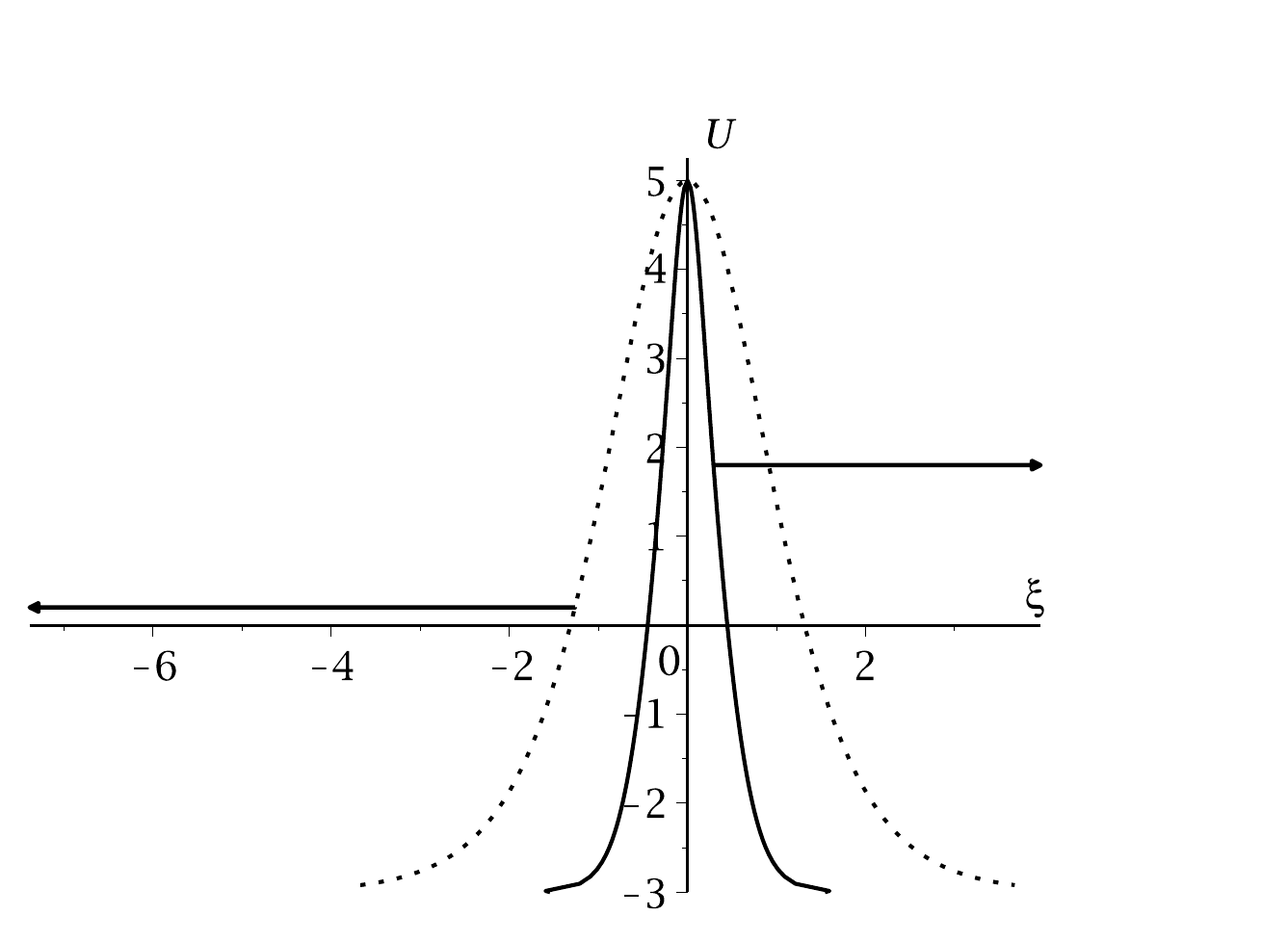}
\caption{gKdV solitary waves on a positive background (top) and a negative background (bottom) for $p=3$ (solid) and $p=1$ (dot). Arrows indicate direction and speed.}\label{gkvdfig-pis3-solitary}
\end{figure}

Expressions \eqref{gkdv-oddp-solitary-bh-c} and \eqref{gkdv-oddp-solitary-bh-w}
for the speed and the width of the solitary wave \eqref{gkdv-pis3-solitarywave}
yield 
\begin{equation}\label{gkdv-pis3-solitary-c-w}
c =
\tfrac{1}{10}h^3 +\tfrac{1}{2}bh^2 +hb^2 +b^3, 
\quad
w = 
2\sqrt{10}/\sqrt{h(h^2+5bh+10b^2)}
\end{equation}
Fig.~\ref{gkvdfig-pis3-solitary} compares this solitary wave
to the KdV solitary wave on the same background with the same height.

\begin{figure}[h]
\centering
\includegraphics[trim=2cm 16cm 6cm 2cm,clip,width=0.55\textwidth]{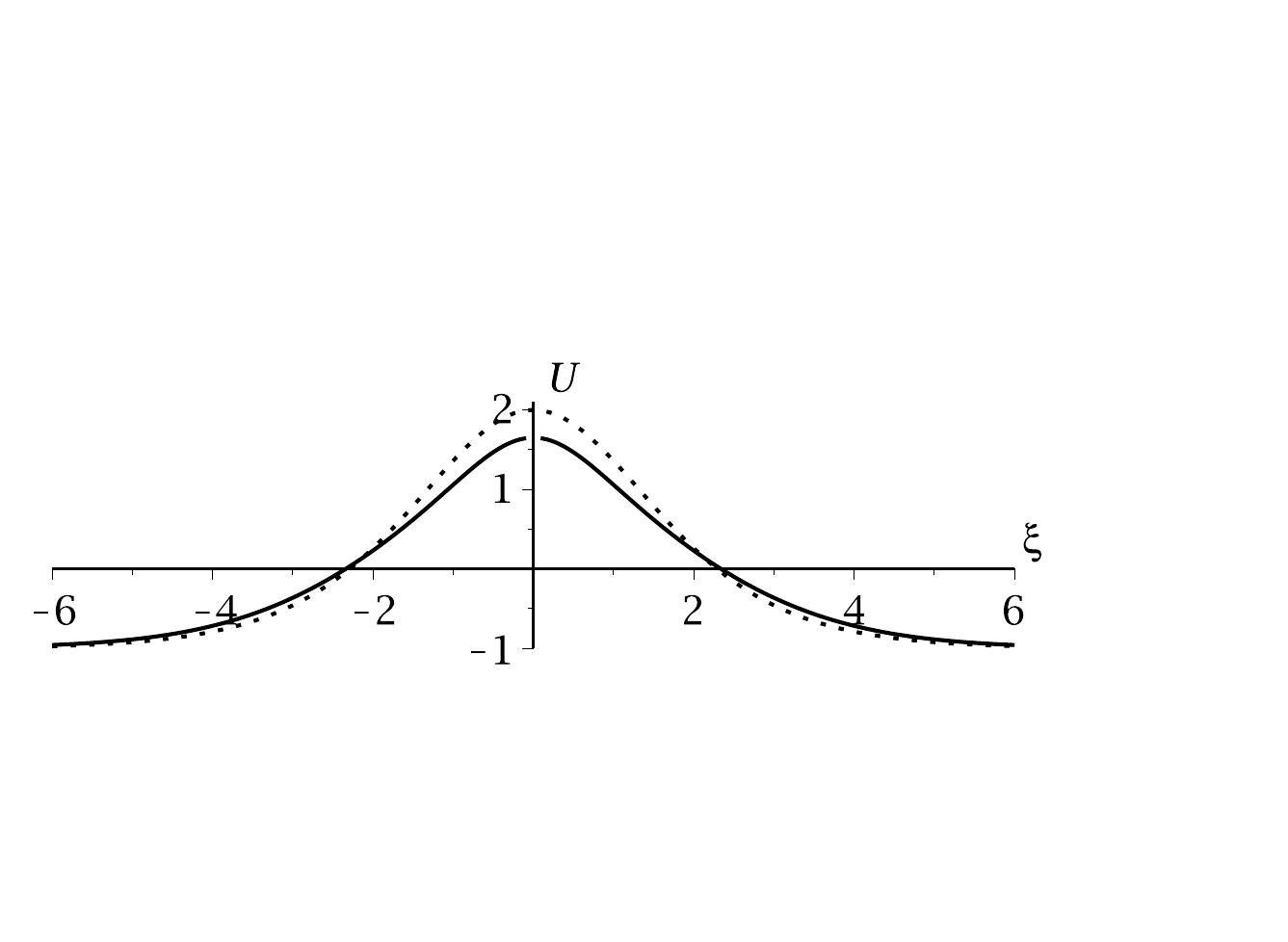}
\quad
\includegraphics[trim=2cm 16cm 6cm 2cm,clip,width=0.41\textwidth]{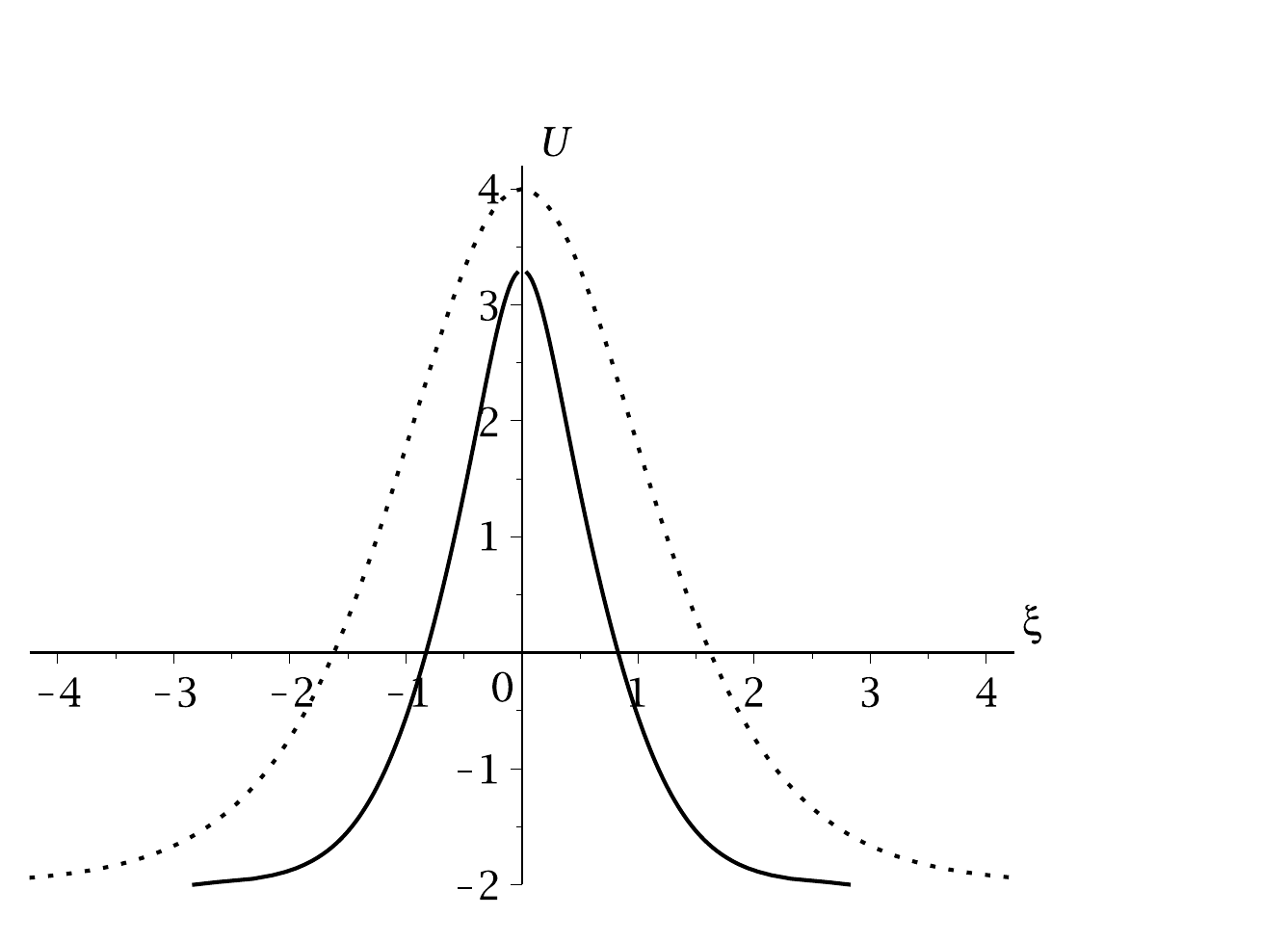}
\caption{gKdV static humps on a negative background for $p=3$ (solid) and $p=1$ (dot). }\label{gkvdfig-pis3-static}
\end{figure}

On negative backgrounds, the static hump with $c=0$ has 
height
$h=\tfrac{1}{3}(\sqrt[3]{15\sqrt{6}+35} -\sqrt[3]{15\sqrt{6}-35} +5)|b|$
and width
$w=2/\sqrt{|b|^3}$.
Fig.~\ref{gkvdfig-pis3-static} compares this solution to the KdV static hump solution. 

\section{Even-power gKdV travelling waves with non-zero boundary conditions in the defocusing case}\label{sec:gkdv-even-p-defocus}

As shown by the analysis in section~\ref{sec:gkdv-classify}, 
when the nonlinearity power $p$ is an even integer 
then the gKdV equation supports solitary waves on non-zero backgrounds
in both the focusing and defocusing cases.
In addition,
heavy-tail waves are supported in a certain limit in the focusing case,
while kink waves are supported in the defocusing case. 
No other kind of bounded non-periodic waves exist. 

We will now study the kinematic properties of these three types of solutions 
and derive analytical formulas in the cases $p=2,4$. 
The discussion will use the same physical parameterization introduced in the previous section for odd $p$. 

The results for $p=4$ are new. 
In the mKdV case, $p=2$, 
the solitary wave solution with a non-zero background
and the heavy-tailed wave solution as well as the kink wave 
are known in the literature \cite{JefKak,Au-YeuFunAu1984,KamSpiKon,Mar},
but their kinematics using a physical parameterization in terms of speed $c$, background $b$, and height/depth $h$ has not been studied previously. 

In this section, the defocusing case is considered. 
The focusing case will be considered in the next section.

\subsection{Defocusing-gKdV solitary waves}

For any even power $p\geq 2$, 
solitary waves arise when the mass parameter $M$ in the defocusing gKdV potential \eqref{gkdv-evenp-defocus-potentialwell}
is not zero. 
This potential has three critical points \eqref{gkdv-evenp-defocus-critpoints-Mnot0}, 
consisting of two local maximums at $U_1$ and $U_3$, and a local minimum at $U_2$,
which have the properties \eqref{gkdv-evenp-defocus-critpoints-rel}. 
The solitary wave solutions $U(\xi)$ are obtained from 
the resulting nonlinear oscillator equation \eqref{scaled-oscil-ode}
by taking $E=V_\max=V(U_1)$
which is the maximum of least height. 
The potential \eqref{gkdv-evenp-defocus-potentialwell}
then has the factorization 
\begin{equation}\label{gkdv-evenp-defocus-energyeqn}
V(U_1) - V(U)
=(U-U_1)^2(U-U_+)(U-U_-) W(U), 
\end{equation}
where $U_+$ and $U_-$ are roots of the energy equation, 
and where $W(U)$ is a polynomial that has even degree $p-2$.

\begin{figure}[h]
\centering
\includegraphics[trim=1cm 12cm 5cm 2cm,clip,width=0.5\textwidth]{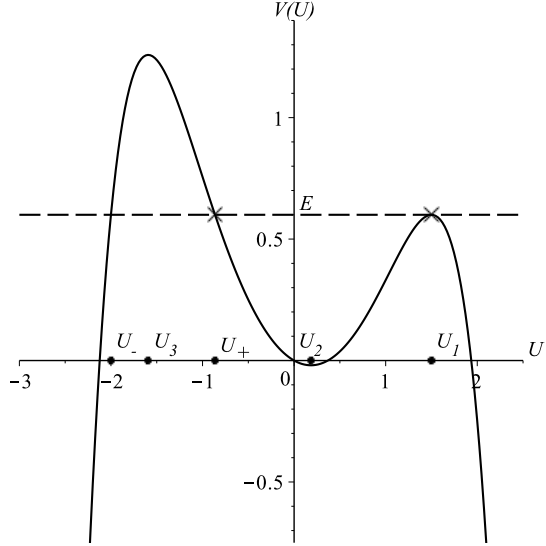} 
\caption{gKdV defocusing potential well for $p=4$.}\label{gkvdfig-potentialwell-pis4-defocus}
\end{figure}

There are two different cases, depending on the sign of $M$. 
They are related by the reflection symmetry \eqref{even-p-V-reflect} of the potential. 
For $\sgn(M)=1$, 
the roots obey 
\begin{equation}\label{evenp-defocus-Mpos-roots}
U_- <U_+ <U_1 ,
\quad
U_1>U_2>0,
\quad
U_-<U_3<0
\end{equation}
where $U_+$ is a turning point which is the left side rim in the potential well,
and $U_1$ is an asymptotic turning point which is the right side rim. 
In this case $W(U)$ is positive on $U_+\leq U\leq U_1$. 
For $\sgn(M)=-1$, 
the roots obey 
\begin{equation}\label{evenp-defocus-Mneg-roots}
U_1 <U_- <U_+,
\quad
U_1<U_2< 0,
\quad
U_+ > U_3 >0
\end{equation}
where $U_-$ is a turning point which is the right side rim in the potential well,
and $U_1$ is an asymptotic turning point which is the left side rim. 
In this case $W(U)$ is positive on $U_1\leq U\leq U_-$.
In both cases, note that $\sgn(M)=\sgn(U_1)$. 
This potential well is shown in Fig.~\ref{gkvdfig-potentialwell-pis4-defocus}. 

For the case \eqref{evenp-defocus-Mpos-roots}, 
the quadrature \eqref{quadrature} for $U(\xi)$ 
is given by 
\begin{equation}\label{gkdv-evenp-defocus-Mpos-solitary-quadrature}
\int^{U}_{U_-} \frac{dU}{(U_1-U)\sqrt{(U-U_-)(U-U_+)W(U)}} = \sqrt{2}|\xi|,
\quad
U_-\leq U\leq U_1 . 
\end{equation}
Here $U_\pm$ and $U_1$ are implicitly given in terms of $(c,M)$ by the polynomial equations
\begin{equation}\label{gkdv-evenp-defocus-solitary-U1-UpUm}
\tfrac{1}{p+1}U_1^{p+1} + cU_1 +M=0,
\quad
\tfrac{1}{(p+1)(p+2)}(U_\pm^{p+2} - U_1^{p+2}) 
+\tfrac{1}{2}c(U_\pm^2 - U_1^2)
+M(U_\pm -U_1) =0
\end{equation}
coming from $V'(U_1)=0$ and $V(U_1)=V(U_\pm)$. 

For the case \eqref{evenp-defocus-Mneg-roots},
the quadrature is similar:
\begin{equation}\label{gkdv-evenp-defocus-Mneg-solitary-quadrature}
\int^{U_-}_{U} \frac{dU}{(U-U_1)\sqrt{(U_--U)(U_+-U)W(U)}} = \sqrt{2}|\xi|,
\quad
U_1\leq U\leq U_- . 
\end{equation}

These integrals can be evaluated in terms of elementary functions when $p=2$,
and elliptic functions when $p=4$. 
When $p\geq 6$,
the integrals cannot be evaluated explicitly in general.
Due to the reflection symmetry \eqref{even-p-V-reflect} of the potential,
hereafter we will concentrate on solitary wave solution given by the first quadrature \eqref{gkdv-evenp-defocus-Mpos-solitary-quadrature},
which has $M>0$. 
The solitary wave solution arising from the second quadrature, which has $M<0$, 
is obtained by applying the reflection symmetry to the previous solution.

\subsection{Kinematics}\label{sec:solitarywave-evenp-defocus}

For any even power $p\geq 2$, 
the quadrature \eqref{gkdv-evenp-defocus-Mpos-solitary-quadrature} 
for the case $M>0$
implicitly defines a family of the solitary wave solutions $U(\xi)$ 
which are parameterized by $(c,M)$.
This family has not been previously studied when the background is non-zero. 
We now will look at the main kinematic properties.

In this solution family, 
the wave speed is $c<0$, due to condition \eqref{gkdv-evenp-Mcond}. 
The wave peak is $U_-<0$ and the background (asymptote) is $b=U_1>0$,
and so the wave has depth $h=U_1-U_->0$. 
The width is proportional to $w=2/\sqrt{U_1^p-|c|}$,
which is obtained from an asymptotic expansion of the quadrature as $U\to U_1$,
combined with $V''(U_1)=-2(U_1-U_+)(U_1-U_-)W(U_1)=-U_1^p-c$
from the factorization \eqref{gkdv-evenp-defocus-energyeqn}.
So that $w^2 >0$, 
the background is required to satisfy $b>|c|^{1/p}$. 
The background also has a maximum 
$b_\max=((p+1)|c|)^{1/p}$,
which arises from the limit $M\to 0$ given by the potential well \eqref{gkdv-evenp-defoc-critpoints-Mis0}. 
Hence, $b$ belongs to the positive interval 
$|c|^{1/p} < b < ((p+1)|c|)^{1/p}$. 
Alternatively,
the background $b$ can be chosen to have any positive value,
and then the speed is required to satisfy the kinematic condition
\begin{equation}
\tfrac{1}{p+1}b^p< -c < b^p . 
\end{equation}
The solution family $U(\xi)$ describes 
a \emph{dark solitary wave on a positive background}.
(Alternatively, it could be viewed as a \emph{bright solitary hole in a positive background}.)
It is an even function of $\xi$. 

Note that $(c,b)$ provides an implicit physical parameterization of the solution family, 
similar to case \eqref{gkdv-oddp-solitary-b-c}, 
after $M$ is eliminated from the polynomial equations \eqref{gkdv-evenp-defocus-solitary-U1-UpUm}:
\begin{equation}\label{gkdv-evenp-defocus-solitary-b-c}
M = |c|b -\tfrac{1}{p+1}b^{p+1}, 
\quad
U_1=b,
\quad
U_\pm = bz, 
\quad
\tfrac{1}{p+1} |b|^p (\tfrac{1}{p+2}S_{p+2}(z)  -1) -\tfrac{1}{2}|c|(z -1) =0,
\end{equation}
where $S_{p+2}(z)$ is the polynomial \eqref{Spoly}.

An explicit physical parameterization can be obtained by 
writing out the quadrature \eqref{gkdv-evenp-defocus-Mpos-solitary-quadrature} 
in terms of $(b,h)$. 
As shown in the Appendix, 
this yields 
\begin{equation}\label{gkdv-evenp-defocus-solitary-bispos-h-quadrature}
\int^{b}_{U} \frac{dU}{(b-U)\sqrt{(U+h-b)(U+g+h-b) W(U)}} = \sqrt{2}\, |\xi|,
\quad
b-h\leq U\leq b 
\end{equation}
with $g$ given by the relation 
\begin{equation}\label{gkdv-evenp-defocus-solitary-bispos-h-g}
R_{p+1}(1 -(g+h)/b) = R_{p+1}(1 -h/b), 
\quad
g>0, 
\end{equation}
and with 
\begin{equation}\label{gkdv-evenp-defocus-solitary-bispos-W}
\begin{aligned}
W(U) =
\tfrac{1}{(p+1)(p+2)}|b|^{p-2}b & \big(
R_{p-1}(1-(g+h)/b,1-h/b) + R_{p-2}(1-(g+h)/b,1-h/b)(U/b)
\\&\qquad
+ \cdots + R_{1}(1-(g+h)/b,1-h/b)(U/b)^{p-2}
\big) . 
\end{aligned}
\end{equation}
Here $R_n(z)$ is the polynomial \eqref{Rpoly},
and \begin{equation}\label{doubleRpoly}
R_n(y,z) = \sum_{j=0}^{n-1} (n-j) \sum_{i=0}^{j} y^{i}z^{j-i}
= (R_{n+1}(y)-R_{n+1}(z))/(y-z),
\quad
n\in\mathbb{Z}^+
\end{equation}  
is a polynomial of degree $n-1$ in $y,z$.

The speed and the width of the solitary wave 
are given by the explicit expressions
\begin{equation}\label{gkdv-evenp-defocus-solitary-bispos-c}
c = -\tfrac{2}{(p+1)(p+2)} |b|^p R_{p+1}(1 - h/b)
= \tfrac{2}{(p+1)(p+2)} (b^{p+2} -(p+2)h b^{p+1} -(b -h)^{p+2})/h^2
\end{equation}
and 
\begin{equation}\label{gkdv-evenp-defocus-solitary-bispos-w}
\begin{aligned}
w & = 2/\sqrt{|b|^p(1-\tfrac{2}{(p+1)(p+2)} R_{p+1}(1 - h/b))}
\\
&= \sqrt{2(p+1)(p+2)}h/\sqrt{(p+1)(p+2)h^2b^p -2(p+2)h b^{p+1} -(b -h)^{p+2}-b^{p+2}}
\end{aligned}
\end{equation}
which satisfy the relation
\begin{equation}\label{gkdv-evenp-defocus-solitary-cwrel}
w^2(b^p-|c|)=4 . 
\end{equation}

Additional kinematic features of the solitary waves 
can be understood from the properties of the polynomial $R_{p+1}(1- \eta)$,
as a function of $\eta$, 
discussed in section~\ref{sec:solitarywave-oddp}. 
Those properties applied to the polynomial relation \eqref{gkdv-evenp-defocus-solitary-bispos-h-g}
show that $g$ is determined in terms of $(h,b)$,
with $h$ having a maximum of $2b$ that corresponds to $g$ having a minimum of $0$. 

Consequently,
the parameters $(h,b)$ in the solitary wave solution family \eqref{gkdv-evenp-defocus-solitary-bispos-h-quadrature}
obey 
\begin{equation}
b>0,
\quad
0\leq h\leq 2b . 
\end{equation}
The maximum depth $h=2b$ of the solitary wave occurs
when $|c|$ is a minimum, $|c|_\min=\tfrac{1}{p+1}b^p$,
while the width is $w=2\sqrt{\tfrac{p+1}{p}}\big/b^{p/2}$.
When $|c|$ reaches a maximum $|c|_\max=b^p$,
the depth of the wave becomes $h\to 0$ and its width $w$ goes to infinity,
namely $U\to b$. 

This shows the advantage of having the explicit physical parameterization $(b,h)$,
rather than the implicit parameterization $(c,M)$, 
for understanding the kinematic features of these solitary waves. 

Finally, we consider the case $M<0$. 
The resulting family of solitary wave solutions can be obtained 
by taking $b\to -b$ and $U\to -U$ in the previous family, 
due to the reflection symmetry \eqref{even-p-V-reflect}. 
This yields
\begin{equation}\label{gkdv-evenp-defocus-solitary-bisneg-h-quadrature}
\int^{b}_{U} \frac{dU}{(U-b)\sqrt{(h+b-U)(g+h+b-U) W(U)}} = \sqrt{2}\, |\xi|,
\quad
b\leq U\leq b+h 
\end{equation}
where $W(U)$ has the same expression \eqref{gkdv-evenp-defocus-solitary-bispos-W},
with the parameters $(h,b)$ now obeying
\begin{equation}
b<0,
\quad
0\leq h\leq 2|b|
\end{equation}
and with $g$ again determined by the polynomial relation \eqref{gkdv-evenp-defocus-solitary-bispos-h-g}.
This solution family describes 
a \emph{bright solitary wave on a negative background}. 
Its kinematic features are same as those of the previous solution family.

\subsection{Defocusing-gKdV kink waves}\label{subsec:gkdv-evenp-defocus-kink}

Kink wave solutions $U(\xi)$ arise when the mass parameter in the gKdV potential \eqref{gkdv-evenp-defocus-potentialwell}
is $M=0$, 
which produces a symmetric potential well \eqref{gkdv-evenp-defocus-Mis0-potentialwell}. 
Its two maximum points become $U_1$ and $U_3=-U_1$,
and the potential at these points has equal height $V_\max=V(U_1)=V(-U_1)$.
Moreover, 
the roots $U=U_\pm$ of the energy equation \eqref{gkdv-evenp-defocus-energyeqn} 
coalesce into a repeated root which coincides with one of the maximum points,
namely 
$|U_{\pm}|= |U_1|$. 
This energy equation thereby has the factorization 
\begin{equation}\label{gkdv-evenp-defocus-energyeqn-Mis0}
V_\max - V(U)
=(U-U_1)^2(U+U_1)^2 W(U), 
\end{equation}
where $W(U)$ is a polynomial that has even degree $p-2$
and that is positive on $-|U_1|\leq U\leq |U_1|$. 

The quadrature for $U(\xi)$ has a similar form to the one for solitary waves, 
except that the endpoint needs to be changed. 
A convenient choice is the minimum point $U_2=0$ of the potential. 
This yields the symmetric quadrature 
\begin{equation}\label{gkdv-evenp-defocus-kink-quadrature}
\int^{0}_{U} \frac{dU}{(U_1^2-U^2)\sqrt{W(U)}} = \sqrt{2} \xi,
\quad
-|U_1|\leq U\leq |U_1| . 
\end{equation}
As shown later, this integral can be evaluated in terms of 
elementary functions when $p=2,4$, and elliptic functions when $p=6$. 
For $p\geq 8$, it cannot be evaluated explicitly in general.

The quadrature \eqref{gkdv-evenp-defocus-kink-quadrature}
implicitly defines a family of \emph{kink (shock) wave} solutions $U(\xi)$,
parameterized by the wave speed $c<0$.
These solutions are odd functions of $\xi$, 
and as $\xi\to \pm \infty$, the asymptotes are $U\to\pm |U_1|$,
which can be seen from an asymptotic expansion of the quadrature in $U$.

A more explicit parameterization is obtained by using $b=|U_1|>0$,
as shown in the Appendix. 
In this parameterization,
the quadrature is given by 
\begin{equation}\label{gkdv-evenp-defocus-kink-b-quadrature}
\int^{0}_{U} \frac{dU}{(b^2-U^2)\sqrt{W(U)}} = \sqrt{2} \xi,
\quad
-b\leq U\leq b
\end{equation}
with 
\begin{equation}\label{gkdv-evenp-defocus-kink-W}
W(U) = \tfrac{1}{4(p+1)(p+2)}b^{p-2}\big( R_{p}(U/b)+R_{p}(-U/b) \big)
\end{equation}
where $R_n(z)$ is the polynomial \eqref{Rpoly}.
Note that the polynomial $W(U)$ is symmetric in $U$ 
and hence it has an even degree $p-2$. 
The speed and the width of the kink wave 
are given by the expressions
\begin{equation}\label{gkdv-evenp-defocus-kink-c}
c = -\tfrac{2}{(p+1)(p+2)} b^p R_{p+1}(-1) 
= -\tfrac{1}{p+1} b^p
\end{equation}
and 
\begin{equation}\label{gkdv-evenp-defocus-kink-w}
w = 2/\sqrt{b^p(1-\tfrac{2}{(p+1)(p+2)} R_{p+1}(-1))}
= 2\sqrt{1+\tfrac{1}{p}}\big/b^{p/2}
\end{equation}
which satisfy the relation
\begin{equation}\label{gkdv-evenp-defocus-kink-cwrel}
w^2|c|=\tfrac{4}{p} .
\end{equation}
Unlike the solitary wave solutions,
the kink wave solutions have no restriction on $b>0$,
or alternatively, no restriction on $c<0$.

\subsection{Defocusing-mKdV ($\boldsymbol{p=2}$) solitary waves on a background and kink waves}

In the mKdV case, $p=2$,
the three critical points of the potential $V(U)$ 
can be found explicitly as the roots of a cubic. 
The corresponding factorization of the potential \eqref{gkdv-evenp-defocus-energyeqn} 
has $W=1$. 

The physically parameterized quadrature \eqref{gkdv-evenp-defocus-solitary-bispos-h-quadrature}
for the case $M>0$
yields the solitary wave solution 
\begin{equation}\label{mkdv-defocus-solitary-Mispos}
U(\xi) =b-\frac{h(4b-h)}{(2b-h)\cosh\big(\tfrac{1}{\sqrt{6}}\sqrt{h(4b-h)}\,\xi\big)+2b}
\end{equation}
in terms of the depth $h>0$ and the background $b>\tfrac{1}{2}h$, 
where
\begin{equation}
c= -\tfrac{1}{6}(2b^2 + (2b-h)^2) <0
\end{equation}
is the speed. 
Thus, this solution family describes a \emph{dark solitary wave on a positive background},
parameterized by $(b,h)$. 
(Alternatively, it could be viewed as a \emph{bright solitary hole in a positive background}.)

The equivalent physical parameterization \eqref{gkdv-evenp-defocus-Mpos-solitary-quadrature} and \eqref{gkdv-evenp-defocus-solitary-b-c}
also has an explicit form
\begin{equation}\label{mkdv-defocus-solitary-physical-bispos}
U(\xi)= b-\frac{6(b^2-|c|)}{\sqrt{6|c|-2b^2}\,\cosh(\sqrt{b^2-|c|}\,\xi)+2b} ,
\quad
b>0,
\quad
c<0. 
\end{equation}
The wave peak is $-b + \sqrt{6|c|-2b^2}$, 
and the wave depth is $h=2b-\sqrt{6|c|-2b^2}$, 
while the width of the wave is proportional to $w=2/\sqrt{b^2-|c|}$. 
Clearly, 
if the background $b$ is taken to be any positive value, 
then the speed is required to satisfy the kinematic condition 
\begin{equation}\label{mkdv-defocus-crange}
-b^2<c<-\tfrac{1}{3}b^2 . 
\end{equation}
Alternatively, 
the speed $c$ can be taken to be any negative value, 
and then the background is required to satisfy 
$\sqrt{3|c|}>b>\sqrt{|c|}$. 
The sign property $c<0$ is in contrast to the KdV case, 
where the wave speed $c$ could have any sign.

Going back to the remaining case $M<0$, 
we can apply the reflection symmetry $U\rightarrow -U$ and $b\rightarrow -b$ 
to the previous solution \eqref{mkdv-defocus-solitary-physical-bispos}, 
yielding 
\begin{equation}\label{mkdv-defocus-solitary-physical-bisneg}
U(\xi)= b +\frac{6(b^2-|c|)}{\sqrt{6|c|-2b^2}\,\cosh(\sqrt{b^2-|c|}\,\xi)-2b},
\quad
b<0,
\quad
c<0. 
\end{equation}
The wave peak is $|b| -\sqrt{6|c|-2b^2}$, 
and the wave height is $h=2|b| -\sqrt{6|c|-2b^2}$, 
while the width of the wave is proportional to $w=2/\sqrt{b^2-|c|}$. 
If the background $b$ is taken to be any negative value, 
then the speed is required to satisfy the kinematic condition \eqref{mkdv-defocus-crange}.
Alternatively, 
if the speed $c$ is taken to be any negative value, 
then the background is required to satisfy 
$-\sqrt{3|c|}<b<-\sqrt{|c|}$. 
In terms of the height $h>0$ and the background $b<-\tfrac{1}{2}h$, 
the solution family \eqref{mkdv-defocus-solitary-physical-bisneg} 
has the equivalent form 
\begin{equation}\label{mkdv-defocus-solitary-Misneg}
U(\xi) =b +\frac{h(4|b|-h)}{(2|b|-h)\cosh\big(\tfrac{1}{\sqrt{6}}\sqrt{h(4|b|-h)}\,\xi\big) +2|b|} 
\end{equation}
where
\begin{equation}
c= -\tfrac{1}{6}(2b^2 + (2|b|-h)^2) <0
\end{equation}
is the speed. 
Physically, this family describes 
a \emph{bright solitary wave on a negative background},
parameterized by $(b,h)$. 

\begin{figure}[h]
\centering
\includegraphics[trim=2cm 17cm 8cm 5cm,clip,width=0.45\textwidth]{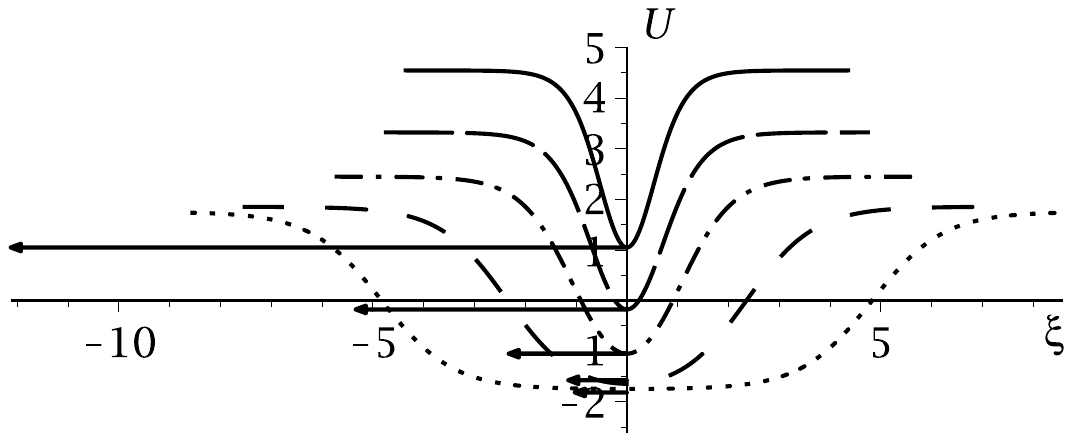} 
\quad
\includegraphics[trim=2cm 17cm 8cm 5cm,clip,width=0.40\textwidth]{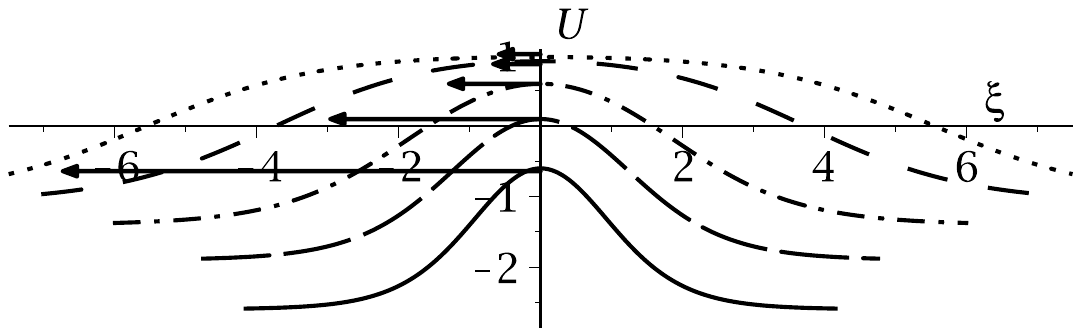} 
\caption{Defocusing-mKdV solitary waves on different positive backgrounds (left) and negative backgrounds (right). Arrows indicate direction and speed of the waves.}\label{gkdvfig-mkdv-profile-defocus-backgrounds}
\end{figure}

The kinematical properties of both solution families \eqref{mkdv-defocus-solitary-physical-bispos} and \eqref{mkdv-defocus-solitary-physical-bisneg}
have not been studied previously using a physical parameterization. 
See Fig.~\ref{gkdvfig-mkdv-profile-defocus-backgrounds}. 

\begin{figure}[h]
\centering
\includegraphics[trim=2cm 17cm 8cm 4.5cm,clip,width=0.5\textwidth]{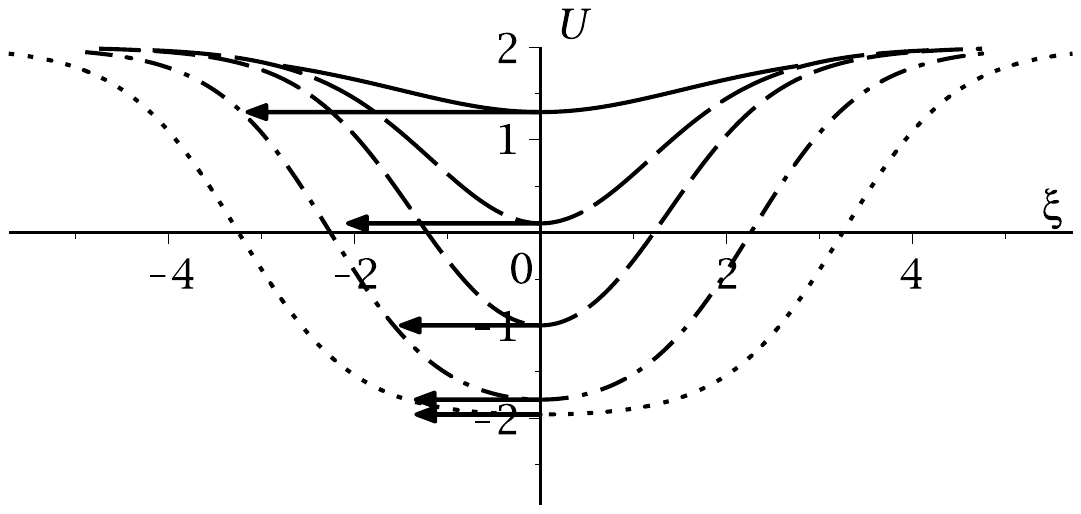} 
\quad
\includegraphics[trim=2cm 17cm 7cm 3cm,clip,width=0.43\textwidth]{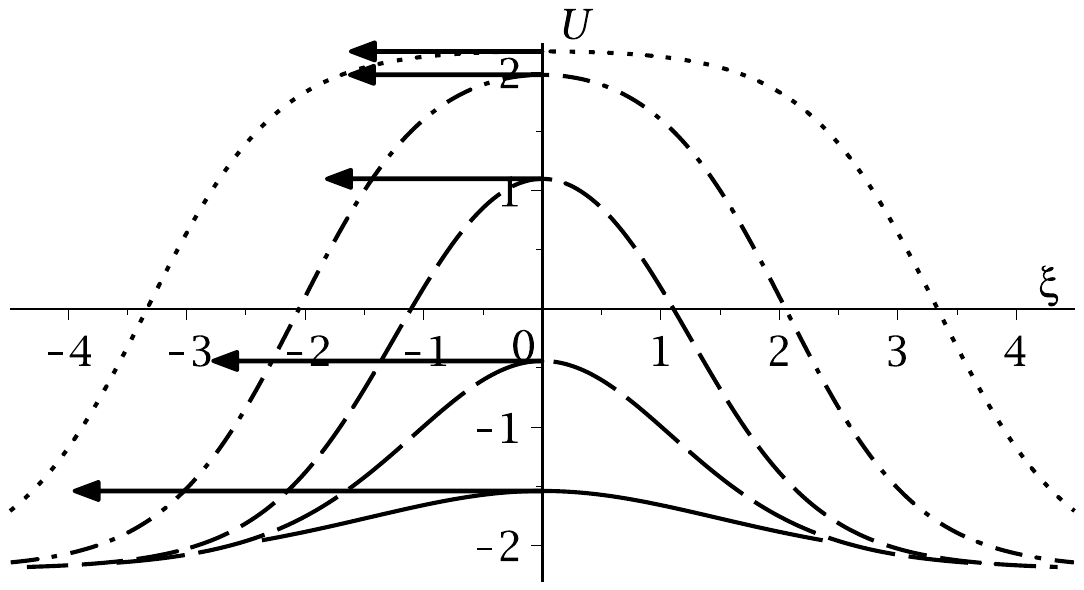}
\caption{Defocusing-mKdV dark solitary waves with different depths on a positive background (left)
and bright solitary waves with different heights on a negative background (right). Arrows indicate direction and speed of the waves. }\label{gkdvfig-mkdv-profile-defocus-heights}
\end{figure}

For any fixed negative speed, 
as the size of the background $|b|$ decreases, 
the depth/height of the wave decreases while the width increases. 
In the limit $|b|\to\sqrt{|c|}$, the wave flattens to become $U\to b$. 
But in the opposite limit $|b|\to\sqrt{3|c|}$, 
the wave widens and has the approximate form 
\begin{equation}\label{mkdv-defocus-solitary-apprx}
U \simeq \sgn(b)\sqrt{3|c|}\bigg( 1 -\frac{2}{\textstyle{\sqrt{1-|b|/\sqrt{3|c|}}}\,\cosh(\xi\sqrt{2|c|})+1} +O(1-|b|/\sqrt{3|c|}) \bigg) . 
\end{equation}
The limiting depth/height is $h=2|b|=2\sqrt{3|c|}$,
and the limiting width is $w=\sqrt{6}/|b|=\sqrt{2/|c|}$.
See Fig.~\ref{gkdvfig-mkdv-profile-defocus-heights}. 

Finally, 
the conserved integrals \eqref{U-mass}--\eqref{U-ener} 
for mass, momentum, and energy 
of the two solitary waves \eqref{mkdv-defocus-solitary-physical-bispos} and \eqref{mkdv-defocus-solitary-physical-bisneg},
are readily evaluated: 
\begin{align}
\mathcal{M}_\pm=\pm4\sqrt{6}\,\arctanh\bigg(\frac{\sqrt{6|c|-2b^2}\mp 2b}{\sqrt{6(b^2-|c|)}}\bigg) , 
\quad
\mathcal{P} =-6\sqrt{b^2-|c|} , 
\quad
\mathcal{E} =2|c|\sqrt{b^2-|c|} . 
\end{align}

For the kink (shock) wave solution, 
in the mKdV case, $p=2$, 
the quadrature \eqref{gkdv-evenp-defocus-kink-quadrature} 
straightforwardly yields
\begin{equation}\label{mkdv-defocus-kink}
U(\xi) = \sqrt{3|c|}\, \tanh\big(\tfrac{1}{\sqrt{2}}\sqrt{|c|}\,\xi\big),
\quad
c<0 . 
\end{equation}
This solution has the following features. 
Its speed can be either positive or negative. 
As $\xi\to \pm \infty$, the asymptotes are $U\to\pm \sqrt{3|c|}=\pm b$. 
Hence the net change in amplitude (height) is $h=2\sqrt{3|c|}=2b$. 
The width is proportional to $w=\sqrt{2}/\sqrt{|c|}=\sqrt{6}/b$.

By a straightforward to evaluation, 
the conserved integrals \eqref{U-mass}--\eqref{U-ener}
for mass, momentum, and energy yield 
\begin{equation}
\mathcal{M}=0, 
\quad
\mathcal{P} =-3\sqrt{2|c|} , 
\quad
\mathcal{E} =|c|\sqrt{2|c|} . 
\end{equation}

\subsection{Solitary wave and kink solutions for $\boldsymbol{p=4}$}

We will now evaluate 
the dark solitary wave quadrature \eqref{gkdv-evenp-defocus-solitary-bispos-h-quadrature} 
in terms of elliptic functions when $p=4$,
and we will obtain the bright solitary wave by applying a reflection symmetry. 
We will also evaluate the kink wave quadrature \eqref{gkdv-evenp-defocus-kink-b-quadrature} explicitly in terms of elementary functions when $p=4$. 
The resulting solutions will be compared to the mKdV case $p=2$.

For the dark solitary wave,
we start from equation \eqref{gkdv-evenp-defocus-solitary-bispos-W},
which gives
\begin{equation}\label{gkdv-pis4-defocus-solitary-W}
W(U)=\tfrac{1}{30}(U^2 +(4b-2h-g)U +10b^2 -5b(2h+g) +3h^2 + g^2 +3gh),
\end{equation}
and 
\begin{equation}\label{gkdv-pis4-defocus-solitary-nontp}
\begin{aligned}
& g = \tfrac{1}{3}({\textstyle \sqrt[3]{p+3\sqrt{3q}} -\sqrt[3]{p-3\sqrt{3q}}}) +2 b-\tfrac{4}{3} h,
\\
& p= 81b^3 -81b^2 h+ 45h^2b -10 h^3,
\\
& q= (3 b^2 -2 bh +h^2) (90 b^4-120 b^3 h+79 b^2 h^2-28 b h^3+4 h^4) . 
\end{aligned}
\end{equation}
The quadrature \eqref{gkdv-evenp-defocus-solitary-bispos-h-quadrature} 
can be evaluated by a standard method \cite{AbrSte,Law}
using a suitable change of variable to transform them into a sum of an elementary integral and two elliptic integrals
which have a straightforward evaluation in terms of the elliptic functions
$\sn$, $\cn$ and $\Pi$.
See \Ref{AbrSte} for notation.
For $\sn^{-1}$ and $\cn^{-1}$, the primary branch is used. 

\begin{figure}[h]
\centering
\includegraphics[trim=2cm 15cm 6cm 2cm,clip,width=0.48\textwidth]{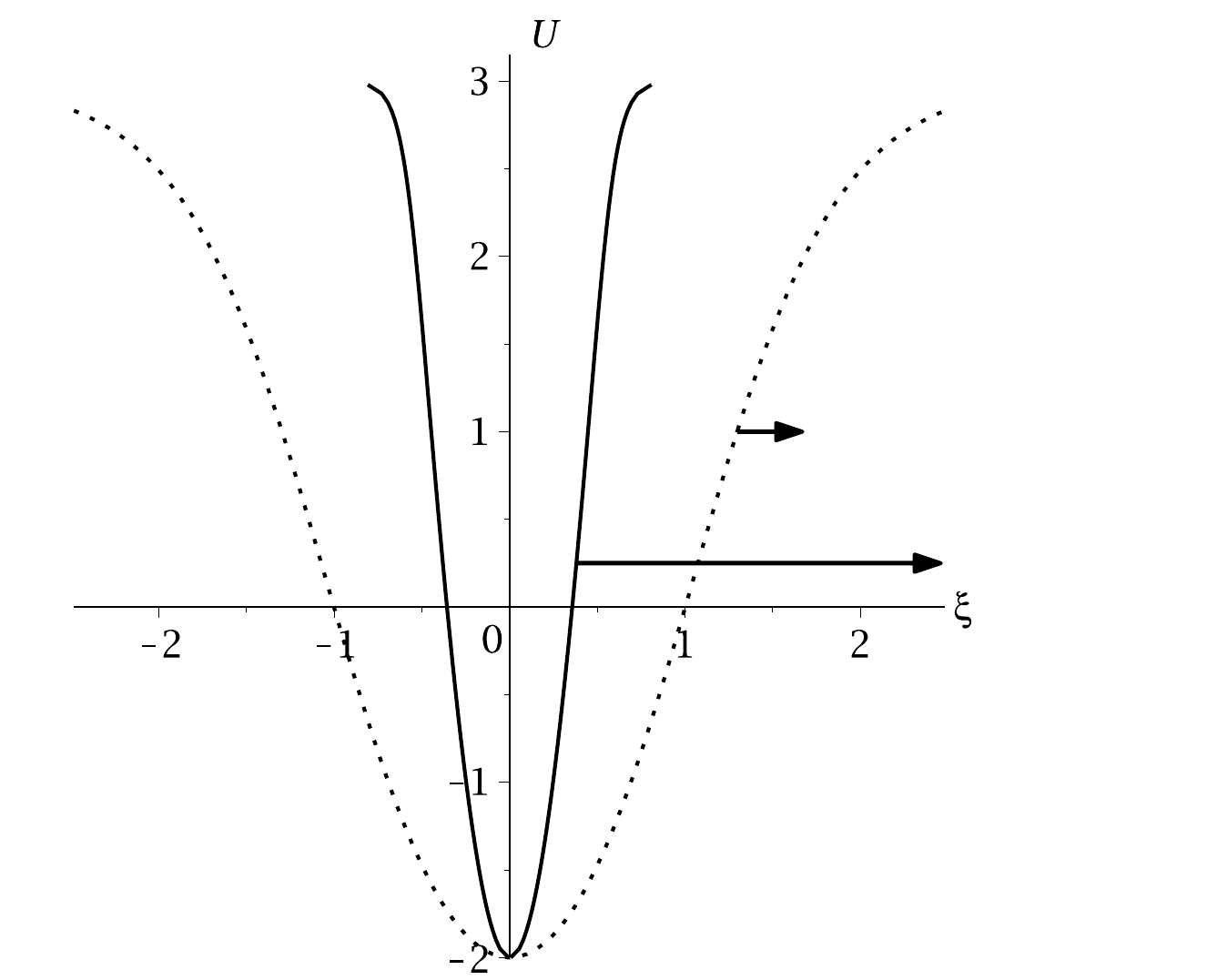}
\quad
\includegraphics[trim=2cm 16cm 6cm 2cm,clip,width=0.48\textwidth]{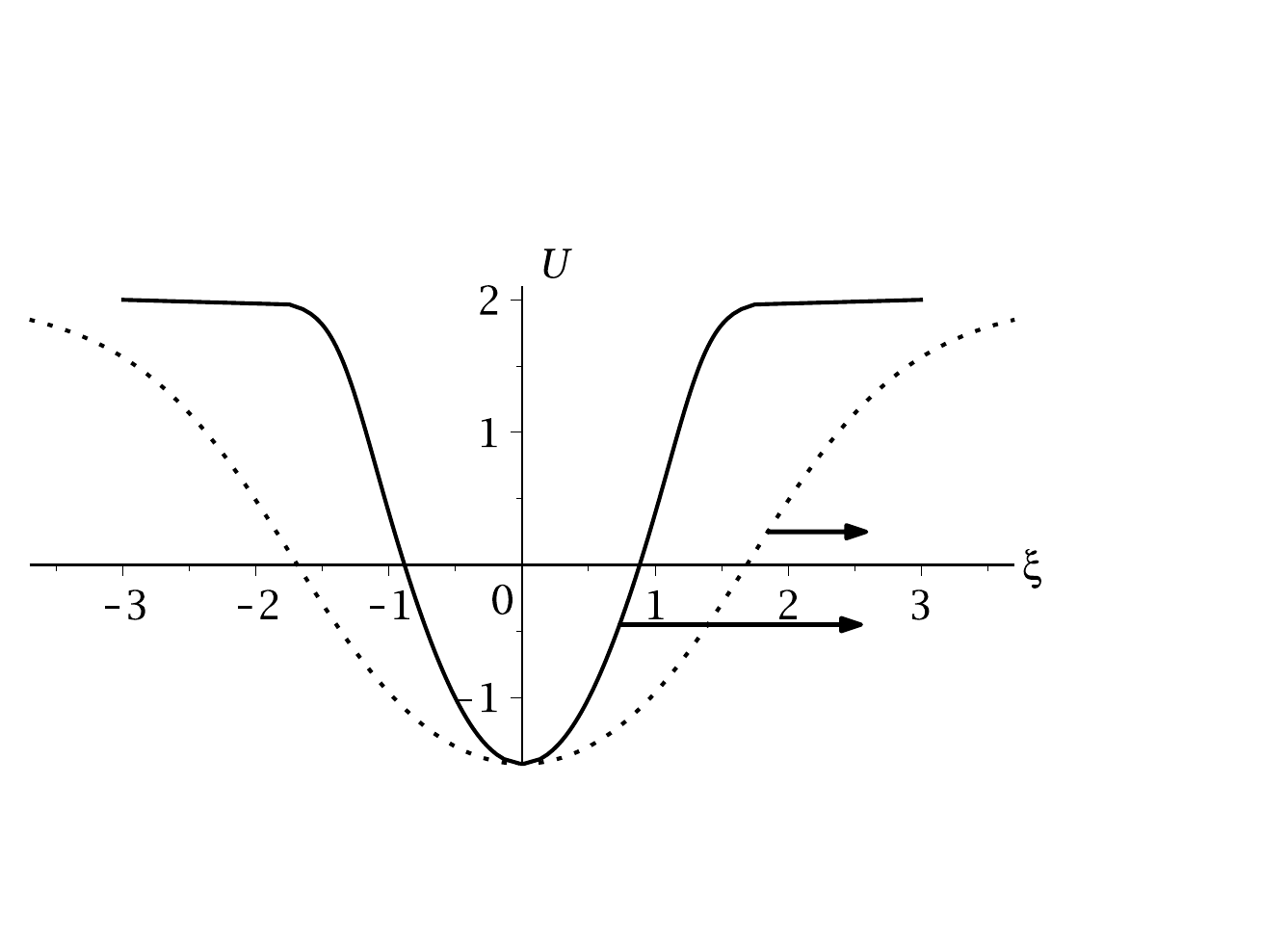}
\quad
\includegraphics[trim=2cm 18cm 8cm 6cm,clip,width=0.48\textwidth]{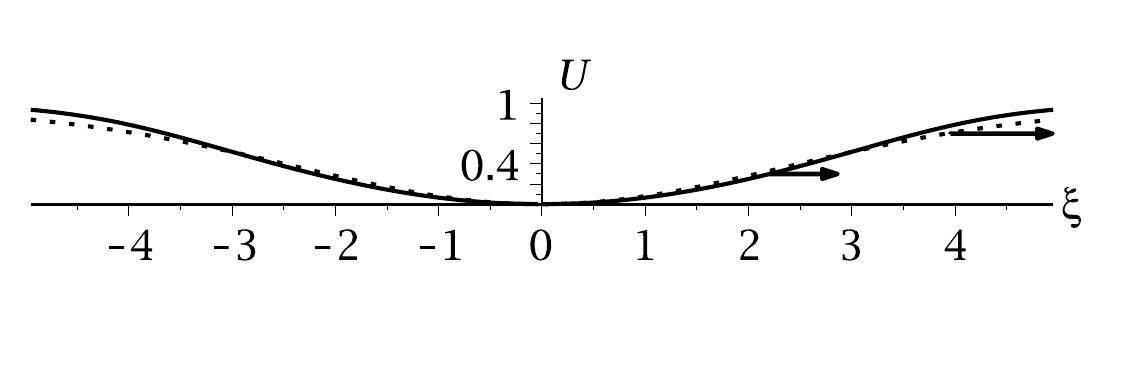}
\caption{defocusing-gKdV dark solitary waves on a positive background for $p=4$ (solid) and $p=2$ with same background and same depth (dot). Arrows indicate direction and speed.}\label{gkvdfig-pis4-defocus-dark-solitary}
\end{figure}

These steps yield
\begin{subequations}\label{gkdv-pis4-defocus-solitary-dark}
\begin{align}
&\begin{aligned}
&
\frac{\epsilon_-\mu_-\kappa_+}{\nu (h+g) h \kappa_-}
\Pi\bigg( \frac{\epsilon_+^2\kappa_-^2}{h(h+g)}, \sn^{-1}\bigg(\frac{\nu\sqrt{U-b+h+g}\sqrt{U-b+h}}{\epsilon_+(2\nu(b-U) -\kappa_-)},\frac{\tau^2 \epsilon_+^2}{\nu^2}\bigg),\frac{\tau^2 \epsilon_+^2}{\nu^2} \bigg)
\\&\qquad
+\frac{2 \nu}{\sqrt{\mu_1 \mu_2} \kappa_-}
\cn^{-1}\bigg( \frac{\mu_-(2\nu(b-U) -\kappa_+)}{\mu_+(2\nu(b -U) -\kappa_-)}, \frac{\tau^2 \epsilon_+^2}{\nu^2} \bigg)
\\&\qquad
+ \frac{\sqrt{8\mu_1\mu_2}}{b \rho \sqrt{(h+g) h}}
\arctanh\bigg( \frac{b\rho \sqrt{U-b+h+g}\sqrt{U-b+h}}{4\sqrt{15}\sqrt{\mu_1\mu_2} \sqrt{(h+g) h} \sqrt{W(U)}} \bigg)
=\frac{1}{\sqrt{15}}|\xi|,
\end{aligned}
\\
&\begin{aligned}
b-h\leq U\leq b,
\quad
b>0
\end{aligned}
\\
&\begin{aligned}
&
\mu_1 = \sqrt{15 b^2-6 b g-18 b h+g^2+4 g h+6 h^2},
\\
&
\mu_2 =\sqrt{15 b^2-12 b g-18 b h+3 g^2+8 g h+6 h^2},
\\
&
\mu_\pm = \mu_1\pm\mu_2,
\quad
\nu = 2 h+g -3 b,
\quad
\epsilon_\pm = \mu_\pm/(4\sqrt{\mu_1\mu_2}),
\quad
\kappa_\pm = 2\nu h -\mu_\pm \mu_2 ,
\\
& 
\rho = \sqrt{6 b(2b-\nu) +\mu_1^2+\mu_2^2},
\quad
\tau=\sqrt{4\nu^2 -\mu_-^2},
\end{aligned}
\end{align}
\end{subequations}
giving an implicit algebraic expression for the dark solitary wave solution $U(\xi)$
on a background $b>0$ with height $h\leq 2b$. 
Expressions \eqref{gkdv-evenp-defocus-solitary-bispos-c} and \eqref{gkdv-evenp-defocus-solitary-bispos-w}
for the speed and the width of the solitary wave yield
\begin{equation}\label{gkdv-pis4-defocus-solitary-bispos-c-w}
c=-b^4+\tfrac{4}{3}b^3h-b^2h^2 +\tfrac{2}{5}bh^3 -\tfrac{1}{15}h^4,
\quad
w=\sqrt{60/(h(20b^3 -15b^2h +6bh^2 -h^3))} .
\end{equation}
Fig.~\ref{gkvdfig-pis4-defocus-dark-solitary} 
compares the solution to the defocusing-mKdV dark solitary wave solution. 

To obtain the bright solitary wave, we apply the reflection
$b\to -b$ and $U\to -U$, 
yielding
\begin{subequations}\label{gkdv-pis4-defocus-solitary-bright}
\begin{align}
&\begin{aligned}
&
\frac{\epsilon_-\mu_-\kappa_+}{\nu (h+g) h \kappa_-}
\Pi\bigg( \frac{\epsilon_+^2\kappa_-^2}{h(h+g)}, \sn^{-1}\bigg(\frac{\nu\sqrt{h-|b|+g-U}\sqrt{h-|b|-U}}{\epsilon_+(2\nu(|b|+U) -\kappa_-)},\frac{\tau^2 \epsilon_+^2}{\nu^2}\bigg),\frac{\tau^2 \epsilon_+^2}{\nu^2} \bigg)
\\&\qquad
+\frac{2 \nu}{\sqrt{\mu_1 \mu_2} \kappa_-}
\cn^{-1}\bigg( \frac{\mu_-(2\nu(|b|+U) -\kappa_+)}{\mu_+(2\nu(|b| +U) -\kappa_-)}, \frac{\tau^2 \epsilon_+^2}{\nu^2} \bigg)
\\&\qquad
+ \frac{\sqrt{8\mu_1\mu_2}}{|b| \rho \sqrt{(h+g) h}}
\arctanh\bigg( \frac{|b|\rho \sqrt{h+g-|b|-U}\sqrt{h-|b|-U}}{4\sqrt{15}\sqrt{\mu_1\mu_2} \sqrt{(h+g) h} \sqrt{W(U)}} \bigg)
=\frac{1}{\sqrt{15}}|\xi|,
\end{aligned}
\\
&\begin{aligned}
b\leq U\leq b+h,
\quad
b<0
\end{aligned}
\\
&\begin{aligned}
&
\mu_1 = \sqrt{15 b^2 -6|b| g -18|b| h+g^2+4 g h+6 h^2},
\\
&
\mu_2 =\sqrt{15 b^2 -12|b| g -18|b| h+3 g^2+8 g h+6 h^2},
\\
&
\mu_\pm = \mu_1\pm\mu_2,
\quad
\nu = 2 h +g -3|b|,
\quad
\epsilon_\pm = \mu_\pm/(4\sqrt{\mu_1\mu_2}),
\quad
\kappa_\pm = 2\nu h -\mu_\pm\mu_2,
\\
& 
\rho = \sqrt{6 |b|(2|b| -\nu) +\mu_1^2+\mu_2^2},
\quad
\tau=\sqrt{4\nu^2 -\mu_-^2} .
\end{aligned}
\end{align}
\end{subequations}
This gives an implicit algebraic expression for the bright solitary wave solution $U(\xi)$
on a background $b<0$ with height $h\leq 2|b|$. 
The speed and the width of this solitary wave are given by 
\begin{equation}\label{gkdv-pis4-defocus-solitary-c-w}
c=-b^4 +\tfrac{4}{3}|b|^3h -b^2h^2 +\tfrac{2}{5}|b|h^3 -\tfrac{1}{15}h^4,
\quad
w=\sqrt{60/(h(20|b|^3 -15b^2h +6|b|h^2 -h^3))} .
\end{equation}

\begin{figure}[h]
\centering
\includegraphics[trim=2cm 16cm 8cm 3cm,clip,width=0.34\textwidth]{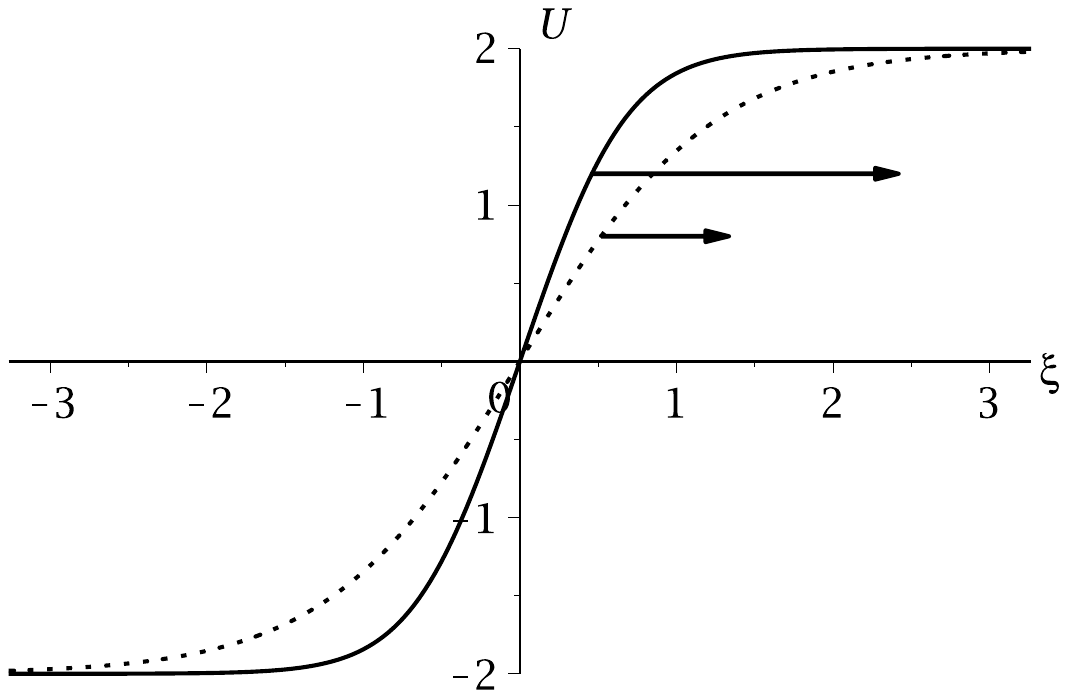}
\;\;
\includegraphics[trim=2cm 18cm 8cm 2cm,clip,width=0.63\textwidth]{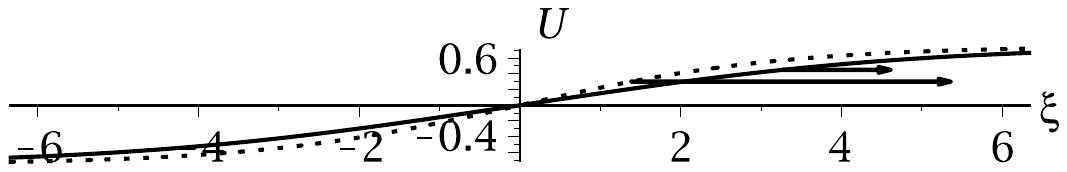}
\caption{gKdV kink waves for $p=4$ (solid) and $p=2$ with same background (dot). Arrows indicate direction and speed.}\label{gkvdfig-pis4-kink-b}
\end{figure}

\begin{figure}[h]
\centering
\includegraphics[trim=2cm 15cm 9cm 2cm,clip,width=0.41\textwidth]{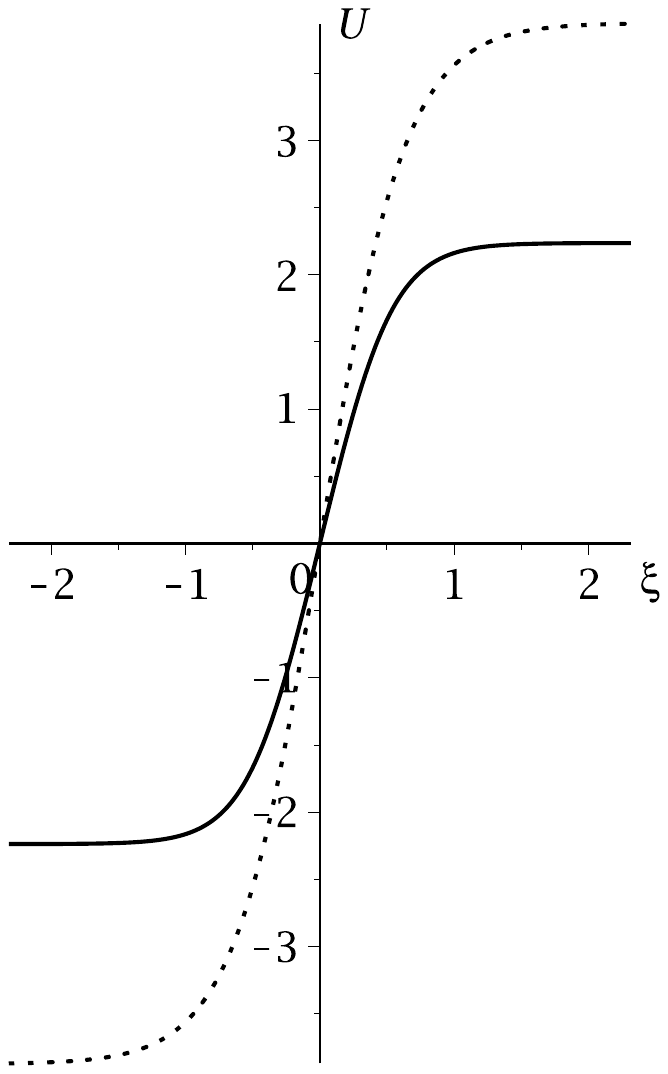}
\;\;
\includegraphics[trim=2cm 16cm 7cm 2cm,clip,width=0.56\textwidth]{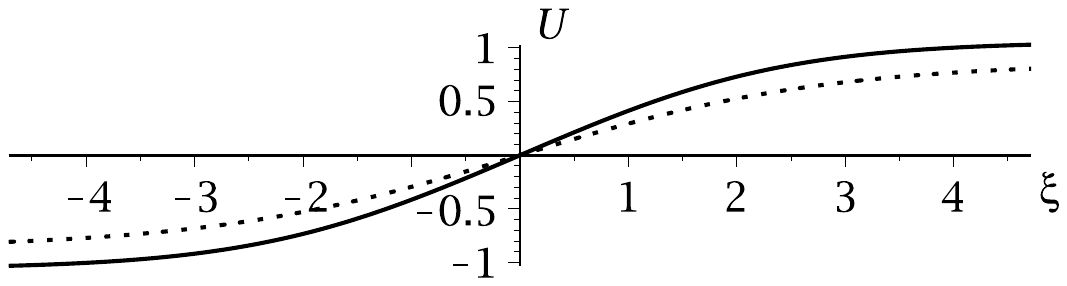}
\caption{gKdV kink waves for $p=4$ (solid) and $p=2$ (dot) with same speed.}\label{gkvdfig-pis4-kink-c}
\end{figure}

Finally, for the kink wave, 
we have $W(U) = \tfrac{1}{30}(U^2+2b^2)$ from equation \eqref{gkdv-evenp-defocus-kink-W},
with the quadrature \eqref{gkdv-evenp-defocus-kink-b-quadrature} yielding 
\begin{equation}
\begin{aligned}
\frac{1}{2\sqrt{3}b^2}\arctanh\bigg( \frac{U(\sqrt{3U^2+6b^2}+3b)(U^2+\sqrt{3U^2+6b^2}b+2b^2)}{(2U^2+b^2)(U^2+2\sqrt{3U^2+6b^2}b+5b^2)} \bigg)
= \frac{1}{\sqrt{15}}\, \xi . 
\end{aligned}
\end{equation}
This algebraic equation can be explicitly solved to obtain the kink wave 
\begin{equation}
U(\xi)= \frac{2b\sinh(\tfrac{1}{\sqrt{5}}b^2\xi)}{\sqrt{2\cosh(\tfrac{2}{\sqrt{5}}b^2 \xi)+4}},
\quad
b>0 . 
\end{equation}
Expressions \eqref{gkdv-evenp-defocus-kink-c} and \eqref{gkdv-evenp-defocus-kink-w}
for the speed and the width of the kink are given by 
\begin{equation}
w=\sqrt{5}/b^2, 
\quad
c = -\tfrac{1}{5}b^4 <0 .
\end{equation}
Figs.~\ref{gkvdfig-pis4-kink-b} and~\ref{gkvdfig-pis4-kink-c}
compare this kink solution to the mKdV kink solution.

\section{Even-power gKdV travelling waves with non-zero boundary conditions in the focusing case}\label{sec:gkdv-even-p-focus}

As explained in section~\ref{sec:gkdv-classify}, 
when the nonlinearity power $p$ is an even integer, 
then the gKdV equation in the focusing case 
supports solitary waves on non-zero backgrounds 
and, in a certain limit, heavy-tail waves. 
No other kind of bounded non-periodic waves are supported. 

We will now study the kinematic properties of these solutions 
and derive analytical expressions in the cases $p=2,4$. 
The discussion will use the same physical parameterization introduced in the previous section. 

\subsection{Focusing-gKdV solitary waves}

\begin{figure}[h]
\centering
\includegraphics[trim=1cm 12cm 5cm 2cm,clip,width=0.5\textwidth]{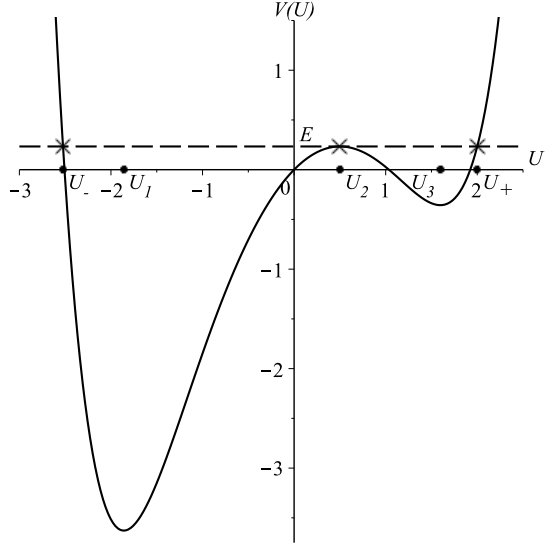} 
\caption{gKdV focusing potential well for $p=4$.}\label{gkvdfig-potentialwell-pis4-focus}
\end{figure}

For any odd power $p\geq 1$, 
two solitary waves arise under the condition \eqref{gkdv-evenp-Mcond} on the mass parameter $M$ 
in the focusing gKdV potential \eqref{gkdv-evenp-focus-potentialwell}. 
This potential has three critical points \eqref{gkdv-evenp-focus-critpoints}
consisting of two local minimums $U_1$ and $U_3$, and a local maximum $U_2$. 
The solitary wave solutions $U(\xi)$ are obtained from 
the resulting nonlinear oscillator equation \eqref{scaled-oscil-ode}
by taking $E=V_\max=V(U_2)$. 
The potential \eqref{gkdv-evenp-focus-potentialwell}
thereby has the factorization 
\begin{equation}\label{gkdv-evenp-focus-energyeqn}
V(U_2) - V(U)
=(U-U_2)^2(U_+-U)(U-U_-) W(U), 
\end{equation}
where $U_+>U_3>0$ and $U_-<U_1<0$ are turning points, 
with 
\begin{equation}
U_- <U_2 < U_+, 
\end{equation}
and where $W(U)$ is a polynomial that has even degree $p-2$ 
and that is positive on $U_-\leq U\leq U_+$. 
We therefore have two turning points and an asymptotic turning point, 
which yield two different solitary wave solutions $U(\xi)$ 
corresponding to the adjacent pairs of turning points
$(U_-,U_2)$ and $(U_2,U_+)$. 
See the potential well in Fig.~\ref{gkvdfig-potentialwell-pis4-focus}. 
Due to the reflection property \eqref{even-p-V-reflect} of the potential,
these points transform as 
$(U_-,U_2,U_+)\to (-U_+,-U_2,-U_-)$
under $M\to-M$. 
Also, as shown by the properties \eqref{gkdv-evenp-focus-critpoints-rel}, 
$M$ and $U_2$ have opposite signs,
while $U_+>|U_-|$ when $M>0$, and $U_+<|U_-|$ when $M<0$. 

The quadrature \eqref{quadrature} for $U(\xi)$ is given by 
\begin{equation}\label{gkdv-evenp-focus-solitary1-quadrature}
\int^{U_+}_{U} \frac{dU}{(U-U_2)\sqrt{(U_+-U)(U-U_-)W(U)}} = \sqrt{2}\, |\xi|,
\quad
U_2\leq U\leq U_+
\end{equation}
for the first solution, 
and 
\begin{equation}\label{gkdv-evenp-focus-solitary2-quadrature}
\int^{U}_{U_-} \frac{dU}{(U_2-U)\sqrt{(U_+-U)(U-U_-)W(U)}} = \sqrt{2}\, |\xi|,
\quad
U_-\leq U\leq U_2
\end{equation}
for the second solution. 
When $p=4$, both integrals can be evaluated in terms of elliptic functions as shown later, 
but for $p\geq 6$,
there is no explicit evaluation in general.
For $p=2$, $W(U)$ is a positive constant, 
and so the integral has a straightforward evaluation in terms of elementary functions.

\subsection{Kinematics}\label{sec:solitarywave-evenp-focus}

For any even power $p\geq 2$, 
the two quadratures \eqref{gkdv-evenp-focus-solitary1-quadrature} and \eqref{gkdv-evenp-focus-solitary2-quadrature}
each implicitly define a family of the solitary wave solutions $U(\xi)$, 
parameterized by $(c,M)$.
These solution families have not been studied previously when the background is non-zero. 
They have the following main kinematic properties.

Due to condition \eqref{gkdv-evenp-Mcond}, 
the wave speed for both families is $c>0$. 
They also share the same background (asymptote) $b=U_2$,
and the same width proportional to $w=2/\sqrt{c-|U_2|^p}$,
which is obtained from an asymptotic expansion of the quadratures as $U\to U_2$,
combined with $V''(U_2)=2(U_+-U_2)(U_2-U_-)W(U_2)=U_2^p-c$ from the factorization \eqref{gkdv-evenp-focus-energyeqn}. 

The first solution family \eqref{gkdv-evenp-focus-solitary1-quadrature} 
has a wave peak $U_+>0$ and a wave height $h_+=U_+-U_2>0$, 
while the second solution family \eqref{gkdv-evenp-focus-solitary2-quadrature} 
has a wave peak $U_-<0$ and a wave depth $h_-=U_2-U_->0$. 
In both families, $U(\xi)$ is an even function of $\xi$. 

The background $b=U_2$ in these solution families can be positive or negative, 
with the speed $c$ being larger than $c_\min = |b|^p$,
so that $w^2 >0$. 
Alternatively, the speed can be taken to be any positive value, 
with the background satisfying $|b|_\max = c^{1/p}$.
Physically, the first solution family \eqref{gkdv-evenp-focus-solitary1-quadrature}
describes a \emph{bright solitary wave on a positive/negative background},
parameterized by $(h_+,b)$; 
the second solution family \eqref{gkdv-evenp-focus-solitary2-quadrature}
describes a \emph{dark solitary wave on a positive/negative background},
parameterized by $(h_-,b)$. 

In both solution families, 
the peaks $U_\pm$ and the background $b=U_2$ are implicitly given in terms of $(c,M)$ by the polynomial equations
\begin{equation}
\tfrac{1}{p+1}U_2^{p+1} - cU_2 -M=0,
\quad
\tfrac{1}{(p+1)(p+2)}(U_\pm^{p+2} -U_2^{p+2}) 
-\tfrac{1}{2}c(U_\pm^2 - U_2^2) -M(U_\pm -U_2) =0
\end{equation}
which come from $V'(U_2)=0$ and $V(U_2)=V(U_\pm)$. 
An equivalent implicit physical parameterization in terms of $(c,b)$ is provided by 
\begin{equation}\label{gkdv-evenp-focus-solitary-b-c}
M = \tfrac{1}{p+1}b^{p+1} -cb,
\quad
U_1=b,
\quad
U_\pm = bz, 
\quad
\tfrac{1}{p+1} |b|^p (\tfrac{1}{p+2}S_{p+2}(z)  -1) -\tfrac{1}{2}c(z -1) =0,
\end{equation}
where $S_{p+2}(z)$ is the polynomial \eqref{Spoly}.

Similarly to the defocusing case,
a more useful explicit physical parameterization of the two solution families
can be obtained by writing out the quadratures \eqref{gkdv-evenp-focus-solitary1-quadrature}--\eqref{gkdv-evenp-focus-solitary2-quadrature}
in terms of $(b,h_\pm)$. 
As shown in the Appendix, 
this yields
\begin{align}
& \int^{b+h_+}_{U} \frac{dU}{(U-b)\sqrt{(h_++b-U)(U +h_--b)W(U)}} = \sqrt{2}\, |\xi|,
\quad
b\leq U\leq b+h_+, 
\label{gkdv-evenp-focus-solitary1-bh-quadrature}
\\
& \int^{U}_{b-h_-} \frac{dU}{(b-U)\sqrt{(h_++b-U)(U +h_--b)W(U)}} = \sqrt{2}\, |\xi|,
\quad
b-h_-\leq U\leq b,
\label{gkdv-evenp-focus-solitary2-bh-quadrature}
\end{align}
with
\begin{equation}\label{gkdv-evenp-focus-solitary-bh-W}
\begin{aligned}
W(U) =
\tfrac{1}{(p+1)(p+2)}|b|^{p-2}b & \big( 
R_{p-1}(1+h_+/b,1-h_-/b) + R_{p-2}(1+h_+/b,1-h_-/b)(U/b)
\\&\qquad
+ \cdots + R_{1}(1+h_+/b,1-h_-/b)(U/b)^{p-2}
\big) ,
\end{aligned}
\end{equation}
where $h_+$ and $h_-$ are related by 
\begin{equation}\label{gkdv-evenp-focus-solitary-hp-hm}
R_{p+1}(1 + h_+/b)  = R_{p+1}(1 - h_-/b) ,
\end{equation}
and where $R_n(z)$ and $R_n(y,z)$ are given by the polynomials \eqref{Rpoly} and \eqref{doubleRpoly}. 

The speed and the width of both the bright and the dark solitary waves
are then given by the expressions
\begin{equation}\label{gkdv-evenp-focus-solitary-bh-c}
c = \tfrac{2}{(p+1)(p+2)} |b|^p R_{p+1}(1 \pm h_\pm/b)
= \tfrac{2}{(p+1)(p+2)} ((b\pm h_\pm)^{p+2} \mp(p+2)h_\pm b^{p+1} -b^{p+2})/h_\pm^2
\end{equation}
and 
\begin{equation}\label{gkdv-evenp-focus-solitary-bh-w}
\begin{aligned}
w & = 2/\sqrt{|b|^p(\tfrac{2}{(p+1)(p+2)} R_{p+1}(1 \pm h_\pm/b) -1)}
\\
&= \sqrt{2(p+1)(p+2)}h_\pm/(2(b\pm h_\pm)^{p+2}-b^{p+2}) \mp 2(p+2)h_\pm b^{p+1} -(p+1)(p+2)h_\pm^2b^p) . 
\end{aligned}
\end{equation}
A kinematic relation \eqref{gkdv-oddp-solitary-bh-cwrel} holds between $c$ and $w$,
which is the same as in the odd power case.

Note that the polynomial relation \eqref{gkdv-evenp-focus-solitary-hp-hm}
determines $h_-$ in terms of the parameters $(h_+,b)$ for the first solution family \eqref{gkdv-evenp-focus-solitary1-bh-quadrature},
and likewise determines $h_+$ in terms of the parameters $(h_-,b)$ for the second solution family \eqref{gkdv-evenp-focus-solitary2-bh-quadrature}. 

The kinematic features of these two solitary wave solution families 
are consequently tied to the properties of the polynomials $R_{p+1}(1\pm \eta)$
as a function of $\eta$ for $\eta\geq 0$. 
Specifically,
$R_{p+1}(1 + \eta)$ is positive and increasing,
with $R_{p+1}(1)=\tfrac{(p+1)(p+2)}{2}$; 
$R_{p+1}(1 - \eta)$ is positive and convex,
has its minimum value $\tfrac{p+2}{2}$ at $\eta=2$,
and reaches the value $R_{p+1}(1)=\tfrac{(p+1)(p+2)}{2}$
at $\eta=0$ and $\eta=q_{p+1}$
where $1< q_{p+1} \leq 4$
(with $q_{3}=4$ when $p=2$).
In particular, $q_{p+1}$ is the positive root of the polynomial equation 
\begin{equation}\label{q-root}
R_{p+1}(1 -q_{p+1}) = \tfrac{(p+1)(p+2)}{2} . 
\end{equation}  

As a result, each of the solitary wave solution families \eqref{gkdv-evenp-focus-solitary1-bh-quadrature}--\eqref{gkdv-evenp-focus-solitary2-bh-quadrature}
is well-defined with $h_\pm$ and $b$ obeying
\begin{equation}
h_+ >
\begin{cases}
0, & b>0
\\
q_{p+1}|b|, & b<0
\end{cases},
\quad
h_- >
\begin{cases}
q_{p+1}b, & b>0
\\
0 & b<0 
\end{cases}
\end{equation}
for both solution families.

In the case $\pm b>0$, 
when $c$ decreases to $c_\min=|b|^p$,
the respective height/depth of the solitary waves goes to zero
and their width goes to infinity,
corresponding to $U\to b$.
Interestingly,
the limiting behaviour in the other case $\pm b<0$ is very different,
as the height/depth goes to a non-zero value.
This limit turns out to describe a heavy-tail wave,
which will be derived in the next subsection. 
In both cases,
as $h_\pm$ increases,
the wave speed $c$ is positive and increases while $w$ decreases to $0$. 

Just like in the odd power case,
the well-known solitary waves on a zero background, $b=0$, for even $p$ 
can be obtained from the quadratures \eqref{gkdv-evenp-focus-solitary1-quadrature} and \eqref{gkdv-evenp-focus-solitary2-quadrature}
in the limit $b\to0$. 
This leads to the solutions 
\begin{equation}\label{gkdv-evenp-focus-soliton}
U(\xi) = \pm h\, \sech^{2/p}\Big(\tfrac{p}{\sqrt{2(p+1)(p+2)}}h^{p/2} \xi\Big),
\quad
p\text{ is even }
\end{equation}
where the $\pm$ sign represents an up/down orientation. 
The speed and width are given by the expressions \eqref{gkdv-oddp-soliton-cw},
which are unchanged compared to the odd power case.

\subsection{Focusing even-power gKdV heavy-tail waves on a non-zero background}\label{subsec:gkdv-even-tailwave}

Heavy-tail wave solutions $U(\xi)$ arise from
the limiting case \eqref{gkdv-evenp-foc-Mlimit-critpoints} of
the focusing gKdV potential well \eqref{gkdv-evenp-focus-potentialwell}
where one of the local maximums $U_1$ or $U_3$ coincides with the local minimum $U_2$,
yielding an inflection point.
There are two separate cases, corresponding to $\sgn(M)=\pm1$,
where $M = \pm\tfrac{p}{p+1} c^{1+1/p}$ is the mass parameter in the potential,
and $c>0$ is the wave speed. 
They are related by the reflection symmetry \eqref{even-p-V-reflect}.

\begin{figure}[h]
\centering
\includegraphics[trim=1cm 12cm 5cm 2cm,clip,width=0.5\textwidth]{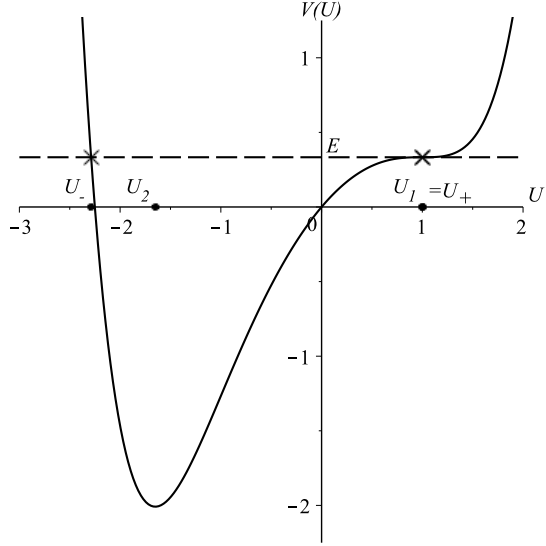} 
\caption{gKdV focusing potential well for $p=4$, with inflection point.}\label{gkvdfig-potentialwell-pis4-focus-inflect}
\end{figure}

In the case $\sgn(M)=1$, 
the turning points in the potential well are 
$U_-=U_2=-c^{1/p}$ and $U_+>U_0$,
which correspond to the factorization 
$V_\inflect -V(U)= (U-U_-)^3(U_+-U)W(U)$. 
In the other case $\sgn(M)=-1$, 
the turning points are 
$U_+=U_2=c^{1/p}$ and $U_-<U_0$,
corresponding to the factorization 
$V_\inflect -V(U)= (U_+-U)^3(U-U_-)W(U)$. 
Here, in both cases,
$W(U)$ is a positive polynomial with even degree $p-2$ 
on $U_-\leq U\leq U_+$,
and the inflection point is given by
\begin{equation}
V_\inflect =V(U_2) = \tfrac{p}{2(p+2)}c^{1+p/2} . 
\end{equation}
The potential well is shown in Fig.~\ref{gkvdfig-potentialwell-pis4-focus-inflect}. 

Using the previous quadratures \eqref{gkdv-evenp-focus-solitary1-bh-quadrature} and \eqref{gkdv-evenp-focus-solitary2-bh-quadrature}, 
we can respectively take $(h_+,h_-)=(h,0)$ with $b<0$, and $(h_+,h_-)=(0,h)$ with $b>0$,
since this corresponds to a local maximum coinciding with the local minimum
in the gKdV potential.
Hence, the respective quadratures become
\begin{subequations}
\begin{equation}\label{gkdv-evenp-focus-heavytail1-bh-quadrature}
\int^{b+h}_{U} \frac{dU}{\sqrt{(U -b)^3(h+b-U)W(U)}} = \sqrt{2}\, |\xi|,
\quad
b\leq U\leq b+h,
\quad
b=-c^{1/p}<0
\end{equation}
and 
\begin{equation}\label{gkdv-evenp-focus-heavytail2-bh-quadrature}
\int^{U}_{b-h} \frac{dU}{\sqrt{(b-U)^3(U -b+h)W(U)}} = \sqrt{2}\, |\xi|,
\quad
b-h\leq U\leq b,
\quad
b=c^{1/p} >0
\end{equation}
\end{subequations}
with
\begin{equation}\label{gkdv-evenp-heavytail-W}
\begin{aligned}
W(U) =
\tfrac{1}{(p+1)(p+2)}|b|^{p-1}b/h & \big(
(R_{p}(1)-R_{p}(1-h/|b|)) + (R_{p-1}(1)-R_{p-1}(1+h/|b|))(U/b)
\\&\qquad
+ \cdots + (R_{2}(1)-R_{2}(1+h/|b|))(U/b)^{p-2} \big)
\end{aligned}
\end{equation}
which is a polynomial of degree $p-2$,
where $R_n(z)$ is the polynomial \eqref{Rpoly},
and $R_n(1)=\tfrac{n(n+1)}{2}$. 
The previous relation \eqref{gkdv-evenp-focus-solitary-bh-c} for the wave speed $c$
now yields
\begin{equation}\label{gkdv-evenp-focus-heavytail-h}
R_{p+1}(1 -h/|b|) = \tfrac{(p+1)(p+2)}{2}
\end{equation}
which determines the wave height/depth $h$ in terms of $b$:
specifically, $h = q_{p+1}|b|$ where $q_{p+1}$ is the positive root of the polynomial equation \eqref{q-root}.

The quadratures \eqref{gkdv-evenp-focus-heavytail1-bh-quadrature}--\eqref{gkdv-evenp-focus-heavytail2-bh-quadrature}
implicitly define a family of the heavy-tail wave solutions $U(\xi)$, 
parameterized by $b$, where $b\gtrless 0$. 
Alternatively, they can be parameterized by $c>0$ or $h>0$. 
Their tails have the rational form 
\begin{equation}
U\simeq b - 12/(pb^{p-1}\xi^2)
\end{equation}
as $|\xi|\to\infty$, 
which is obtained from an asymptotic expansion of the quadrature as $U\to b$,
combined with $V'''(b)=6hW(b)=pb^{p-1}$ 
from the factorization of $V_\inflect-V(U)$. 
Thus, since the wave does not decay to $b$ exponentially in $|\xi|$, 
it is not localized in the sense of a solitary wave.
Note that the focusing gKdV potential is different for the two different signs of $b$.
In particular,
$V(U) = \tfrac{1}{(p+1)(p+2)}U^{p+2} - \tfrac{1}{2}|b|^p U^2 +\tfrac{p}{p+1}|b| b U$, 
respectively for the solution families \eqref{gkdv-evenp-focus-heavytail1-bh-quadrature} and \eqref{gkdv-evenp-focus-heavytail2-bh-quadrature}. 

This solution family is new for $p\geq 4$. 
Physically, it describes
a \emph{dark/bright heavy-tail wave on a positive/negative background}. 
When $p=4$, the solutions can be expressed in terms of elliptic functions
as shown later, 
but for $p\geq 6$,
they have no explicit expression in general.
For $p=2$, they are rational functions.

\subsection{Focusing-mKdV ($\boldsymbol{p=2}$) solitary waves and heavy-tail waves on a background}\label{subsec:mkdvsolitary-tailwaves}

In the mKdV case, $p=2$, 
the critical points $U_1,U_2,U_3$, and the turning points $U_\pm$,
can be found explicitly as roots of a cubic. 
The corresponding factorization of the potential \eqref{gkdv-evenp-focus-energyeqn} 
has $W=1$. 

The two quadratures \eqref{gkdv-evenp-focus-solitary1-bh-quadrature} and \eqref{gkdv-evenp-focus-solitary2-bh-quadrature}
for solitary waves yield the physically parameterized solutions
\begin{equation}\label{mkdv-focus-solitary}
U^\pm(\xi)= b+ \frac{h_\pm(h_\pm \pm 4b)}{2b +(2b\pm h_\pm)\cosh(\tfrac{1}{\sqrt{6}}\sqrt{h_\pm(h_\pm \pm 4b)})}
\end{equation}
in terms of the height/dept $h_\pm>0$ and the background $b$,
where the $+/-$ cases respectively correspond to the first and second solutions. 
The first solution family $U^+(\xi)$ physically is 
a \emph{bright solitary wave on a positive/negative background}, 
and the second solution family $U(\xi)^-$ physically is
a \emph{dark solitary wave on a positive/negative background}. 
These solution families were first derived with a non-physical parameterization \cite{JefKak,Au-YeuFunAu1984},
and they have been studied recently in the context of wave modulation theory \cite{KamSpiKon,Mar}. 
Their kinematic properties have not been studied to-date. 

The speed of the waves is 
\begin{equation}
c= \tfrac{1}{6}(2b^2 + (2b \pm h_\pm)^2) >0  . 
\end{equation}
Thus, the wave height $h_+$ and the wave depth $h_-$ obey
$0<h_+<h_-$ when $b>0$, and $h_+>h_->0$ when $b<0$. 

\begin{figure}[h]
\centering
\includegraphics[trim=2cm 15.5cm 8cm 2.5cm,clip,width=0.44\textwidth]{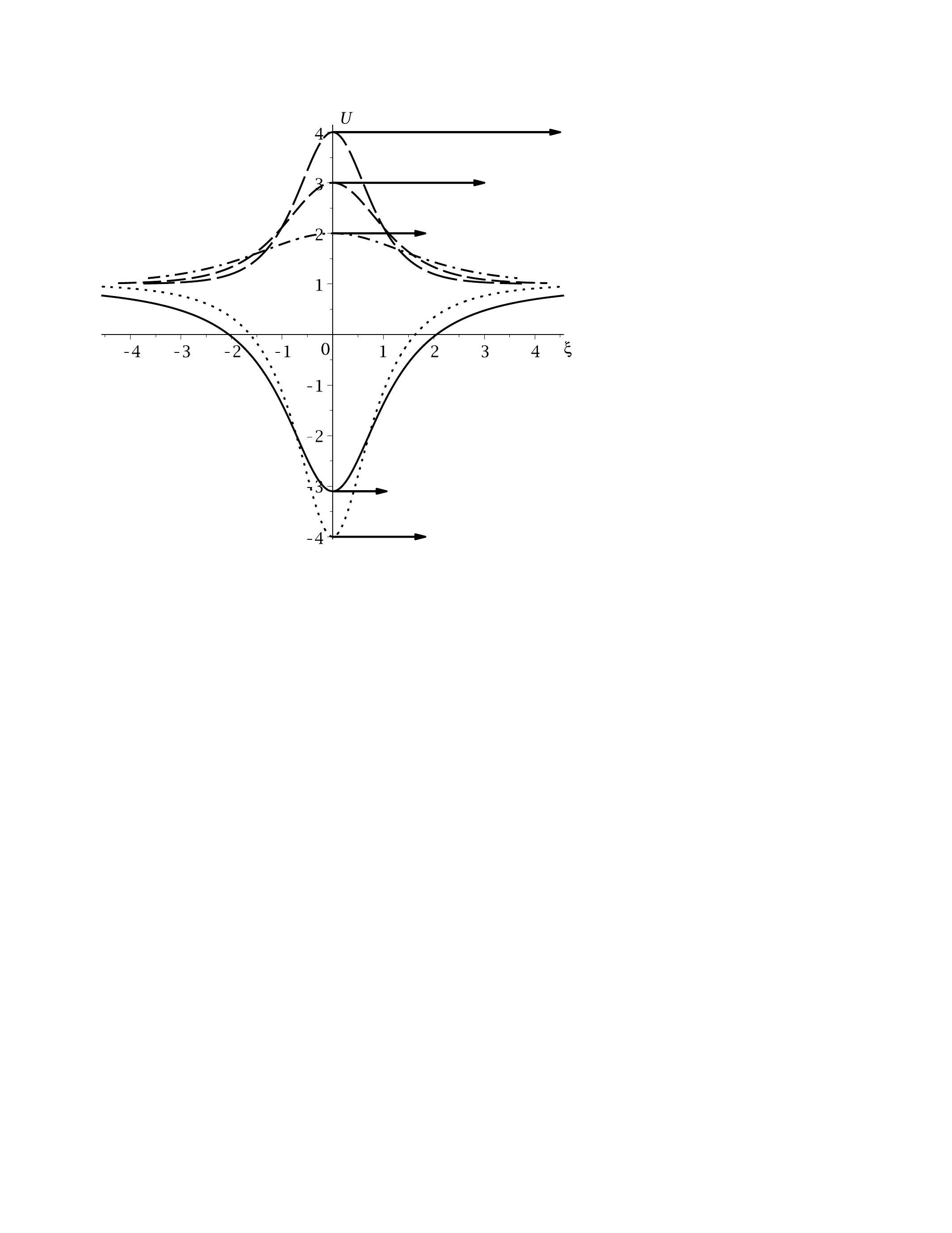}
\quad
\includegraphics[trim=2cm 15.5cm 8cm 2.5cm,clip,width=0.44\textwidth]{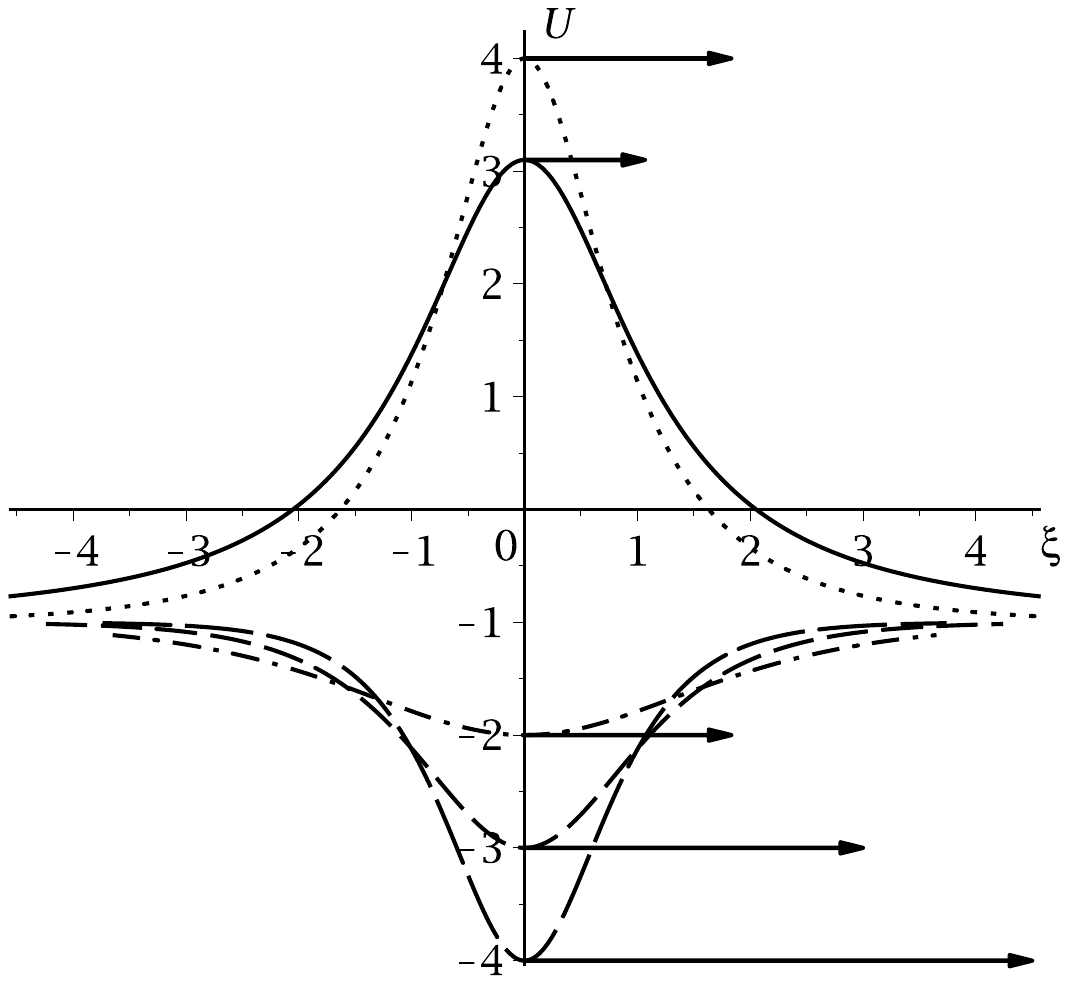}
\caption{Focusing-mKdV solitary waves with different heights on a positive background (left) and a negative background (right). Arrows indicate direction and speed of the waves.}\label{gkdvfig-mkdv-profile-focus-heights}
\end{figure}

\begin{figure}[h]
\centering
\includegraphics[trim=2cm 16cm 6cm 3.5cm,clip,width=0.45\textwidth]{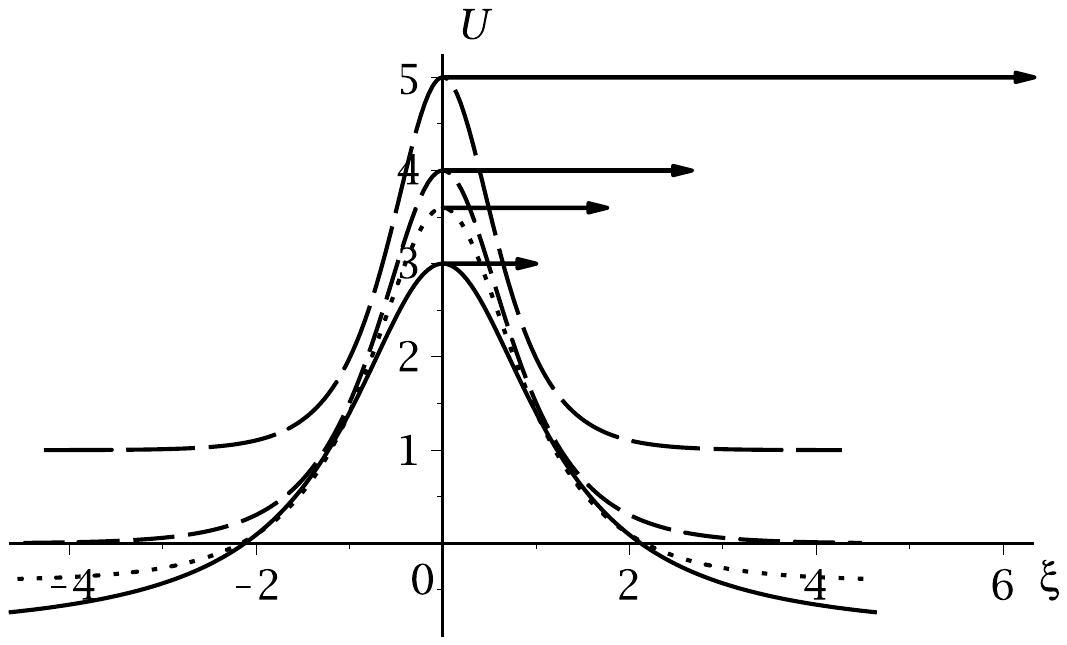}
\quad
\includegraphics[trim=2cm 17cm 8cm 3.5cm,clip,width=0.43\textwidth]{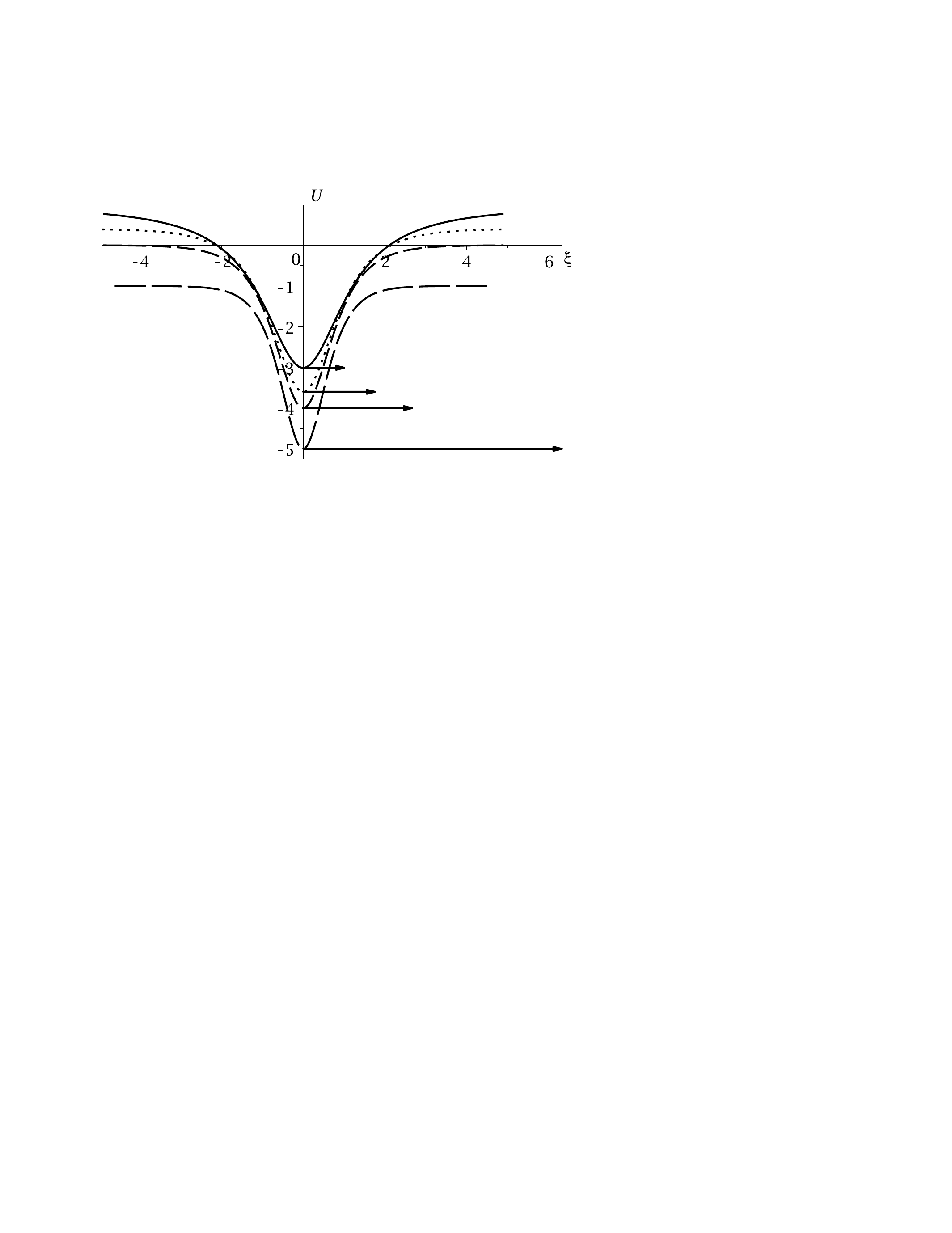}
\caption{Focusing-mKdV bright solitary waves (left) and dark solitary waves (right)}\label{gkdvfig-mkdv-profile-focus-backgrounds}
\end{figure}

The equivalent physical parameterization \eqref{gkdv-evenp-focus-solitary1-quadrature}--\eqref{gkdv-evenp-focus-solitary2-quadrature} and \eqref{gkdv-evenp-focus-solitary-b-c}
for the two quadratures also has an explicit form
\begin{equation}\label{mkdv-focus-solitary-physical}
U^\pm(\xi)= b+ \frac{6(c-b^2)}{2b\pm\sqrt{6c-2b^2}\,\cosh(\sqrt{c-b^2}\,\xi)} ,
\quad
c>b^2\geq 0 . 
\end{equation}
The wave peak is $-b\pm\sqrt{6c-2b^2}$, 
and the wave height/depth is $h_\pm=\sqrt{6c-2b^2} \mp 2b$
while the width of the wave is proportional to $w=2/\sqrt{c-b^2}$.
These quantities satisfy the scaling relation 
\begin{equation}
h_+ h_- w^2= 24 . 
\end{equation}
As the wave speed $c$ decreases, 
the height/depth of the wave decreases while the width increases.
See Figs.~\ref{gkdvfig-mkdv-profile-focus-heights} and~\ref{gkdvfig-mkdv-profile-focus-backgrounds}.
In the limit $c\to c_\min=b^2$, the wave flattens to become $U\to b$. 
The situation for opposite-sign backgrounds, $\pm b<0$, 
is very different,
since the wave height/depth approaches $4|b|\neq 0$.
The limiting wave consists of a heavy-tail wave. 

Finally, 
we evaluate the conserved integrals \eqref{U-mass}--\eqref{U-ener} 
for mass, momentum, and energy
of the solitary waves \eqref{mkdv-focus-solitary-physical}:
\begin{align}
\mathcal{M}_\pm =\pm4\sqrt{6}\,\arctan\bigg(\frac{\sqrt{6c-2b^2} \mp 2b}{\sqrt{6(c-b^2)}}\bigg) , 
\quad
\mathcal{P}_\pm =6\sqrt{c-b^2} , 
\quad
\mathcal{E}_\pm =2c\sqrt{c-6b^2} . 
\end{align}
For $b=0$, these conserved integrals reduce to the well-known mass, momentum, and energy of the focusing mKdV zero-background soliton,
$\mathcal{M}_\pm=\pm\pi \sqrt{6}$, $\mathcal{P}_\pm=6\sqrt{c}$, $\mathcal{E}_\pm=2\sqrt{c^3}$,
where $\pm$ corresponds to the soliton's up/down orientation. 

\begin{figure}[h]
\centering
\includegraphics[trim=2cm 17cm 8cm 3cm,clip,width=0.45\textwidth]{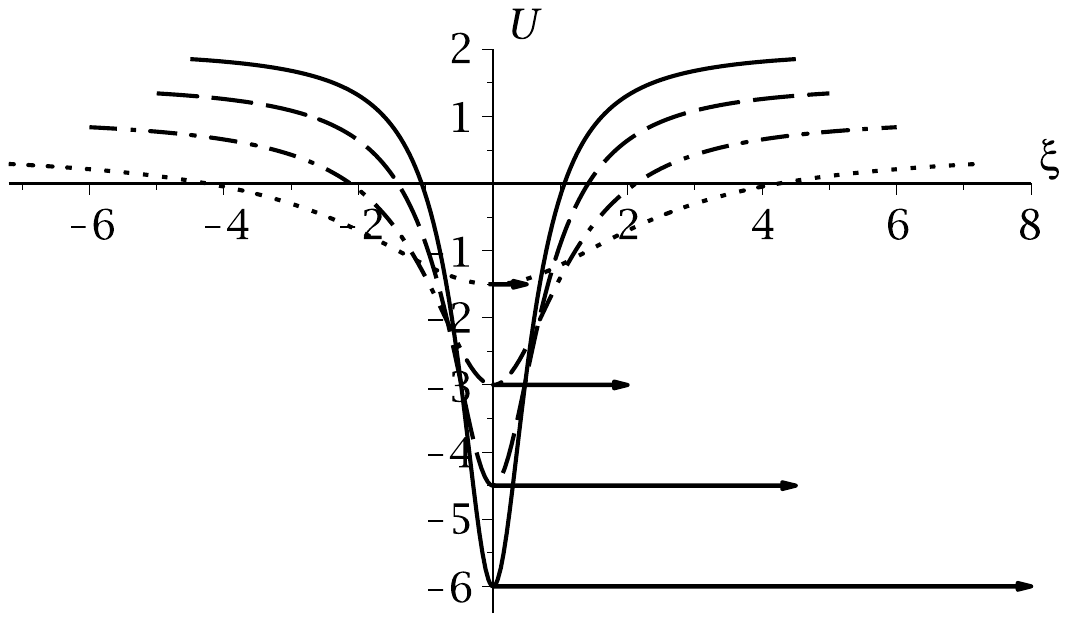} 
\quad
\includegraphics[trim=2cm 17cm 8cm 3cm,clip,width=0.45\textwidth]{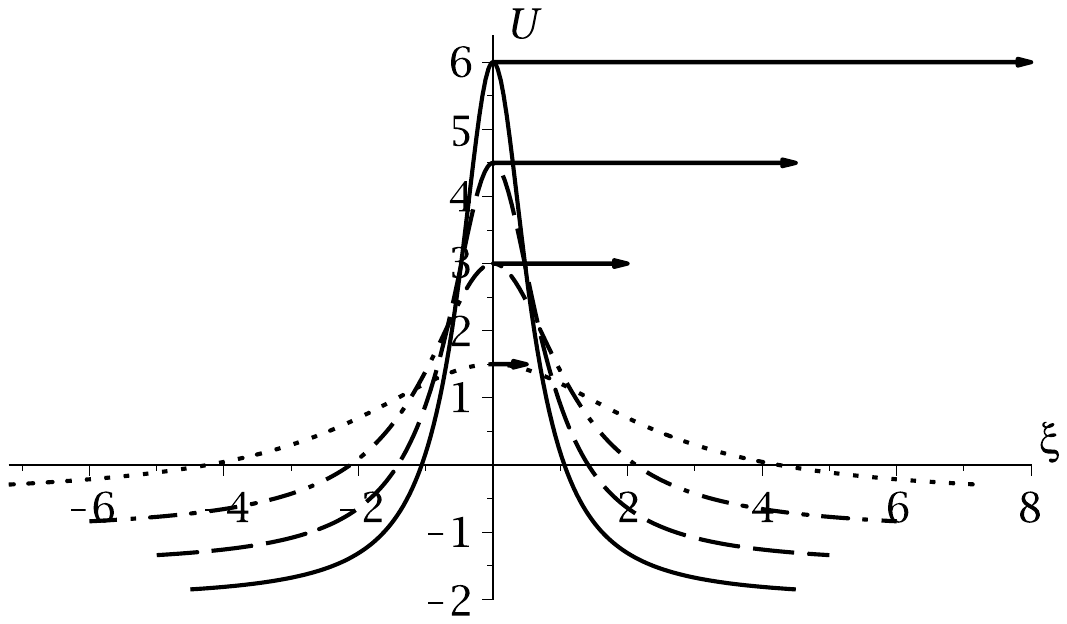} 
\caption{Focusing-mKdV heavy-tail waves on different positive backgrounds (left) and negative backgrounds (right). Arrows indicate direction and speed of the waves.}\label{gkdvfig-mkdv-profile-focus-heavytail}
\end{figure}

For heavy-tail waves in the mKdV case, $p=2$, 
the two quadratures \eqref{gkdv-evenp-focus-heavytail1-bh-quadrature}--\eqref{gkdv-evenp-focus-heavytail2-bh-quadrature}
yield
\begin{equation}\label{mkd-focus-rat}
U(\xi)^{\pm} =\pm \sqrt{c}\left( 1-\frac{12}{2c \xi^2 +3} \right), 
\quad 
c>0 . 
\end{equation}
This a family of rational solutions which decay like $1/\xi^2$. 
Their background (asymptote) is $b=\pm\sqrt{c}$
and their wave peak is $\mp3\sqrt{c}$. 
Hence, the wave height/depth is $h=4\sqrt{c}$. 
See Fig.~\ref{gkdvfig-mkdv-profile-focus-heavytail}.
They were first derived in \Ref{JefKak}. 
Physically, they describe 
a \emph{dark/bright heavy-tail wave} on a positive/negative background.

Finally, 
evaluation of the conserved integrals \eqref{U-mass}--\eqref{U-ener} 
for mass, momentum, and energy of these heavy-tailed waves 
yields 
\begin{equation}
\mathcal{M}_\pm =\mp 2\sqrt{6}\pi, 
\quad
\mathcal{P}_\pm =0,
\quad
\mathcal{E}_\pm =0.
\end{equation}

\subsection{Solitary wave and heavy-tail wave solutions for $\boldsymbol{p=4}$}

The bright and dark solitary wave quadratures \eqref{gkdv-evenp-focus-solitary1-bh-quadrature} and \eqref{gkdv-evenp-focus-solitary2-bh-quadrature}, 
along with the heavy-tail wave quadratures \eqref{gkdv-evenp-focus-heavytail1-bh-quadrature} and \eqref{gkdv-evenp-focus-heavytail2-bh-quadrature}, 
will now be evaluated in terms of elliptic functions,
when $p=4$. 
The resulting solutions will be compared to the mKdV case $p=2$.

See \Ref{AbrSte} for the notation of the elliptic functions.
For the sequel,
we note that the positive root of the polynomial equation \eqref{q-root}
for $p=4$ is given by 
\begin{equation}
q_5=\textstyle{\sqrt[3]{\sqrt{10}+3} - \sqrt[3]{\sqrt{10}-3} +2} . 
\end{equation}

We start from equations \eqref{gkdv-evenp-focus-solitary-bh-W} and \eqref{gkdv-evenp-focus-solitary-hp-hm} for the bright and dark solitary waves,
which give 
\begin{equation}\label{gkdv-pis4-focus-solitary-W}
W(U)=\tfrac{1}{30}(U^2 +(4b +h_+ -h_-)U +10b^2 +5b(h_+-h_-) +h_+^2 +h_-^2 -h_+ h_-)
\end{equation}
where, for the bright solitary wave, 
\begin{equation}\label{gkdv-pis4-focus-brightsolitary-nontp}
\begin{aligned}
& h_-  = \tfrac{1}{3}({\textstyle \sqrt[3]{p+3\sqrt{3q}} +\sqrt[3]{p-3\sqrt{3q}}}) +2 b +\tfrac{1}{3} h_+,
\\
& p= 81b^3 +81b^2 h_++ 45bh_+^2 +10 h_+^3,
\\
& q= (3b^2 -2bh_+ +h_+^2)(90b^4 +120b^3h_+ +79b^2h_+^2 +28bh_+^3 +4h_+^4) ,
\end{aligned}  
\end{equation}
and, for the dark solitary wave,
\begin{equation}\label{gkdv-pis4-focus-darksolitary-nontp}
\begin{aligned}
& h_+  = -\tfrac{1}{3}({\textstyle \sqrt[3]{p+3\sqrt{3q}} -\sqrt[3]{p-3\sqrt{3q}}}) -2 b +\tfrac{1}{3} h_-,
\\
& p= 81b^3 -81b^2 h_-+ 45bh_-^2 -10 h_-^3,
\\
& q= (3b^2 + 2bh_- +h_-^2)(90b^4 -120b^3h_- +79b^2h_-^2 -28bh_-^3 +4h_-^4) . 
\end{aligned}  
\end{equation}
The respective quadratures \eqref{gkdv-evenp-focus-solitary1-bh-quadrature} and \eqref{gkdv-evenp-focus-solitary2-bh-quadrature}
can be evaluated as elliptic integrals similarly to the ones in the focusing case.
See \Ref{AbrSte} for the notation of the elliptic functions.

Under the reflection symmetry \eqref{even-p-V-reflect},
the first quadrature in the case $b\gtrless 0$ corresponds to the second quadrature in the case $b\lessgtr 0$.
Consequently, we will consider just the case $b>0$ for both quadratures. 

Evaluating the first quadrature \eqref{gkdv-evenp-focus-solitary1-bh-quadrature}, 
and making substantial simplifications, 
we obtain
\begin{subequations}\label{gkdv-pis4-focus-brightsolitary}
\begin{align}
&\begin{aligned}
&
\frac{-\kappa_-\epsilon_+(\mu_1+\mu_2)}{\kappa_+\nu h_+ h_-}
\Pi\bigg( \frac{\epsilon_+^2\kappa_+^2}{h_+ h_-},
\sn^{-1}\bigg( \frac{\nu\sqrt{U-b +h_-}\sqrt{b+h_+-U}}{\epsilon_-(2\nu(b -U) +\kappa_+)}, \frac{\tau^2\epsilon_-^2}{\nu^2} \bigg),
\frac{\tau^2\epsilon_-^2}{\nu^2} \bigg)
\\&\qquad
+\frac{2\nu}{\sqrt{\mu_1 \mu_2} \kappa_+}
\cn^{-1}\bigg( \frac{(\mu_2+\mu_1)(2\nu(b-U) +\kappa_-)}{(\mu_2-\mu_1)(2\nu(b -U) +\kappa_+)}, \frac{\tau^2\epsilon_-^2}{\nu^2} \bigg)
\\&\qquad
+\frac{2}{\rho\sqrt{h_+ h_-}}
\arctanh\bigg( \frac{\rho \sqrt{U-b+h_-}\sqrt{b+h_+-U}}{2\sqrt{30}\sqrt{h_+ h_-} \sqrt{W(U)}} \bigg)
=\frac{1}{\sqrt{15}}|\xi|,
 \end{aligned}
\\
&\begin{aligned}
&
b\leq U\leq b+h_+,
\quad
 b>0 , 
\end{aligned}
\\
&\begin{aligned}
&
\mu_1 = \sqrt{15b^2 +6b(h_+ -2h_-) +h_+^2 +3h_-^2 -2h_+ h_-},
\\
&
\mu_2 =\sqrt{15b^2 +6b(2h_+ -h_-) +3h_+^2 +h_-^2 -2h_+ h_-},
\\
&
\epsilon_\pm = (\mu_1\pm\mu_2)/(4\sqrt{\mu_1\mu_2}),
\quad
\nu = h_- -h_+ -3b,
\quad
\kappa_\pm = 2\nu h_+ +\mu_2(\mu_2 \pm\mu_1), 
\\
&
\rho = \sqrt{6b(2b-\nu) +\mu_1^2+\mu_2^2},
\quad
\tau=\sqrt{(\mu_1+\mu_2)^2 -4\nu^2},
\end{aligned}
\end{align}
\end{subequations}
which is an implicit algebraic expression for the bright solitary wave solution $U(\xi)$
on a background $b>0$ with height $h_+>0$.
Here $h_-$ is given in terms of $(h_+,b)$ by expression \eqref{gkdv-pis4-focus-brightsolitary-nontp}
and satisfies $h_->q_5b$. 

\begin{figure}[h]
\centering
\includegraphics[trim=2cm 16cm 6cm 4cm,clip,width=0.37\textwidth]{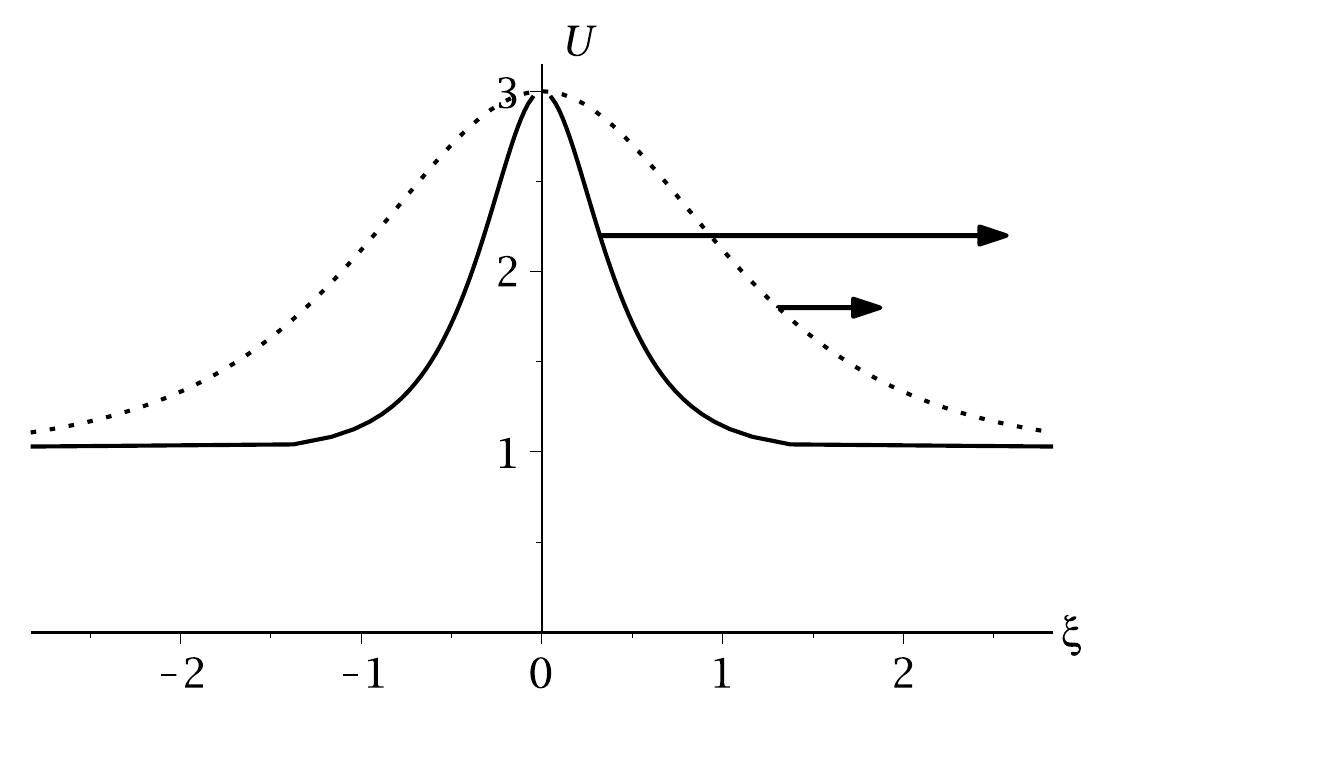}
\quad
\includegraphics[trim=2cm 18cm 6cm 6cm,clip,width=0.59\textwidth]{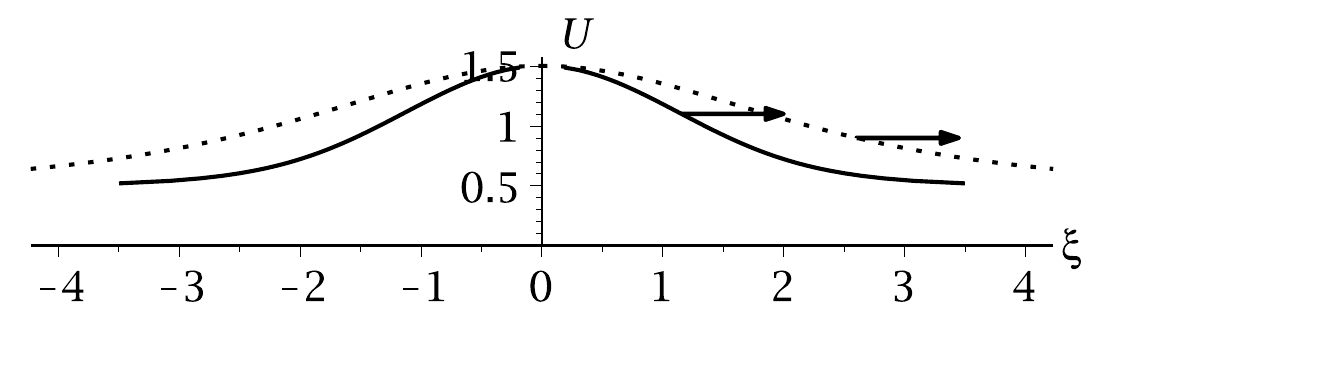}
\quad
\includegraphics[trim=2cm 19cm 8cm 6cm,clip,width=0.9\textwidth]{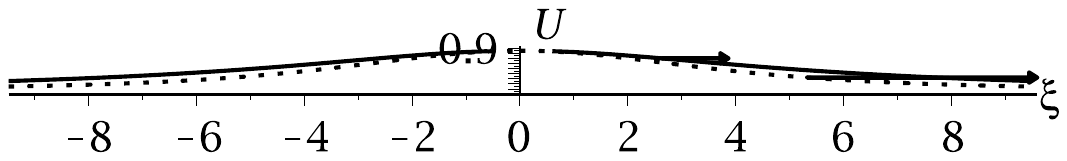}
\caption{Focusing-gKdV bright solitary waves on a positive background for $p=4$ (solid) and $p=2$ with same background and height (dot). Arrows indicate direction and speed.}\label{gkvdfig-pis4-focus-bright-solitary} 
\end{figure}

Expressions \eqref{gkdv-evenp-focus-solitary-bh-c} and \eqref{gkdv-evenp-focus-solitary-bh-w}
for the speed and the width of the solitary wave \eqref{gkdv-pis4-focus-brightsolitary}
yield
\begin{equation}
c = b^4 +\tfrac{4}{3}b^3h_+ +b^2h_+^2 +\tfrac{2}{5}bh_+^3 +\tfrac{1}{15}h_+^4,
\quad
w= 1/\sqrt{15 h_+(h_+^3 +6bh_+^2 +15b^2h_+  +20b^3)} . 
\end{equation}
Fig.~\ref{gkvdfig-pis4-focus-bright-solitary} 
compares this solitary wave solution to the focusing-mKdV bright solitary wave solution. 

The second quadrature \eqref{gkdv-evenp-focus-solitary2-bh-quadrature}
can be evaluated by the same method,
leading to a similar expression involving
$\arctanh$, $\sn^{-1}$, $\cn^{-1}$, and $\Pi$.
However, the $\arctanh$ term and the $\Pi$ term
each have a logarithmic singularity 
at $U =b -(2 h_- -\mu_1(\mu_1-\mu_2)/\nu)(2 h_- -\mu_1(\mu_1+\mu_2)/\nu)/(4 h_--2\mu_1^2/\nu))$ in $b-h_-<U<b$, 
which cancel when they are combined.
These singular terms can be isolated and combined by use of an elliptic identity (cf. 17.7.8 in \Ref{AbrSte}),
yielding a more complicated but singularity-free expression for the quadrature:
\begin{subequations}\label{gkdv-pis4-focus-darksolitary}
\begin{align}
&\begin{aligned}
&
\frac{2}{\rho\sqrt{h_- h_+}}\ln\bigg( 
\frac{(\rho\sqrt{b+h_+ -U}\sqrt{U-b+h_-}+2\sqrt{30}\sqrt{h_+h_-}\sqrt{W(U)})Y(U)}
{4\mu_1 \kappa_+ \nu h_- (b-U)\sqrt{30\mu_-^2 \kappa_+ ^2 W(U)+\tau^2\kappa_-^2 (b+h_+-U)(U-b+h_-)}}
\bigg)
\\&\qquad
+\frac{\kappa_-\epsilon_+ \mu_+}{\kappa_+\nu h_+h_-}\Bigg(
\sn^{-1}\bigg(\frac{\nu\sqrt{b+h_- -U}\sqrt{U-b +h_+}}{\epsilon_- (2\nu(U-b) +\kappa_+)}, \frac{\tau^2 \epsilon_-}{\nu^2}\bigg)
\\&\qquad
-\Pi\bigg( \frac{\tau^2 h_+h_-}{\kappa_+^2}, \sn^{-1}\bigg(\frac{\nu\sqrt{b+h_- -U}\sqrt{U-b +h_+}}{\epsilon_- (2\nu(U-b) +\kappa_+)}, \frac{\tau^2 \epsilon_-}{\nu^2}\bigg), \frac{\tau^2\epsilon_-}{\nu^2} \bigg)
\Bigg)
\\&\qquad
+\frac{2 \nu}{\sqrt{\mu_1\mu_2} \kappa_+}
\cn^{-1}\bigg(
\frac{\mu_+(2\nu(U-b) +\kappa_-)}{\mu_-(2\nu(U-b) +\kappa_+)}, \frac{\tau^2\epsilon_-}{\nu^2}
\bigg)
=\frac{1}{\sqrt{15}}|\xi|,
\end{aligned}
\\
&\begin{aligned}
b-h_-\leq U\leq b,
\end{aligned}
\\
&\begin{aligned}
Y(U) = & 2\sqrt{h_+h_-}(2\nu (U-b) +\kappa_-)\times
\\&\qquad
\sqrt{\kappa_+^2 (2\nu (U-b) +\kappa_+)^2 -2\rho \nu^4 \epsilon^2 \sqrt{\mu_+h_-h_+} (b+h_+-U) (U-b+h_-)}
\\&\qquad
-\rho\kappa_-\mu_+ (2\nu (U-b) +\kappa_+) \sqrt{U-b+h_-}\sqrt{b+h_+-U} ,
\end{aligned}
\\
&\begin{aligned}
& \mu_1 = \sqrt{15 b^2 +6b(h_+ -2h_-) +h_+^2-2h_+ h_- +3h_-^2},
\\
&\mu_2= \sqrt{15b^2 +6b(2h_+ -h_-) +3h_+^2 -2h_+ h_- +h_-^2}, 
\\
&
\mu_\pm = \mu_1 \pm \mu_2,
\quad
\nu =h_--h_+ -3b,
\quad
\rho= \sqrt{ 6b(2b-\nu)+\mu_1^2+\mu_2^2},
\quad
\tau= \sqrt{\mu_+^2-4\nu^2},
\\
&
\kappa_\pm = 2\nu h_- -\mu_1\mu_\pm, 
\quad
\epsilon_\pm = \mu_\pm/(4\sqrt{\mu_1\mu_2}),
\quad
\epsilon = 8\mu_1\mu_2/(\tau\mu_+\mu_-) .
\end{aligned}
\end{align}
\end{subequations}
This is an implicit algebraic expression for the dark solitary wave $U(\xi)$
on a background $b$ with height $h_-$.
Here the height obeys $h_- >q_5 b$, 
while $h_+$ is given in terms of $(h_-,b)$ by expression \eqref{gkdv-pis4-focus-darksolitary-nontp} and satisfies $h_+>0$. 

\begin{figure}[h]
\centering
\includegraphics[trim=2cm 16cm 8cm 4cm,clip,width=0.39\textwidth]{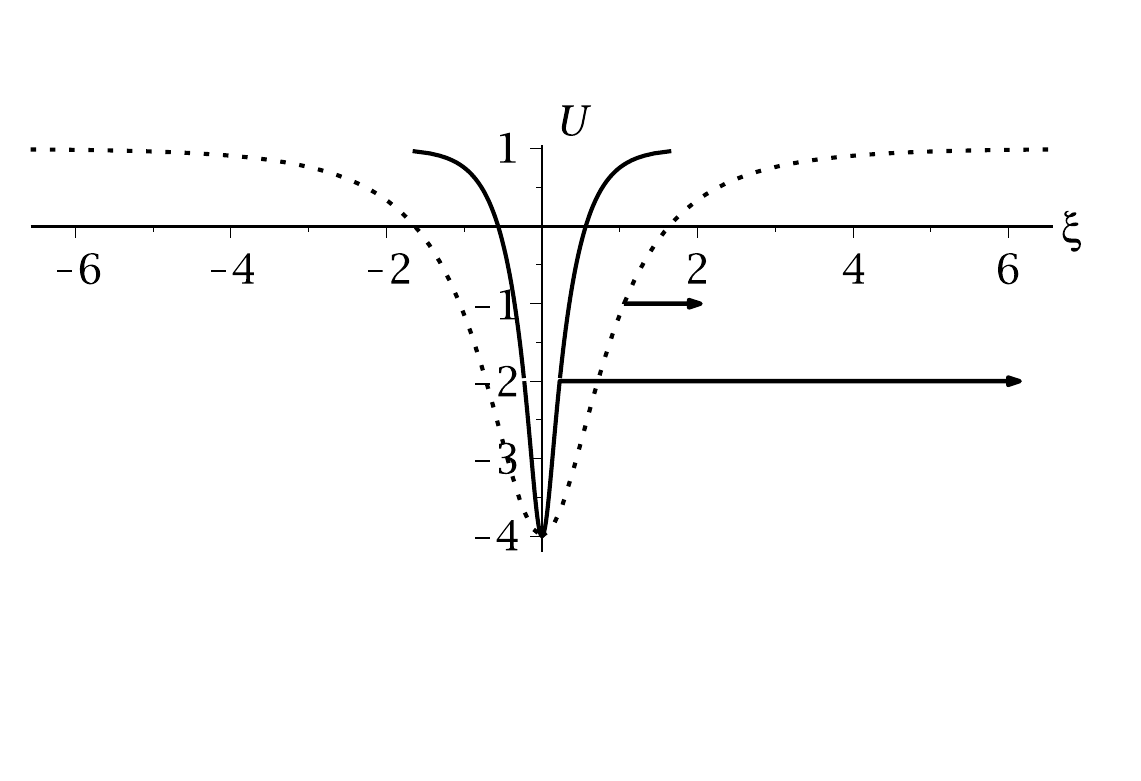}
\;\;
\includegraphics[trim=2cm 18cm 8cm 6cm,clip,width=0.58\textwidth]{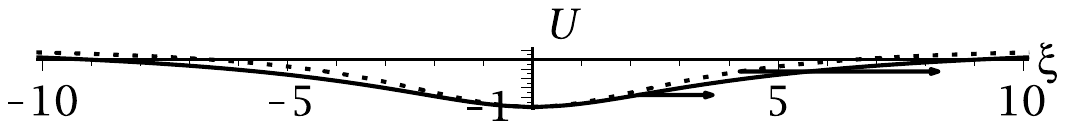}
\quad
\includegraphics[trim=2cm 19cm 8cm 6cm,clip,width=0.6\textwidth]{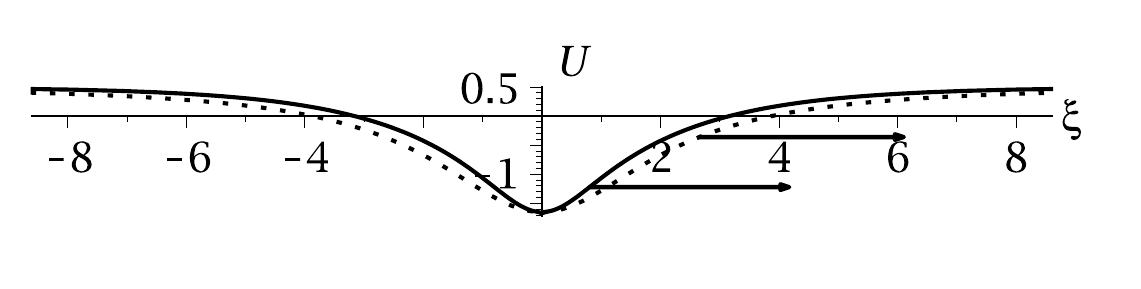}
\caption{Focusing-gKdV dark solitary waves on a positive background for $p=4$ (solid) and $p=2$ with same background and height (dot). Arrows indicate direction and speed.}\label{gkvdfig-pis4-focus-dark-solitary} 
\end{figure}

Expressions \eqref{gkdv-evenp-focus-solitary-bh-c} and \eqref{gkdv-evenp-focus-solitary-bh-w}
for the speed and the width of the solitary wave \eqref{gkdv-pis4-focus-darksolitary} are given by
\begin{equation}
c = b^4 -\tfrac{4}{3}b^3 h_- +b^2 h_-^2 -\tfrac{2}{5}b h_-^3 +\tfrac{1}{15}h_-^4,
\quad
w= 1/\sqrt{15h_- (h_-^3 -6bh_-^2 +15b^2h_- -20b^3)} .
\end{equation}  
Fig.~\ref{gkvdfig-pis4-focus-dark-solitary} 
compares this solitary wave solution to the focusing-mKdV dark solitary wave solution. 

\begin{figure}[h]
\centering
\includegraphics[trim=2cm 16cm 6cm 2cm,clip,width=0.48\textwidth]{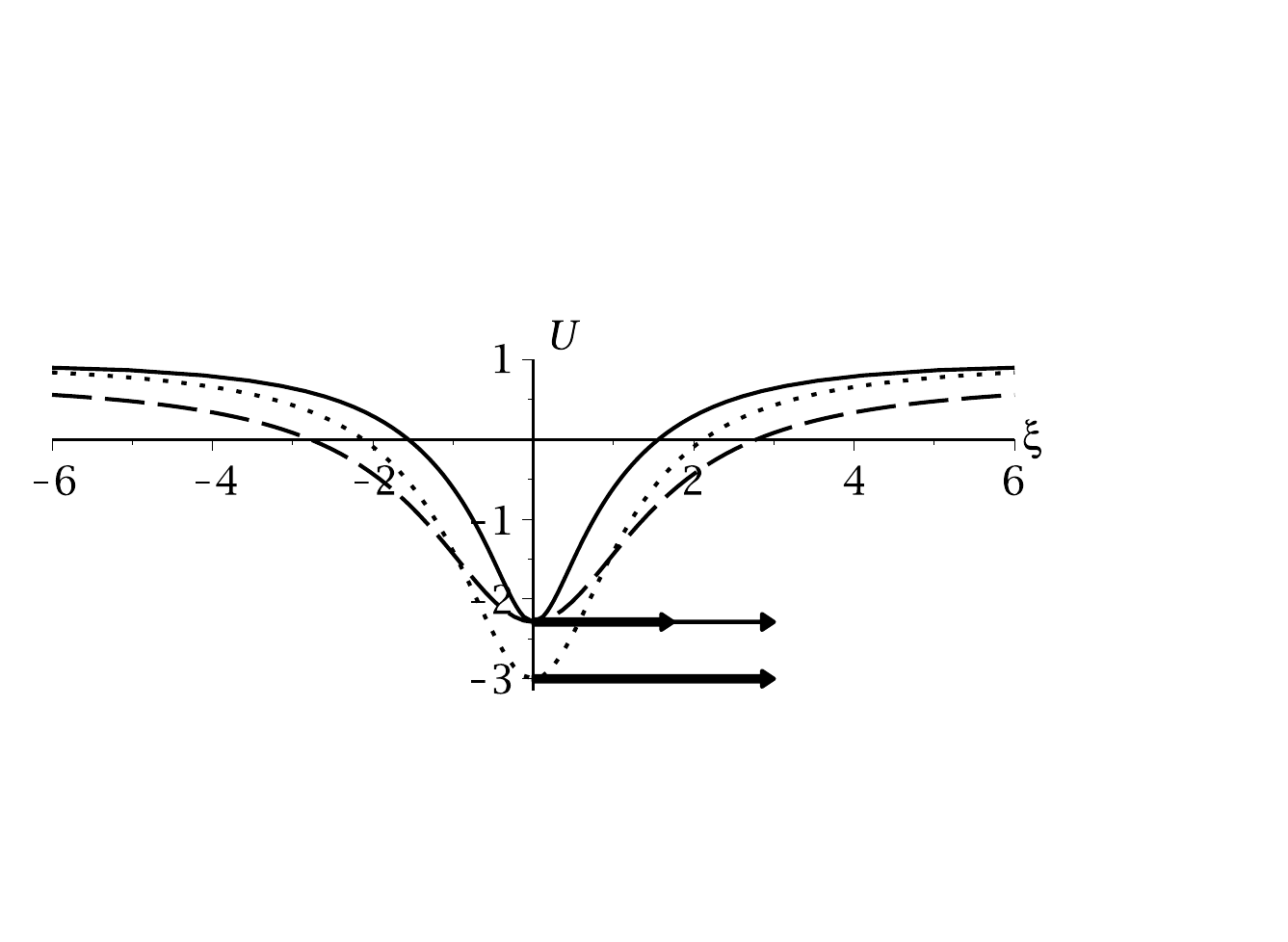}
\quad
\includegraphics[trim=2cm 16cm 6cm 2cm,clip,width=0.48\textwidth]{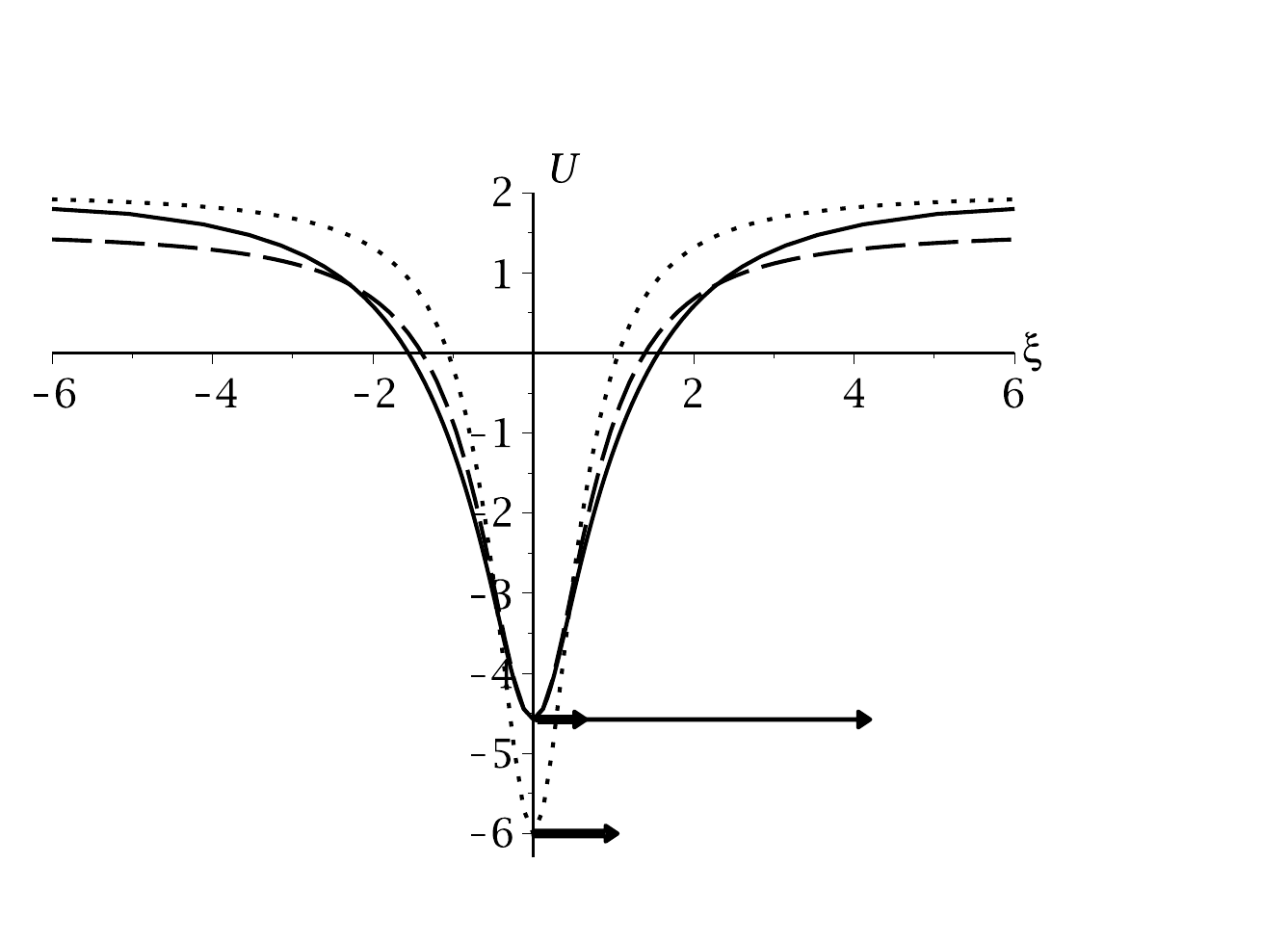}
\quad
\includegraphics[trim=2cm 18cm 6cm 6cm,clip,width=0.48\textwidth]{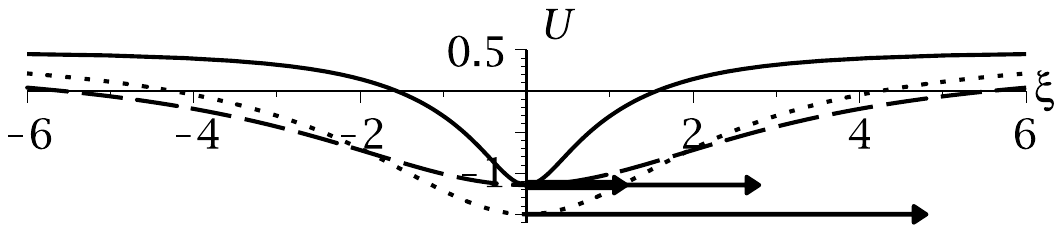}
\caption{gKdV heavy-tail waves on a positive background for $p=4$ (solid) and $p=2$ with same background (dot) and with same depth (dash). Arrows indicate direction and speed.}\label{gkvdfig-pis4-dark-heavytail}
\end{figure}

\begin{figure}[h]
\centering
\includegraphics[trim=2cm 16cm 6cm 2cm,clip,width=0.48\textwidth]{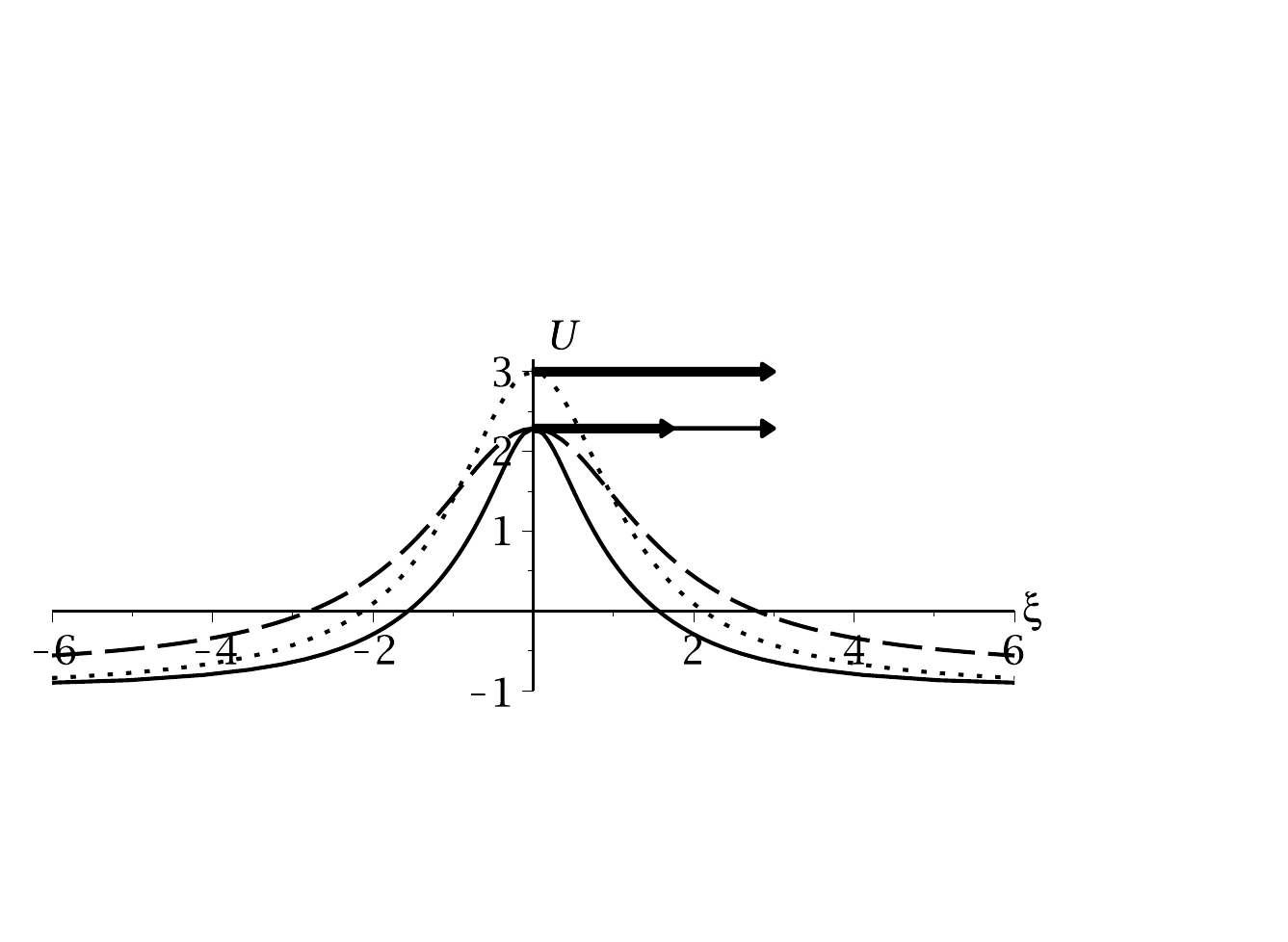}
\quad
\includegraphics[trim=2cm 16cm 6cm 2cm,clip,width=0.48\textwidth]{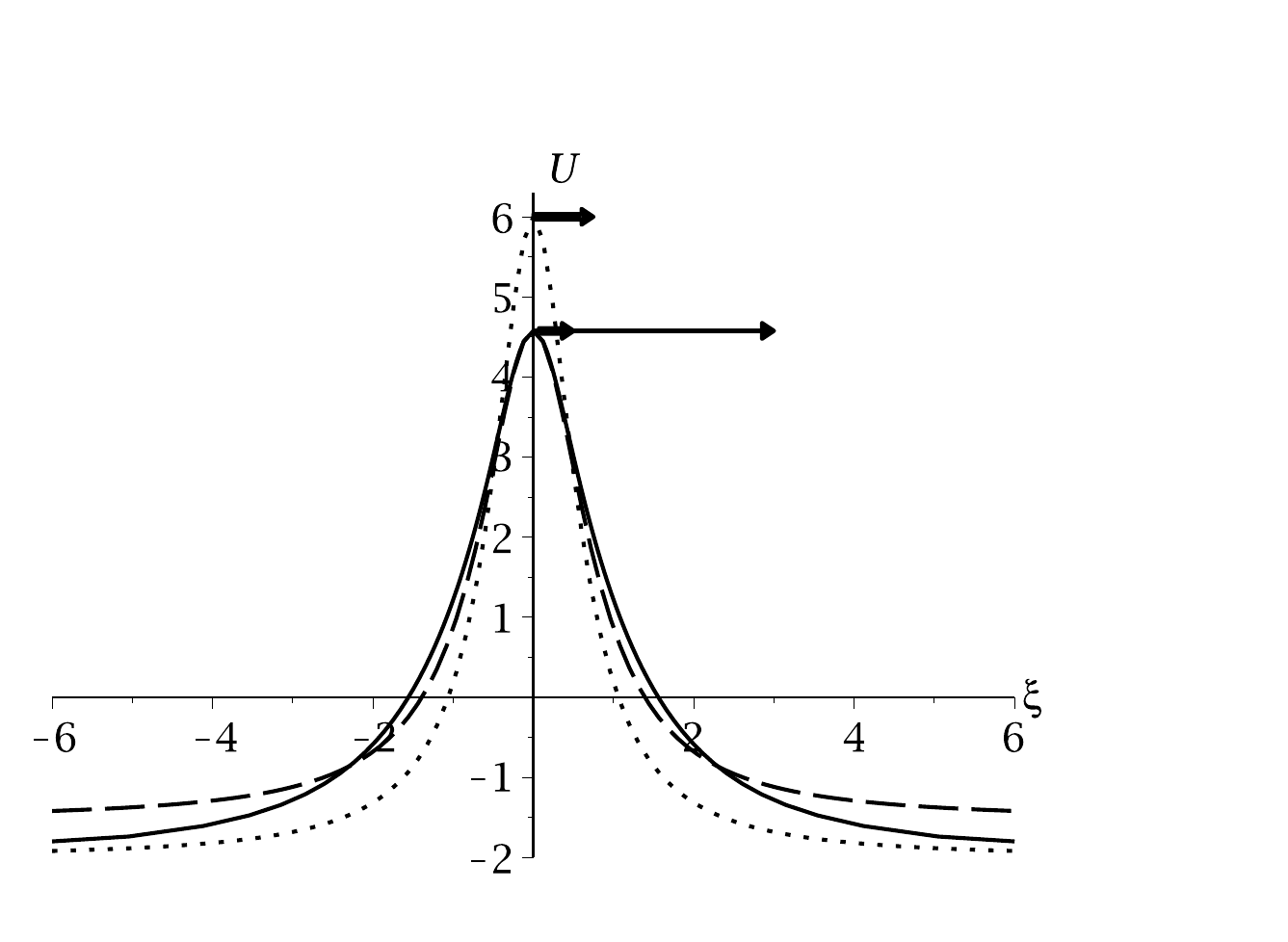}
\quad
\includegraphics[trim=2cm 18cm 6cm 6cm,clip,width=0.48\textwidth]{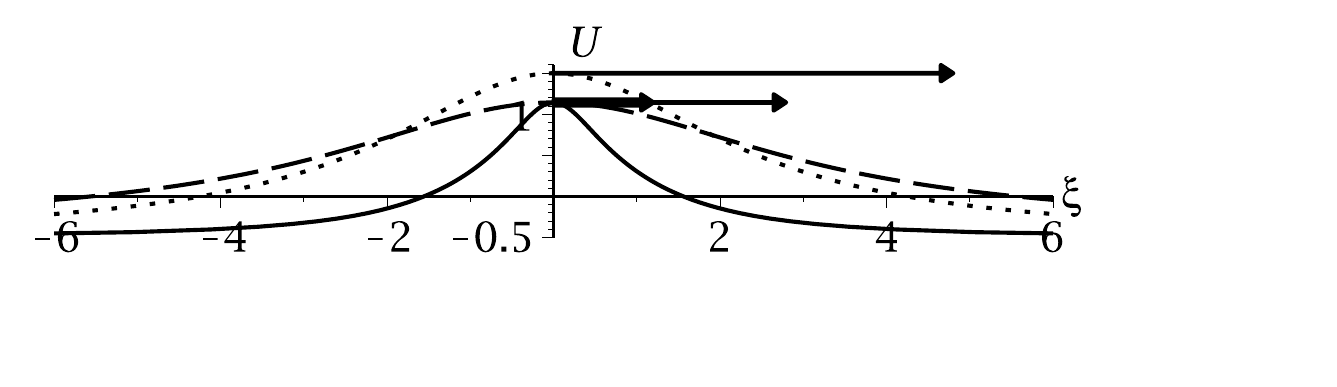}
\caption{gKdV heavy-tail waves on a negative background for $p=4$ (solid) and $p=2$ with same background (dot) and with same depth (dash). Arrows indicate direction and speed.}\label{gkvdfig-pis4-bright-heavytail}
\end{figure}

Next, for the dark/bright heavy-tail waves, we have respectively 
\begin{equation}\label{gkdv-pis4-heavytail-W}
W_\mp (U) = \tfrac{1}{30}(U^2 +(4b \mp h)U +10b^2 \mp 5bh+h^2),
\end{equation}
where
\begin{equation}\label{gkdv-pis4-heavytail-h}
h = (2+{\textstyle \sqrt[3]{\sqrt{10} +3}-\sqrt[3]{\sqrt{10}-3}})|b|,  
\end{equation}  
from equations \eqref{gkdv-evenp-heavytail-W} and \eqref{gkdv-evenp-focus-heavytail-h}.
By the same steps used for the solitary waves,
we can evaluate the respective quadratures \eqref{gkdv-evenp-focus-heavytail1-bh-quadrature} and \eqref{gkdv-evenp-focus-heavytail2-bh-quadrature}:
\begin{subequations}\label{gkdv-pis4-dark-heavytail}
\begin{align}
&\begin{aligned}
&
\frac{\sqrt{2} \rho_- \nu^2}{\tau\mu_2^2 \mu_-}\Bigg(
E\bigg(\sn^{-1}\bigg(\frac{ \mu_+(\kappa_- -\nu(U-b))}{ \mu_- (\kappa_+-\nu(U-b))}, \frac{-\mu_-^2\rho_+}{\mu_+^2\rho_-}\bigg),
\frac{-\mu_-^2\rho_+}{\mu_+^2\rho_-} \bigg)
-E\bigg(\frac{-\mu_-^2\rho_+}{\mu_+^2\rho_-}\bigg)  
\Bigg)
\\&\qquad
+\frac{\sqrt{8} \mu_1 \nu^2}{\rho_-\mu_2 \mu_-(\mu_1 +\mu_2)^2}\Bigg(
K\bigg(\frac{-\mu_-^2\rho_+}{\mu_+^2\rho_-}\bigg)
-\sn^{-1}\bigg( \frac{\mu_+ (\kappa_- -\nu(U-b))}{\mu_- (\kappa_+ -\nu(U-b))}, \frac{-\mu_-^2\rho_+}{\mu_+^2\rho_-} \bigg)
\Bigg)
\\&\qquad
-\frac{\nu}{\mu_2 \sqrt{\mu_1 \mu_2}\mu_+}
\cn^{-1}\bigg( \frac{\mu_+ (\kappa_- -\nu(U-b))}{\mu_- (\kappa_+ -\nu(U-b))}, \frac{\mu_-^2\rho_+}{2\mu_1\mu_2 \nu^2} \bigg)
\\&\qquad
+\frac{2\nu^3}{\tau\mu_2 \mu_-}
\frac{\sqrt{30(U-b+h)W_-(U)}}{\sqrt{b-U}(\nu(b-U)+\mu_2\mu_+)}
=\frac{1}{\sqrt{15}} |\xi|,
\end{aligned}
\\
&\begin{aligned}
& b-h\leq U\leq b,
\quad
b>0  ,
\end{aligned}
\\
&\begin{aligned}
&
\mu_1= \sqrt{3(5 b^2-4 b h+h^2)},
\quad
\mu_2 = \sqrt{15 b^2-6 b h+h^2},
\quad
\mu_\pm = \mu_1\pm\mu_2,
\\
&
\nu = 2h-6b,
\quad
\tau = \mu_1^2 +\mu_2^2 -6b(h-b),
\quad
\kappa_\pm = \mu_\pm \mu_1 -\nu h, 
\quad
\rho_\pm =\mu_1\mu_2 \pm 3b(h-b),
\end{aligned}
\end{align}
\end{subequations}
and
\begin{subequations}\label{gkdv-pis4-bright-heavytail}
\begin{align}
&\begin{aligned}
&
\frac{\sqrt{2} \rho_- \nu^2}{\tau\mu_2^2 \mu_-}\Bigg(
E\bigg(\sn^{-1}\bigg(\frac{\mu_+(\kappa_- -\nu(b-U))}{\mu_-(\kappa_+ -\nu(b-U))}, \frac{-\mu_-^2\rho_+}{\mu_+^2 \rho_-}\bigg), \frac{-\mu_-^2\rho_+}{\mu_+^2 \rho_-} \bigg)
-E\bigg(\frac{-\mu_-^2\rho_+}{\mu_+^2\rho_-}\bigg)  
\Bigg)
\\&\qquad
+\frac{\sqrt{8} \mu_1 \nu^2}{\rho_-\mu_2 \mu_-(\mu_1 +\mu_2)^2}\Bigg(
K\bigg(\frac{-\mu_-^2\rho_+}{\mu_+^2\rho_-}\bigg)
-\sn^{-1}\bigg(\frac{\mu_+(\kappa_- -\nu(b-U))}{\mu_-(\kappa_+ -\nu(b-U))}, \frac{-\mu_-^2\rho_+}{\mu_+^2\rho_-} \bigg)
\Bigg)
\\&\qquad
-\frac{\nu}{\mu_2 \sqrt{\mu_1 \mu_2}\mu_+}
\cn^{-1}\bigg( \frac{\mu_+ (\kappa_- -\nu(b-U))}{\mu_- (\kappa_+ -\nu(b-U))}, \frac{\mu_-^2\rho_+}{2\mu_1\mu_2 \nu} \bigg)
\\&\qquad
+\frac{2\nu^3}{\tau\mu_2 \mu_-}
\frac{\sqrt{30(b+h-U)W_+(U)}}{\sqrt{U-b}(\nu(U-b)+\mu_2\mu_+)}
=\frac{1}{\sqrt{15}} |\xi|,
\end{aligned}
\\
&\begin{aligned}
& b\leq U\leq b+h,
\quad
b<0  ,
\end{aligned}
\\
&\begin{aligned}
&
\mu_1= \sqrt{3(5 b^2 +4 b h+h^2)},
\quad
\mu_2 = \sqrt{15 b^2 +6 b h+h^2},
\quad
\mu_\pm = \mu_1\pm\mu_2,
\\
&
\nu = 2h+6b,
\quad
\tau = \mu_1^2 +\mu_2^2 +6b(h+b),
\quad
\kappa_\pm = \mu_\pm \mu_1 -\nu h, 
\quad
\rho_\pm =\mu_1\mu_2 \mp 3b(h+b),
\end{aligned}
\end{align}
\end{subequations}
involving the additional elliptic function $E$ and the complete elliptic integrals $E$ and $K$.
This yields implicit algebraic expressions for the dark and bright heavy-tail wave solutions $U(\xi)$, respectively, 
on a background $b\gtrless 0$ with depth/height $h>0$. 
For both solutions, the speed is given by 
\begin{equation}\label{gkdv-pis4--heavytail-c}
c=b^4 >0
\end{equation}
and the tails have the asymptotic form
\begin{equation}\label{gkdv-pis4--heavytail-asympt}
U\simeq b - 3/(b^3\xi^2)
\end{equation}
for large $|\xi|$.
Figs.~\ref{gkvdfig-pis4-dark-heavytail} and~\ref{gkvdfig-pis4-bright-heavytail}
compare the solutions \eqref{gkdv-pis4-dark-heavytail} and \eqref{gkdv-pis4-bright-heavytail} 
to the mKdV heavy-tail wave solutions. 

\section{Conclusions}\label{sec:conclusion}

For the $p$-power gKdV equation \eqref{gkdv},
we have presented a complete classification of all types of non-periodic travelling wave solutions with non-zero boundary conditions.
This classification is summarized in Table~\ref{table:types}.

In addition to solitary waves on a non-zero background,
which exist for all even and odd powers, 
there are static humps on a non-zero background for all odd powers,
heavy-tail waves (with power decay) on a non-zero background for all even powers in the focusing case,
and kink (shock) waves for all even powers in the defocusing case. 

We have studied the main kinematic features of each type of travelling wave
by employing a physical parameterization in terms of
the wave speed $c$, the background $b$, and the wave height/depth $h$.
We have also derived analytical formulas for these travelling waves
for $p=3,4$, in terms of elliptic functions,
and compared them to the corresponding waves known for the integrable cases $p=1,2$, 
which are expressions given in terms of elementary functions.

The waves obtained for $p=3,4$, which are new solutions,  
can be used as the starting point 
for an investigation of wave modulation theory (see, e.g., \Ref{Mar})
and also for a study of waves in mean field theory (see, e.g., \Ref{MaiAndFraElHoe})
and soliton gas theory (see, e.g., \Ref{ShuPel,CarDutEl}). 

For future work,
it will be interesting to study interactions of
these non-zero background waves for the KdV and mKdV equations \eqref{gkdv}
respectively with $p=1,2$.
In particular, how KdV solitary waves interact with static humps
on a non-zero background,
and how mKdV solitary waves interact with heavy-tailed mKdV waves
on a non-zero background.
A worthwhile well-known remark is that,
under a shift of $u\to u-b$
followed by a Galilean boost $(x,t)\to (x-\alpha bt,t)$,
KdV solutions on a non-zero background $b\neq 0$
are transformed to zero-background solutions,
while mKdV solutions on a non-zero background $b\neq 0$ are transformed
to zero-background solutions of the Gardner equation 
$u_t +\alpha uu_x +\beta u^2u_x + \gamma u_{xxx}=0$.
However, this transformation sends $c\to c-\alpha b$,
and so for the purpose of understanding physical properties of non-zero background solutions, 
it is easier to work with them by using a parameterization in terms of the physical (untransformed) speed $c$ and background $b$.

\appendix
\section{}

\subsection{Useful polynomial functions}

For $n\in\mathbb{Z}^+$,
consider the polynomials \eqref{Spoly} and \eqref{Rpoly}:
\begin{subequations}
\begin{align}
& S_n(z) =\sum_{j=1}^{n} z^{n-j} =\sum_{j=0}^{n-1} z^{j} 
\\
& R_n(z) = \sum_{j=1}^{n} j z^{n-j} = \sum_{j=0}^{n-1} (n-j) z^{j} .
\end{align}
\end{subequations}
The main properties of these two polynomials
needed for deriving the gKdV solitary waves
will now be summarized.

The two polynomials are related by the identity 
\begin{equation}\label{SR-identity}
S_n(z)-n = (z-1)R_{n-1}(z) . 
\end{equation}
They have the values 
\begin{equation}
S_n(1) = n,
\quad
R_n(1) = \tfrac{1}{2}n(n+1) . 
\end{equation}
For $z\neq 1$, they can be expressed in the rational form
\begin{subequations}
\begin{align}
& S_n(z) = (z^n-1)/(z-1) , 
\\
& R_n(z) = (z^{n+1} -(n+1)z+ n)/(z-1)^2 . 
\end{align}
\end{subequations}

When $n$ is even:
$R_n(z)$ is a monotonic increasing function;
it has a root in the interval $-2\leq z<-1$, with equality holding for $n=2$. 
The related function $R_n(z)-R_n(1)$ has no root other than $z=1$. 

When $n$ is odd:
$R_n(z)$ is a positive convex function;
its minimum is $R_n(-1)=\tfrac{1}{2}(n+1)=\tfrac{1}{n}R_n(1)$. 
The related function $R_n(z)-R_n(1)$ has a negative root in the interval $-3\leq z<-1$,
with equality holding for $n=3$. 

Some multi-variable symmetric generalizations of $R_n$ will also be needed. 
\begin{align}
\label{doubleR-identity}
&\begin{aligned}
R_n(y,z) & = \sum_{j=0}^{n-1} (n-j) \sum_{i=0}^{j} y^{i}z^{j-i}
= \sum_{j=0}^{n-1} y^j R_{n-j}(z)
= \sum_{j=0}^{n-1} z^j R_{n-j}(y) 
\\
& = (R_{n+1}(y)-R_{n+1}(z))/(y-z) , 
\end{aligned}
\\
\label{tripleR-identity}
&\begin{aligned}
R_n(x,y,z) & = \sum_{k=0}^{n-1} (n-k)\sum_{j=0}^{k} \sum_{i=0}^j x^i y^{j-i} z^{k-j}
= \sum_{j=0}^{n-1} x^j R_{n-j}(y,z)
\\
& = (R_{n+1}(x,y)-R_{n+1}(x,z))/(y-z) . 
\end{aligned}
\end{align}
They satisfy
\begin{align}
& R_n(z,z) = (nR_{n+1}(z)-(n+2)zR_{n-1}(z))/(z-1)^3,
\\
& R_n(-1,-1) = \tfrac{1}{4}(1-(-1)^n)(n+1) , 
\end{align}
and
\begin{align}
& \begin{aligned}
R_n(1,y,z) & = \sum_{j=0}^{n-1} y^j R_{n-j}(1,z) = \sum_{j=0}^{n-1} z^j R_{n-j}(1,y)
\\
& = (R_{n+1}(1,y)-R_{n+1}(1,z))/(y-z)
\\
& = \big( (R_{n+2}(y)-R_{n+2}(1))/(y-1) - (R_{n+2}(z)-R_{n+2}(1))/(z-1) \big)/(y-z) , 
\end{aligned}
\\
& \begin{aligned}  
R_n(x,-1,-1) & = \sum_{j=0}^{n-1} \tfrac{1}{4}(1-(-1)^{n-j})(n-j+1) x^j
\\
& =\tfrac{1}{4}(R_{n+1}(x)-(-1)^n R_{n+1}(-x)) . 
\end{aligned}    
\end{align}

\subsection{Factorization of potential}

Consider $V(b)-V(U)$.
First, it can be expressed in the factored form 
\begin{equation}
\begin{aligned}
V(b) - V(U) & =
\tfrac{\sigma}{(p+1)(p+2)} (b^{p+2}-U^{p+2})
-\tfrac{1}{2}c(b^2-U^2) - M(b-U)
\\
& = (b-U)\big( \tfrac{\sigma}{(p+1)(p+2)} b^{p+1}S_{p+2}(U/b) -\tfrac{1}{2}c(b + U) - M \big) . 
\end{aligned}
\end{equation}
Next, the terms in brackets can be simplified
through the critical point equation \eqref{gkdv-V'} for $b$:
\begin{equation}
M + c b - \tfrac{\sigma}{p+1} b^{p+1}=0 . 
\end{equation}
This gives, for the second and third terms in the brackets, 
\begin{equation}
\tfrac{1}{2}c(b + U) + M = \tfrac{1}{2}c(U-b) +\tfrac{\sigma}{p+1} b^{p+1} . 
\end{equation}  
When this expression is combined with the first term in the brackets,
it leads to 
\begin{equation}
\tfrac{\sigma}{(p+1)(p+2)}b^{p+1}\big( S_{p+2}(U/b) -(p+2) \big) +\tfrac{1}{2}c(b-U)
 = \tfrac{\sigma}{(p+1)(p+2)} b^{p}(b-U)\big( \tilde c - R_{p+1}(U/b) \big)
\end{equation}
by identity \eqref{SR-identity}, 
with
\begin{equation}\label{c-rel}
 \tilde c = \tfrac{\sigma(p+1)(p+2)}{2}c/b^p . 
\end{equation}  
This yields the second factorization 
\begin{equation}
V(b) - V(U) =
\tfrac{\sigma}{(p+1)(p+2)}b^p (b-U)^2\big( \tilde c - R_{p+1}(U/b) \big)
\end{equation}
where $\tilde c - R_{p+1}(U/b)$ is a polynomial of degree $p$ in $U/b$.

If $p$ is odd, in which case $\sigma=1$, 
then $\tilde c - R_{p+1}(U/b)$ has a root $U=h+b$ with $h>0$. 
Hence
\begin{equation}
\begin{aligned}
\tilde c - R_{p+1}(U/b)
& = R_{p+1}(1+h/b)- R_{p+1}(U/b)
\\
& = (1+(h-U)/b) R_p(1+h/b,U/b)
\\
& = (h+b-U)b^{-1}\tilde W(U/b)
\end{aligned}
\end{equation}
by identity \eqref{doubleR-identity}, 
where
\begin{equation}
\tilde W(U/b) = R_{p}(1+h/b,U/b)  = \sum_{i=0}^{p-1} R_{p-i}(1+h/b) (U/b)^{i} . 
\end{equation}
Then the final factorization in this case is given by
\begin{equation}\label{oddp-Vfactored}
V(b) - V(U) = (b-U)^2(h+b -U)\tfrac{1}{(p+1)(p+2)}b^{p-1}\tilde W(U/b)
\end{equation}
with the factor $b^{p-1}=|b|^{p-1}$ being positive, since $p$ is odd.
As implied by the shape of $V(U)$ shown in Section~\ref{subsec:gkdv-odd},
here $\tilde W(U/b)$ will be a positive function for $U$ in the interval $[b,b+h]$ iff $V''(b)<0$.
This condition can be expressed as
\begin{equation}
\begin{aligned}
0<-V''(b)
& = h\tfrac{2}{(p+1)(p+2)}b^{p-1}R_{p}(1+h/b,1)
\\
& = \tfrac{2}{(p+1)(p+2)} |b|^{p-1}b\big(R_{p+1}(1+h/b)-R_{p+1}(1)\big)
\\
& = |b|^{p-1}b(\tfrac{2}{(p+1)(p+2)}\tilde c -1)
\end{aligned}
\end{equation}
through the relation \eqref{doubleR-identity},
where
\begin{equation}
\tilde c  = R_{p+1}(1+h/b) . 
\end{equation}
Note the relation \eqref{c-rel} yields
\begin{equation}
c = \tfrac{2}{(p+1)(p+2)}b|b|^{p-1}\tilde c  . 
\end{equation}

If $p$ is even and $\sigma =1$ (focusing case), 
then $\tilde c - R_{p+1}(U/b)$ has a pair of roots $U=b+h_+$ and $U=b-h_-$
with $h_\pm>0$. 
Hence
\begin{equation}\label{focusing-even-p-factored}
\tilde c - R_{p+1}(U/b) = (b+h_+-U)(U +h_--b) b^{-2}\tilde W(U/b)
\end{equation}
where $\tilde W(U/b)$ is obtained by iterating the steps used in the odd $p$ case. 
First, for the root $U=b+h_+$:
\begin{equation}
\tilde c - R_{p+1}(U/b) = (1+(h_+-U)/b) R_p(1+h_+/b,U/b) . 
\end{equation}
Next, for the root $U=b-h_-$:
\begin{equation}
\begin{aligned}
R_p(1+h_+/b,U/b) & = R_p(1+h_+/b,U/b) - R_p(1+h_+/b,1-h_-/b)
\\
& = -(1-(h_-+U)/b) R_{p-1}(1+h_+/b,1-h_-/b,U/b)
\end{aligned}
\end{equation}
by the identity \eqref{tripleR-identity}.
This yields the factorization \eqref{focusing-even-p-factored}
where
\begin{equation}
\tilde W(U/b) = R_{p-1}(1+h_+/b,1-h_-/b,U/b)
\end{equation}
with $h_\pm$ related to $b$ via the polynomial equation
\begin{equation}\label{hphm-rel}
R_{p+1}(1+h_+/b,1-h_-/b)=0 . 
\end{equation}
Then the final factorization is given by
\begin{equation}\label{evenp-focus-Vfactored}
V(b) - V(U) = (b-U)^2 (b+h_+-U)(U +h_--b)\tfrac{1}{(p+1)(p+2)}b^{p-2}\tilde W(U/b)
\end{equation}
with the factor $b^{p-2}=|b|^{p-2}$ being positive, since $p$ is even.
From the shape of $V(U)$ shown in Section~\ref{subsec:gkdv-even},
it follows that $\tilde W(U/b)$ will be a positive function for $U$ in the interval $[b-h_-,b+h_+]$
iff $V''(b)<0$.
This condition can be expressed as 
\begin{equation}
\begin{aligned}
0<-V''(b)
& = h_+h_-\tfrac{2}{(p+1)(p+2)}b^{p-1}R_{p-1}(1+h_+/b,1-h_-/b,1)
\\
& = h_+h_-\tfrac{2}{(p+1)(p+2)} b^{p-1}
\big( (R_{p+1}(1+h_+/b)-R_{p+1}(1))/(h_+/b)
\\&\qquad
- (R_{p+1}(1-h_-/b)-R_{p+1}(1))/(-h_-/b) \big)/((h_++h_-)/b)
\\
& = \tfrac{2}{(p+1)(p+2)} b^{p} \big( R_{p+1}(1\pm h_\pm/b) -R_{p+1}(1) \big)
\\
& = |b|^{p}(\tfrac{2}{(p+1)(p+2)}\tilde c -1) 
\end{aligned}
\end{equation}
through the relation \eqref{doubleR-identity},
where
\begin{equation}
\tilde c = R_{p+1}(1\pm h_\pm/b) . 
\end{equation}
Note the relation \eqref{c-rel} yields
\begin{equation}
c = \tfrac{2}{(p+1)(p+2)}|b|^p\tilde c . 
\end{equation}

If $p$ is even and $\sigma =-1$ (defocusing case), 
then $\tilde c - R_{p+1}(U/b)$ has a pair of roots $U=U_-$ and $U=U_+$
with $U_-<U_+$. 
As a consequence of the shape of $V(U)$ shown in Section~\ref{subsec:gkdv-even},
the relevant factorization of $\tilde c - R_{p+1}(U/b)$
uses only one of these roots:
$U=U_+= b-h_+$ when $\sgn(M)=\sgn(b)=1$,
and
$U=U_-=b+h_-$ when $\sgn(M)=\sgn(b)=-1$,
where $h_\pm>0$,
with 
\begin{equation}
\tilde c = R_{p+1}(U_\pm/b) = R_{p+1}(1\mp h_\pm/b) . 
\end{equation}
Hence, the factorization is given by 
\begin{equation}
\begin{aligned}
\tilde c - R_{p+1}(U/b)
& = R_{p+1}(1\mp h_\pm/b) - R_{p+1}(U/b)
\\
& = (1 +(\mp h_\pm -U)/b) R_{p}(1\mp h_\pm/b,U/b)  . 
\end{aligned}
\end{equation}
Since $\sigma=-1$ and $\sgn(b)=\pm 1$, 
the final factorization is given by
\begin{equation}\label{evenp-defocus-Vfactored}
\begin{aligned}
V(b) - V(U) 
& =-(b-U)^2 \tfrac{1}{(p+1)(p+2)}b^{p-1}(b \mp h_\pm -U) \tilde W(U/b)
\\
& = \tfrac{1}{(p+1)(p+2)}|b|^{p-1}(b-U)^2 (h_\pm \mp(b-U)) \tilde W(U/b)
\end{aligned}
\end{equation}
where
\begin{equation}\label{evenp-defocus-tilW}
\tilde W(U/b) =
R_{p}(1\mp h_\pm/b,U/b)  . 
\end{equation}
As implied by the shape of $V(U)$, 
this function $\tilde W(U/b)$ will be positive for $U$ in the interval
$[b-h_+,b]$ or $[b,b+h_-]$ corresponding to $\sgn(M)=\pm 1$, 
iff $V''(b)<0$. 
This condition can be expressed as
\begin{equation}
\begin{aligned}
0>-V''(b)
& = -h_\pm\tfrac{2}{(p+1)(p+2)}|b|^{p-1} R_{p}(1\mp h_\pm/b,1)
\\
& = -h_\pm\tfrac{2}{(p+1)(p+2)} |b|^{p-1} 
\big( (R_{p+1}(1\mp h_\pm/b)-R_{p+1}(1))/(\mp h_\pm/b) \big)
\\
& = \tfrac{2}{(p+1)(p+2)} |b|^{p} \big( R_{p+1}(1\mp h_\mp/b) -R_{p+1}(1) \big)
\\
& = |b|^{p}(\tfrac{2}{(p+1)(p+2)}\tilde c -1)
\end{aligned}
\end{equation}
through the relation \eqref{doubleR-identity}. 
Note that, in both cases, 
\begin{equation}
c = -\tfrac{2}{(p+1)(p+2)}|b|^{p}\tilde c 
\end{equation}
from the relation \eqref{c-rel}.

A worthwhile remark is that the polynomial $R_{p}(1\mp h_\pm/b,U/b)$
in this last factorization \eqref{evenp-defocus-Vfactored}
has the other root in the pair,
which is given by $U=U_\mp = b \mp h_\mp$ when $\sgn(M)=\sgn(b)=\pm 1$,
due to the reflection symmetry \eqref{even-p-V-reflect} of the potential. 
This yields the further factorization
\begin{equation}
R_p(1\mp h_\pm/b,U/b) = -(1+(\mp h_\mp -U)/b) R_{p-1}(1\mp h_\mp/b,1 \mp h_\pm/b,U/b)
\end{equation}
similarly to the focusing case,
giving 
\begin{equation}
\begin{aligned}
\tilde c - R_{p+1}(U/b)
= -(1 +(\mp h_\pm -U)/b) (1+(\mp h_\mp -U)/b) R_{p-1}(1\mp h_\mp/b,1 \mp h_\pm/b,U/b) 
\end{aligned}
\end{equation}
where $h_\pm$ is related to $b$ by the polynomial equation \eqref{hphm-rel}. 
Then the complete final factorization is given by 
\begin{equation}\label{evenp-defocus-Vfullyfactored}
\begin{aligned}
V(b) - V(U) 
& =\tfrac{1}{(p+1)(p+2)}|b|^{p-2}(b-U)^2(b \mp h_\pm -U) (b \mp h_\mp -U)\hat W(U/b)
\end{aligned}
\end{equation}
where
\begin{equation}
\hat W(U/b) =
R_{p-1}(1\mp h_\mp/b,1 \mp h_\pm/b,U/b) . 
\end{equation}
This function shares the same positivity property as the function \eqref{evenp-defocus-tilW}. 

The factorization \eqref{evenp-defocus-Vfullyfactored} is especially useful
in the case $M=0$, which has $h_+=h_-=2|b|$. In this case,
\begin{equation}
\hat W(U/b) = R_{p-1}(-1,-1,U/b) =\tfrac{1}{4}(R_{p}(U/b)+R_{p}(-U/b)) . 
\end{equation}

\end{document}